\documentclass[fleqn,usenatbib]{mnras}

\usepackage{amssymb}
\usepackage{amsmath}
\usepackage{latexsym}
\usepackage{deluxetable}
\usepackage{scrextend}
\usepackage{scalerel}
\usepackage{hyperref}
\usepackage{tabularx}
\usepackage{xspace}
\usepackage{enumerate}
\usepackage{graphicx}
\usepackage{pdflscape}
\usepackage{url}
\usepackage{longtable}
\usepackage{rotating}
\usepackage{caption}
\usepackage{comment}
\usepackage{ragged2e}

\newcommand{\eal}[2]{\ifmmode{\mathrm{#1\,#2}}\else{#1\textsc{$\,$\lowercase{#2}}}\fi\xspace}
\newcommand{\feal}[2]{\ifmmode{\mathrm{#1\,#2}}\else{[#1\textsc{$\,$\lowercase{#2}}]}\fi\xspace}
\newcommand{\hfeal}[2]{\ifmmode{\mathrm{#1\,#2}}\else{#1\textsc{$\,$\lowercase{#2}}]}\fi\xspace}

\usepackage{xcolor}        
\definecolor{tabblue}{HTML}{1F77B4}
\definecolor{taborange}{HTML}{FF7F0E}
\definecolor{tabgreen}{HTML}{2CA02C}
\definecolor{tabred}{HTML}{D62728}
\definecolor{tabpurple}{HTML}{9467BD}
\definecolor{tabbrown}{HTML}{8C564B}
\definecolor{tabpink}{HTML}{E377C2}
\definecolor{tabgray}{HTML}{7F7F7F}
\definecolor{tabolive}{HTML}{BCBD22}
\definecolor{tabcyan}{HTML}{17BECF}
\newcommand{\kvs}[1]{#1}
\newcommand{\hl}{}
\newcommand{\nullstat}{}

\title[$\gamma$-ray Properties of Novae]
{What determines the $\gamma$-ray luminosities of classical novae?}
\author[P.\ Craig et al.]{Peter Craig$^{1}$\thanks{E-mail: craigpe1@msu.edu},
Elias Aydi$^{2,1}$\thanks{E-mail: eaydi@ttu.edu},
Laura Chomiuk$^{1}$,
Ashley Stone$^{3}$,
Jay Strader$^{1}$,
Atticus Chong$^{1}$,
\newauthor
Kwan-Lok Li$^{4}$,
Jhih-Ling Fan$^{4}$,
Arash Bahramian$^{5}$,
David A.~H.\ Buckley$^6$,
Luca Izzo$^{7}$,
\newauthor
Adam Kawash$^{1}$,
Brian D.\ Metzger$^{8,9}$,
Koji Mukai$^{10,11}$,
Justin D.\ Linford$^{12}$,
Marina Orio$^{13,14}$,
\newauthor
J. L.\ Sokoloski$^{8}$,
Kirill V.\ Sokolovsky$^{15}$,
Evangelia Tremou$^{12}$, 
Frederick M.\ Walter$^{16}$,
\newauthor
Joan Guarro Fl\'o,$^{17}$
Christophe Boussin,$^{18}$
St\'ephane Charbonnel,$^{19}$
Olivier Garde,$^{20}$
\newauthor
Konstantin Belyakov,$^{21,22}$
Libert A. G. Monard,$^{21,23,24}$
Franz-Josef Hambsch,$^{21,25,26,27}$
\newauthor
and Neil Thomas$^{21,28}$\\
\\
$^{1}$Center for Data Intensive and Time Domain Astronomy, Department of Physics and Astronomy, Michigan State University,\\\ East Lansing, MI 48824, USA\\
$^2$Department of Physics and Astronomy, Texas Tech University, Lubbock, TX 79409, USA\\
$^3$Department of Physics and Astronomy, West Virginia University, Morgantown, WV 26506-6315, USA\\
$^{4}$Department of Physics, National Cheng Kung University,
No.\ 1 University Road, Tainan City 70101, Taiwan\\
$^5$International Centre for Radio Astronomy Research, Curtin University, GPO Box U1987, Perth, WA 6845, Australia\\
$^6$South African Astronomical Observatory, P.O.\ Box 9, 7935 Observatory, South Africa\\
$^{7}$INAF-Osservatorio Astronomico di Capodimonte, Salita Moiariello 16, 80131, Napoli, Italy\\
$^8$Columbia Astrophysics Laboratory, Columbia University, New York, NY 10027, USA\\
$^{9}$Center for Computational Astrophysics, Flatiron Institute, 162 5th Ave, New York, NY 10010, USA\\
$^{10}$Center for Space Science and Technology, University of Maryland Baltimore County, Baltimore, MD 21250, USA\\
$^{11}$CRESST and X-ray Astrophysics Laboratory, NASA/GSFC, Greenbelt MD 20771 USA\\
$^{12}$National Radio Astronomy Observatory, P.O. Box O, Socorro, NM 87801, USA\\
$^{13}$INAF Astronomical Observatory of Padova, 36012 Asiago (VI), Italy\\
$^{14}$INAF--Osservatorio di Padova, vicolo dell'Osservatorio 5, I-35122 Padova, Italy\\
$^{15}$Department of Astronomy, University of Illinois at Urbana-Champaign, 1002 W. Green Street, Urbana, IL 61801 USA\\
$^{16}$Department of Physics and Astronomy, Stony Brook University, Stony Brook, New York 11794, USA\\
$^{17}$Piera Remote Observatory, C/. Balmes 2, 08784 PIERA (Barcelona)\\
$^{18}$Observatoire de l'Eridan et de la Chevelure de B\'{e}r\'{e}nice, F-02400 Epaux-B\'{e}zu, France\\
$^{19}$Durtal Observatory, Durtal, France\\
$^{20}$Observatoire de la Tourbi\`{e}re, 45 Chemin du Lac - 38690 Chabons - France\\
$^{21}$AAVSO, 185 Alewife Brook Parkway, Suite 410, Cambridge, MA 02138, USA\\
$^{22}$Parallax LTD, Kazan, Russia\\
$^{23}$Bronberg Observatory, Pretoria, Gauteng, South Africa\\
$^{24}$Kleinkaroo Observatory, Calitzdorp, Western Cape, South Africa\\
$^{25}$Vereniging Voor Sterrenkunde (VVS), Zeeweg 96, B-8200 Brugge, Belgium\\
$^{26}$Bundesdeutsche Arbeitsgemeinschaft f\"{u}r Ver\"{a}nderliche Sterne, Munsterdamm 90, D-12169 Berlin, Germany\\
$^{27}$Groupe Europ\'{e}en d'Observations Stellaires (GEOS), 23 Parc de Levesville, 28300 Bailleau l'Ev\`{e}que, France\\
$^{28}$Department of Astronautical Engineering, United States Air Force Academy, CO 80840, USA
}

\begin{document}
\label{firstpage}
\pagerange{\pageref{firstpage}--\pageref{lastpage}}
\maketitle

\clearpage

\begin{abstract}
Classical novae in the Milky Way have now been well-established as high-energy GeV $\gamma$-ray sources. In novae with main-sequence companions, this emission is believed to result from shocks internal to the nova ejecta, as a later fast wind collides with an earlier slow outflow. To test this model and constrain the $\gamma$-ray production mechanism, we present a systematic study of a sample of recent Galactic novae, comparing their $\gamma$-ray properties ($\gamma$-ray luminosity and duration) with their outflow velocities, peak $V$-band magnitudes, and the decline times of their optical light curves ($t_2$). We uniformly estimate distances in a luminosity-independent manner, using spectroscopic reddening estimates combined with three-dimensional Galactic dust maps. Across our sample, $\gamma$-ray luminosities ($>$100 MeV) vary by three orders of magnitude, spanning $10^{34}-10^{37}$ erg s$^{-1}$. Novae with larger velocity of the fast outflow (or larger differential between the fast and slow outflow) have larger $\gamma$-ray luminosities, but are detectable for a shorter duration. The optical and $\gamma$-ray fluxes are correlated, consistent with substantial thermal emission in the optical from shock-heated gas. Across six novae with $\gamma$-ray and infrared light curves, evidence for dust formation appears soon after the end of the detected $\gamma$-ray emission. Dusty and non-dusty novae appear to have similar $\gamma$-ray luminosities, though novae that have more material processed by the shocks may be more likely to form dust. We find that the properties of the $\gamma$-ray emission in novae depend heavily on the ejecta properties, and are consistent with expectations for internal shocks.
\end{abstract}


\section{Introduction}
\label{sec:intro}

Classical novae are common transient events (around 10--15 are discovered in the Galaxy per year)\footnote{https://asd.gsfc.nasa.gov/Koji.Mukai/novae/novae.html} powered by a thermonuclear runaway on the surface of an accreting white dwarf in an interacting binary system (e.g. \citealt{Starrfield_etal_2008,2016PASP..128e1001S,Chomiuk_etal_2021}). The eruption leads to an increase in the optical brightness of the system by 8 to 15 magnitudes and the ejection of at least part of the accreted envelope ($10^{-7}-10^{-3}$~M$_{\odot}$) with velocities ranging between $\sim$ 200\,--\,10,000\,km\,s$^{-1}$ \citep{Payne-Gaposchkin_1957,Gallaher_etal_1978,Yaron_etal_2005}. Emission from remnant nuclear burning on the surface of the white dwarf diffuses through the ejecta and peaks initially in the visible range. Then, as the ejecta expand, the peak of the spectral energy distribution moves from the visible to the ultraviolet and eventually to the soft X-rays, as the ejecta become transparent to the ongoing nuclear burning on the surface of the white dwarf \citep{Page_etal_2015,Chomiuk_etal_2021}.

The detections of high-energy GeV $\gamma$-ray emission from 21 novae by the Large Area Telescope \citep[LAT;][]{2009ApJ...697.1071A} on the \textit{Fermi} $\gamma$-ray satellite \citep{Ackermann_etal_2014,Cheung_etal_2016,Franckowiak_etal_2018} have demonstrated that this standard picture is not sufficient to fully explain nova emission. This discovery highlighted the presence of energetic shocks during the eruption, responsible for accelerating particles to relativistic speeds and producing the non-thermal high-energy GeV emission. The $\gamma$-ray emission typically begins around the optical peak, indicating that shocks have formed by this point \citep{Chomiuk_etal_2021}. As seen in \cite{Franckowiak_etal_2018}, the strongest $\gamma$-ray emission is often detected within 10 or 15 days of the optical  peak.

Theoretical considerations \citep{Metzger_etal_2015,Metzger_etal_2016}  supported by observational evidence \citep{Li_etal_2017_nature,Aydi_etal_2020} showed that these shocks could also be substantially contributing to the visible luminosity of a nova, via thermal free-free emission from the shock-heated ejecta. Nova shocks are expected to be radiative, such that the majority of the dissipated kinetic energy 
is released in the form of thermal X-ray emission, which in turn is reprocessed into the optical by the ejecta \citep{Metzger_etal_2014}. This new understanding that shocks play an important role in novae makes them valuable laboratories for studying shock processes in more distant and rare transients, 
such as super-luminous supernovae (SNe; \citealt{Moriya_etal_2018}), Type IIn SNe \citep{Chugai_etal_2004}, stellar mergers \citep{Metzger_Pejcha_2017}, and tidal disruption events \citep{Roth_etal_2016}.

Our understanding of how shocks form in novae and what controls their energetics is still in its infancy. Between the 10--15 novae discovered in the Galaxy every year, only 2--3 of them are detected by \textit{Fermi} as GeV sources. Moreover, the $\gamma$-ray luminosities of these \textit{Fermi}-detected novae show a spread of several orders of magnitude \citep{Franckowiak_etal_2018, Chomiuk_etal_2021}. Similarly, some novae are only detected with \emph{Fermi} for a few days, while others persist for several months \citep{Cheung_etal_2016,Gordon_etal_2021}. This poses several questions: do all novae produce GeV $\gamma$-rays? What physical parameters determine the power of shocks in novae and the luminosity and duration of the resulting $\gamma$-ray emission? 

One explanation for the presence of shocks during nova eruptions is a collision between the ejected mass and external material, either in the form of a pre-existing circumstellar medium (CSM) or with a strong wind from the companion. 
\kvs{In novae erupting in binaries with red giant secondaries (symbiotic systems),} 
the evolved companions source winds strong enough to create such an external
shock.
\kvs{This was the case for}
the first $\gamma$-ray detected nova, V407 Cyg \kvs{\citep{2010Sci...329..817A}}. 
However, most novae with detected $\gamma$-ray emission have main sequence companions \kvs{\citep{Ackermann_etal_2014}}, and are not expected or observed to have significant CSM. 

\kvs{The shocks in dwarf-donor (non-symbiotic) novae may form}
as a result of collisions between multiple phases of mass ejection within a single nova eruption. 
\kvs{The shocks must be internal to the ejecta and must not depend on the presence of dense external medium.
The initial impulsive ejection associated with the thermonuclear runaway may 
not actually be strong enough to unbind a large fraction of the white dwarf
envelope on its own \citep{Shen_Quataert_2022}. 
Instead, the nova envelope ejection may be 
an extended process involving multiple physical mechanisms \citep{Chomiuk_etal_2021},
including interaction between the expanded envelope and the binary companion 
\citep[producing outflow with velocity comparable to the orbital velocity, $\sim100$\,km\,s$^{-1}$;][]{1983ApJ...273..280K,Livio_etal_1990,2021ApJ...914....5S} 
and radiation-driven wind sustained by near-Eddington luminosity of nuclear-burning white dwarf 
\citep[wind velocity comparable to the white dwarf escape velocity, $\sim1000$\,km\,s$^{-1}$;][]{1990LNP...369..244F,Kato_Hachisu_1994,2004BaltA..13..116F,Shaviv_2001,2002ASPC..261..585S}.}

\kvs{There are strong observational indications that mass loss in novae is not confined to one-time impulsive ejection. 
The evolution of intrinsic X-ray absorbing column in slow novae V959~Mon and V906~Car 
suggests that the majority of their mass loss occurred 40 and 24 days after eruption
onset, respectively \citep{2021MNRAS.500.2798N,2020MNRAS.497.2569S}.
X-ray and radio observations of the recurrent nova T~Pyx were interpreted in
terms of two distinct mass loss episodes by \cite{2014ApJ...788..130C}. 
%
The commonly observed emergence of multiple distinct velocity components in nova optical spectra 
\citep{Russell36,McLaughlin_1944,Friedjung_1966_I,Friedjung_1966_II,Friedjung_1966_III,Friedjung_1987} 
is a strong indication that prolonged and multi-phased mass ejection is common in novae.} 
Recent work has found evidence that nova eruptions consistently show two distinct outflows: 
one slow flow that gets launched before the optical peak with speeds of several hundreds of km s$^{-1}$, 
and another faster component that gets launched near the peak at $\sim$thousands of km s$^{-1}$ \citep{Aydi_etal_2020b}. 
\kvs{The observed velocities aligned with the expectations for  the binary-driven outflow and white dwarf wind discussed above}. 
Collisions between these flows may be the site of the $\gamma$-ray-producing shocks. 
\kvs{Since shock energy scales with the velocities and densities of colliding flows, 
the $\gamma$-ray emission from novae should likewise depend on these outflow properties} 
\citep{Metzger_etal_2015}. 

The goal of this paper is to determine which nova properties are related to the strength and duration of internal shocks in novae. The shock strength and duration are traced by the $\gamma$-ray luminosity and the duration of observed $\gamma$-ray emission. Relationships between optical properties of novae, such as ejecta velocities and peak absolute magnitudes, and the observed $\gamma$-ray properties can provide clues about the nature of the shocks responsible for the $\gamma$-ray emission. The relative velocity between ejecta outflow components and the ejecta mass are expected to be the primary factors that set the energy of the resulting shocks, and therefore the $\gamma$-ray luminosity \citep{Metzger_etal_2015}.

In section \S \ref{sec:obs} of this paper, we describe both the optical and $\gamma$-ray observations of the novae in our sample, along with the construction of the nova sample used here. In \S \ref{sec:gamma-ray-dependence} we discuss the dependence of the $\gamma$-ray properties of these novae on optically-derived nova parameters. We summarize our findings in \S \ref{sec:conclusions}.

\section{Observations}
\label{sec:obs}

\begin{table*}

\setlength{\tabcolsep}{2pt}

\captionsetup{justification=centering, singlelinecheck=false} 
\caption{The sample of non-symbiotic novae detected in $\gamma$-rays by \emph{Fermi}/LAT.}\label{table:characteristics1}
\centering
\def\arraystretch{1.25}
\begin{tabular}{lcccccccccccc}
\hline
\hline
\rule{0pt}{2ex} Name & $t_{\mathrm{max}}$ & $t_2$ & $V_{\mathrm{peak}}$ & $F_{\gamma}$ (avg) & $F_{\gamma}$ (max) & Duration & $v_1$ & $v_2$ & $A_{V}$ & $d$ & Dust & Ref.\\
& (UT date) & (days) & (mag) & \multicolumn{2}{c}{($10^{-10}$ erg cm$^{-2}$ s$^{-1}$)} & (days) & (km s$^{-1}$) &(km s$^{-1}$) & (mag) & (kpc) &\\
\hline
V1674~Her & 2021 Jun 13.0$^{+0.1}_{-0.1}$ & 0.9$^{+0.1}_{-0.7}$ & 6.1  & $21.2\pm6.7$ &$34.0\pm10.3$ &0.5& $3500\pm100$ & $6000\pm200$ & $2.0 \pm 0.3$ & 6.3$^{+3.6}_{-2.0}$ &No& 1 \\
V1405~Cas & 2021 May 10.3$^{+0.2}_{-0.2}$ & 164.5$^{+0.2}_{-0.2}$ & 5.1  & $5.1\pm1.6$ &$9.3\pm2.9$ & 10 & $1100\pm400$ & $2100\pm100$ & $2.1 \pm 0.3$ & 1.8$^{+0.04}_{-0.2}$ &N/A& 2,3,4\\
YZ~Ret & 2020 Jul 17.1$^{+4.2}_{-8.9}$ & 17.0$^{+4.3}_{-9.2}$ & $<5.1$  & $3.5\pm0.5$ &$8.3\pm1.5$ &14& $1200\pm400$ & $3000\pm300$  &  $0.06 \pm 0.02$ & 2.4$^{+0.3}_{-0.2}$&N/A& 5\\
V1707~Sco & 2019 Sep 15.6$^{+0.3}_{-0.7}$ & 5.0$^{+0.7}_{-0.9}$\tablenotemark{a} & 11.8  & $2.9\pm1.0$ & $2.9\pm1.3$ &3& $2500\pm200$ & $4700\pm200$  & $4.2 \pm 0.4$ & 9.6$^{+2.6}_{-1.9}$ &N/A& 6,7\\
V392~Per & 2018 Apr 29.8$^{+0.1}_{-8.3}$ & 3.0$^{+0.1}_{-8.3}$ & $<6.3$ &$2.8\pm0.4$& $13.4\pm3.2$ &8& $2500\pm200$ & $5500\pm300$ & $2.8 \pm 0.4$ & 3.4$^{+0.8}_{-0.3}$ &N/A& 8\\
V906~Car & 2018 Mar 28.5$^{+1.0}_{-0.9}$ & 43.7$^{+1.3}_{-1.1}$ & 5.9  &$15.6\pm0.1$& $24.3\pm0.1$ &23& $250\pm20$ & $2500\pm200$ & $1.1 \pm 0.2$ & 3.0$^{+0.9}_{-0.8}$ &Yes& 9\\
V357~Mus & 2018 Jan 16.4$^{+1.1}_{-11.1}$ & 22.9$^{+1.6}_{-11.1}$ & $<6.5$  & $1.7\pm0.2$ &$5.5\pm1.7$ &27& $750\pm300$ & $2500\pm200$ & $1.6 \pm 0.2$ & 3.8$^{+0.8}_{-1.3}$ &Yes& 10\\
V549~Vel & 2017 Oct 17.7$^{+4.0}_{-2.3}$ & 90.0$^{+4.2}_{-2.5}$ & 9.1  & $0.5\pm0.2$ &$1.0\pm0.3$&33& $700\pm100$ & $2500\pm300$  &  $3.3 \pm 0.3$ &7.3$^{+3.6}_{-3.1}$ &No& 11\\
V5856~Sgr & 2016 Nov 08.0$^{+0.4}_{-0.2}$ & 10.7$^{+0.5}_{-0.6}$\tablenotemark{b} & 5.4\tablenotemark{c}  & $5.9\pm0.5$ &$13.3\pm1.7$&15& $500\pm100$ & $3200\pm300$ & $1.7 \pm 0.2$ & 7.9$^{+1.4}_{-1.5}$ &N/A& 12\\
V5855~Sgr & 2016 Oct 24.4$^{+1.0}_{-1.0}$ & 19.2$^{+1.0}_{-1.9}$\tablenotemark{b} & 7.9  & $3.8\pm0.8$ &$7.0\pm0.9$&26& $500\pm100$ &$2700\pm300$ & $0.8 \pm 0.8$ & 2.1$^{+0.4}_{-0.9}$ &N/A& 13\\
V5668~Sgr & 2015 Mar 21.4$^{+0.3}_{-0.3}$ & 74.7$^{+3.3}_{-3.8}$ & 4.3 & $1.4\pm0.2$ & $6.8\pm2.9$&53& $600\pm100$ & $1100\pm100$ & $1.3 \pm 0.2$ & 3.9$^{+0.3}_{-0.7}$ &Yes& 14,15\\
V1369~Cen & 2013 Dec 14.7$^{+2.0}_{-1.9}$ & 37.7$^{+2.4}_{-4.1}$ & 3.3 & $3.2\pm0.4$ &$9.2\pm4.1$&38& $800\pm100$ & $1400\pm100$ & $0.22 \pm 0.05$ & 0.6$^{+0.4}_{-0.1}$ &Yes& 14,15\\
V339~Del & 2013 Aug 16.7$^{+0.1}_{-0.5}$ & 11.3$^{+0.4}_{-0.5}$ & 4.4 & $2.9\pm0.3$ &$7.6\pm1.4$&27& $900\pm100$ & $2600\pm200$ & $0.65 \pm 0.15$ & 5.1$^{+2.4}_{-1.7}$ &Yes& 15,16\\
V1324~Sco & 2012 Jun 20.0$^{+0.4}_{-0.3}$ & 24.2$^{+0.4}_{-0.3}$ & 9.9 & $7.6\pm0.9$ &$15.7\pm3.7$&17& $650\pm100$ & $3200\pm300$ & $3.1 \pm 0.3$ & 8.8$^{+1.7}_{-1.4}$ &Yes& 15,16\\
V679~Car & 2008 Nov 29.7$^{+0.1}_{-4.4}$ & 16.2$^{+0.6}_{-4.4}$ & 8.1 & $2.4\pm0.5$ & $2.7 \pm 0.8$& 35 & -- & -- & $2.0 \pm 0.6$ &  4.5$^{+0.6}_{-0.5}$ &N/A& 15\\
\hline
\tablenotetext{}{\justifying The parameters of the novae in our sample that have been detected by \emph{Fermi}/LAT. The columns of data include the name of the nova, date of optical peak ($t_{\rm max}$), the rate of decline ($t_2$), peak $V$-band apparent magnitude, average $>$100 MeV $\gamma$-ray flux over the duration of \emph{Fermi} detection, maximum daily $>$100 MeV $\gamma$-ray flux, duration of the significant \emph{Fermi} detection, velocities (see \S \ref{sec:specprop}), $V$ band extinction ($A_V$), distance to the nova, whether the nova is dust forming, and reference for the $\gamma$-ray data. Many systems lack IR light curves requisite for determining the dust forming characteristics, and therefore do not have reliable dust classifications (marked as N/A in the table).}
\tablenotetext{}{\justifying 1= \citet{Sokolovsky_etal_2023}; 2=\citet{Gong_Li2021}; 3=\citet{Buson_etal_2021}; 4=Aydi et al. 2025, submitted; 5= \citet{Sokolovsky_etal_2022}; 6=\citet{Li_etal_2019}; 7=Li et al.\ 2025, in preparation; 8=\citet{Albert_etal_2022}; 9=\citet{Aydi_etal_2020}; 10=\citet{Gordon_etal_2021}; 11=\citet{Li_etal_2020}; 12=\citet{Li_etal_2017_nature}; 13=\citet{Nelson_etal_2019}; 14=\citet{Cheung_etal_2016}; 15=\citet{Franckowiak_etal_2018}; 16=\citet{Ackermann_etal_2014}}
\tablenotetext{a}{\justifying The timing of $t_2$ is measured from a $g$-band light curve (ASAS-SN).}
\tablenotetext{b}{\justifying The timing of $t_2$ is measured from a Visual light curve (AAVSO).}
\tablenotetext{c}{\justifying Peak magnitude is not measured in the $V$-band, but from Visual estimates (AAVSO).}
\end{tabular}
\end{table*}

\begin{table*}
\caption{\label{table:characteristics2} The sample of novae not detected in $\gamma$-rays.}
\def\arraystretch{1.25}
\begin{tabular}{lcccccccccc}
\hline
\hline
\rule{0pt}{2ex} Name & $t_{\mathrm{max}}$ & $t_2$ & $V_{\mathrm{peak}}$ & $F_{\gamma}$ (avg) & $v_1$ & $v_2$ & $A_{V}$ & $d$ \\
& (UT date) & (days) & (mag) & ($10^{-10}$ erg cm$^{-2}$ s$^{-1}$)  & (km s$^{-1}$) &(km s$^{-1}$) & (mag) & (kpc) & \\
\hline
V1710~Sco & 2021 Apr 13.6$^{+0.5}_{-0.2}$ & 4.5$^{+0.6}_{-0.7}$ & 8.5 & $<$2.8 & -- & $4100\pm300$ &  $3.6 \pm 0.7$  & 10.2$^{+4.5}_{-2.8}$\\ 
V6595~Sgr & 2021 Apr 06.2$^{+0.4}_{-0.5}$ & 5.2$^{+0.4}_{-0.5}$ & 7.4 & $<$0.7 & $1100\pm200$ & $3000\pm400$ & $1.9 \pm 0.4$ & 8.2$^{+2.0}_{-1.1}$\\ 
V1112~Per & 2020 Nov 28.2$^{+0.1}_{-0.1}$ & 28.1$^{+0.1}_{-0.1}$ & 8.3 & $<$0.2 & $800\pm200$ & $1300\pm200$ & $2.9 \pm 0.6$ &4.4$^{+1.9}_{-2.0}$\\  
V659~Sct & 2019 Oct 31.1$^{+0.7}_{-0.3}$ & 7.9$^{+1.0}_{-1.1}$ & 8.4 & $<$1.7 & $1100\pm400$ & $3200\pm400$ &  $3.8 \pm 0.7$ & 9.3$^{+5.6}_{-2.6}$\\ 
V3666~Oph &  2018 Aug 11.9$^{+0.9}_{-0.1}$ & 21.6$^{+1.8}_{-0.2}$ & 8.8 & $<$0.9 & $1100\pm300$ & $2300\pm200$ & $2.0 \pm 0.3$ & 1.7$^{+3.6}_{-0.1}$\\  
V408~Lup & 2018 Jun 03.6$^{+0.9}_{-12.3}$ & 42.3$^{+1.9}_{-12.4}$ & 9.0 & $<$0.1 & $500\pm100$ & $2000\pm500$ & $1.6 \pm 0.4$ &  3.6$^{+0.7}_{-1.1}$\\ 
FM~Cir & 2018 Jan 28.3$^{+1.0}_{-0.9}$ & 133.7$^{+1.0}_{-2.2}$ & 6.0 & $<$0.1 & $550\pm200$ & $1300\pm200$ &  $1.0 \pm 0.3$ & 3.6$^{+0.7}_{-1.2}$\\ 
V612~Sct & 2017 Jul 30.0$^{+0.9}_{-1.0}$ & 132.2$^{+78.2}_{-12.5}$ & 8.4 & $<$0.1 & $270$ & $1150$ & $3.0 \pm 0.6$ & 9.6$^{+4.4}_{-3.3}$\\ 
V407~Lup & 2016 Sep 25.1$^{+0.9}_{-1.1}$ & 4.6$^{+0.9}_{-1.1}$ & $<6.3$\tablenotemark{a} & $<$3.6 & $1500\pm100$ & $3500\pm300$ & $0.7 \pm 0.3$ & 2.9$^{+0.8}_{-0.8}$\\ 
V5669~Sgr & 2015 Sep 28.5$^{+0.9}_{-0.4}$ & 33.3$^{+8.7}_{-3.4}$ & 8.9 & $<$1.0 & $700\pm100$ & $1400\pm300$ & $1.2 \pm 0.7$ & 2.7$^{+0.8}_{-0.4}$\\ 
V2944~Oph & 2015 Apr 13.6$^{+1.0}_{-1.0}$ & 32.5$^{+1.1}_{-1.2}$ & 9.3 & $<$0.6 & $1050\pm300$ & $2400\pm300$ & $1.6 \pm 0.3$ & 1.8$^{+0.3}_{-0.4}$\\ 
V5667~Sgr & 2015 Feb 15.7$^{+2.0}_{-0.9}$ & 64.0$^{+2.0}_{-1.3}$ & 9.3 & $<$0.8 & -- & $1500\pm300$ & $1.0 \pm 0.3$ & 7.7$^{+0.7}_{-0.7}$\\ 
V2659~Cyg & 2014 Apr 10.5$^{+0.5}_{-0.7}$ & 115.5$^{+1.0}_{-1.1}$ & 9.4 & $<$0.9 & $370\pm150$ & $1200\pm150$ & $2.8 \pm 1.7$ & 8.2$^{+3.5}_{-3.6}$\\ 
V962~Cep & 2014 Mar 13.9$^{+0.5}_{-0.8}$ & 31.5$^{+1.7}_{-1.5}$ & 11.1 & $<$0.6 & $600\pm300$ & $1900\pm300$ &  $3.5 \pm 0.7$ &4.4$^{+2.3}_{-1.7}$\\ 
V1533~Sco & 2013 Jun 04.5$^{+0.7}_{-3.8}$ & 7.9$^{+3.1}_{-3.9}$ & $<13.5$ & $<$2.8 & -- & $1000\pm200$ & $6.1 \pm 0.7$ & 9.6$^{+3.0}_{-1.9}$\\  
V809~Cep & 2013 Feb 04.7$^{+0.1}_{-0.6}$ & 20.3$^{+0.1}_{-1.2}$ & 11.2 & $<$0.4 & $800\pm300$ & $1600\pm150$ &  $4.7 \pm 0.7$ &7.4$^{+3.9}_{-2.9}$\\ 
V5593~Sgr & 2012 Jul 22.4$^{+7.1}_{-0.2}$ & 43.6$^{+7.1}_{-0.3}$ & 11.1 & $<$1.3 & $800\pm300$ & $2400\pm300$ & $6.4 \pm 0.3$ & 9.0$^{+4.8}_{-2.7}$\\ 
V5591~Sgr & 2012 Jun 27.3$^{+0.2}_{-0.7}$ & 2.3$^{+0.3}_{-0.7}$ & $<9.7$ & $<$1.7 & -- & $5000\pm500$ & $4.2 \pm 0.7$ & 7.7$^{+2.8}_{-1.4}$\\ 
V2677~Oph & 2012 May 21.3$^{+0.2}_{-2.6}$ & 6.7$^{+0.3}_{-2.7}$ & $<11$ & $<$1.8 & -- & $4000\pm500$ & $3.7 \pm 0.7$ & 8.1$^{+1.2}_{-1.4}$\\  
V5589~Sgr & 2012 Apr 22.5$^{+0.1}_{-1.6}$ & 2.3$^{+0.6}_{-1.6}$ & 9.1 & $<$1.7 & -- & $4500\pm500$ & $2.5 \pm 0.6$ & 8.0$^{+1.8}_{-1.4}$\\ 
V1428~Cen & 2012 Apr 07.8$^{+0.5}_{-0.4}$ & 10.8$^{+0.9}_{-0.5}$ & 9.4 & $<$1.9 & -- & $3000\pm400$ & $4.2 \pm 0.3$ & 9.4$^{+4.2}_{-3.2}$\\ 
V2676~Oph & 2012 Apr 04.5$^{+1.9}_{-1.0}$ & 83.8$^{+2.1}_{-1.4}$ & 10.7 & $<$1.4 & -- & -- & $2.9 \pm 0.7$ & 8.1$^{+1.6}_{-1.6}$\\ 
V834~Car & 2012 Mar 01.4$^{+0.1}_{-0.2}$ & 20.1$^{+0.1}_{-0.2}$ & 10.2 & $<$1.2  & -- & $1700\pm400$ & $1.7 \pm 0.7$ & 7.1$^{+3.4}_{-2.7}$\\
V1313~Sco & 2011 Sep 07.5$^{+0.5}_{-0.4}$ & 8.1$^{+0.6}_{-0.6}$ & 10.4 & $<$1.5  & -- & $3400\pm400$ &  $4.3 \pm 0.7$ & 8.6$^{+4.5}_{-2.7}$\\ 
PR~Lup & 2011 Aug 14.4$^{+0.7}_{-0.2}$ & 14.7$^{+0.7}_{-0.6}$ & 8.6 & $<$0.6 & -- & -- & $2.3 \pm 0.7$ & 8.3$^{+4.3}_{-3.2}$\\ 
V1312~Sco & 2011 Jun 02.3$^{+0.2}_{-3.3}$ & 12.9$^{+1.3}_{-3.4}$ & 10.4 & $<$0.8  & -- & -- &  $2.8 \pm 0.3$ & 2.6$^{+0.2}_{-0.5}$\\ 
T~Pyx & 2011 May 12.0$^{+0.3}_{-0.2}$ & 50.4$^{+0.7}_{-0.4}$ & 6.2 & $<$0.8 & $1400\pm300$ & $2600\pm300$ &  $1.3 \pm 0.4$ & 3.6$^{+0.3}_{-0.2}$\\ 
V5588~Sgr & 2011 Apr 07.1$^{+1.0}_{-1.0}$ & 46.0$^{+1.4}_{-1.0}$ & 11.3 & $<$1.4 & -- & $3000\pm300$ & $4.9 \pm 0.7$ & 8.1$^{+4.5}_{-1.8}$\\ 
V1311~Sco & 2010 Apr 26.3$^{+0.8}_{-1.6}$ & 2.2$^{+1.3}_{-1.9}$ & 8.6 & $<$1.8 & -- & $5200\pm500$ & $3.5 \pm 0.7$ & 9.1$^{+3.0}_{-2.5}$\\ 
V2674~Oph & 2010 Feb 21.2$^{+11.3}_{-0.7}$ & 20.3$^{+27.4}_{-7.0}$ & 9.3 & $<$1.3 & -- & -- & -- & --\\ 
V2673~Oph & 2010 Jan 18.3$^{+0.9}_{-1.4}$ & 10.3$^{+1.1}_{-2.0}$ & 8.5 & $<$0.8 & -- & -- & -- & --\\ 
KT~Eri & 2009 Nov 14.6$^{+0.1}_{-0.1}$ & 6.5$^{+0.1}_{-0.1}$ & 5.4\tablenotemark{b} & $<$0.4 & -- & $3600\pm400$ &  $0.17 \pm 0.06$ &4.2$^{+0.5}_{-0.3}$\\ 
V496~Sct & 2009 Nov 18.5$^{+1.0}_{-1.0}$ & 58.3$^{+17.3}_{-28.3}$ & 7.1 & $<$2.6 & $380\pm100$ & -- & $1.3 \pm 1.0$ & 10.4$^{+4.3}_{-4.1}$\\ 
V2672~Oph & 2009 Aug 16.5$^{+0.1}_{-2.4}$ & 2.8$^{+0.2}_{-2.4}$ & 11.4 & $<$2.6 & -- & $6100\pm500$ & $3.9 \pm 0.7$ & 2.0$^{+0.1}_{-0.2}$\\ 
V5583~Sgr & 2009 Aug 06.4$^{+0.1}_{-0.1}$ & 5.4$^{+0.7}_{-3.3}$ & 7.2 & $<$0.8 & -- & $3300\pm400$ &  $0.6 \pm 0.7$ &2.4$^{+0.7}_{-0.2}$\\ 
V1213~Cen & 2009 May 08.2$^{+2.9}_{-4.1}$ & 11.4$^{+2.9}_{-4.1}$ & 8.5 & $<$1.2 & -- & -- & $1.3 \pm 0.7$ & 5.4$^{+2.2}_{-0.9}$\\ 
QY~Mus & 2008 Oct 14.0$^{+5.0}_{-4.6}$ & 84.0$^{+38.5}_{-5.1}$ & 8.7 & $<$0.8 & -- & $1450\pm200$ & -- & --\\ 
\hline
\tablenotetext{}{\justifying The parameters of the novae in our sample that are not detected by \emph{Fermi}/LAT. The columns of data include the name of the nova, $t_{\rm max}$, $t_2$, peak $V$-band apparent magnitude, 95 per cent upper limit on the $>$100 MeV $\gamma$-ray flux, velocities, $V$ band extinction ($A_V$), and distance to the nova.}
\tablenotetext{a}{\justifying Peak magnitude is not measured in the $V$-band, but from Visual estimates (AAVSO).}
\tablenotetext{b}{\justifying Peak magnitude is not measured in the $V$-band, but from unfiltered SMEI light curves \citep{Hounsell_etal_2010}.}
\end{tabular}
\end{table*}

\subsection[]{Sample Selection and $\gamma$-Ray Properties}
\label{sec:sample_selection}

\subsubsection[]{The $\gamma$-Ray Detected Novae}\label{sec:gamma}
We selected all novae detected by \emph{Fermi}/LAT at $\geq 5 \sigma$ significance (likelihood ratio test statistic $TS > 25$), during the time period Aug 2008 (the start of \emph{Fermi} operations) to the end of 2021 (Table \ref{table:characteristics1}). We exclude novae with red giant companions, as the shocks in these systems may be fundamentally different: while classical novae with dwarf companions host shocks internal to the ejecta, novae with giant companions drive shock interaction with external circumstellar material, blown from the giant as a wind \citep{Gordon_etal_2021}. We identified those novae in symbiotic binaries using the \citet{Schaefer_2010} list of Galactic recurrent novae and the Galactic nova compilation of \citet{Ozdonmez_etal_2018}; we also searched the literature for additional constraints on the natures of the companion stars. Based on this, we exclude the $\gamma$-ray detected novae V407 Cyg, RS~Oph, and V3890 Sgr (V745 Sco, another symbiotic recurrent nova, was a marginal \emph{Fermi} detection that does not meet our significance threshold; \citealt{Franckowiak_etal_2018}). One $\gamma$-ray detected nova, V959 Mon, is excluded from our $\gamma$-ray sample due to a lack of optical coverage during the $\gamma$-ray emitting phase. Among non-symbiotic novae, FM Cir and V407 Lup are marginal \emph{Fermi}/LAT detections \citep{Gordon_etal_2021, Wang_etal_2024} and 
\kvs{do not meet our adopted significance threshold;} 
these novae are included in our comparison sample of novae without significant \emph{Fermi}/LAT detections (see \S \ref{sec:ray}). The total number of $\gamma$-ray detected novae in our sample is 15.

For the $\gamma$-ray detected novae, we list in Table \ref{table:characteristics1} the peak $\gamma$-ray flux (in 1 day bins), the duration of significant $\gamma$-rays (generally when $TS > 4$ or $> 2 \sigma$ in daily binning of \emph{Fermi}/LAT data, 
following \citealt{Ackermann_etal_2014,Cheung_etal_2016,Franckowiak_etal_2018}), and the average $\gamma$-ray flux during this time. For converting from $\gamma$-ray photon flux to energy flux, we assumed that the shape of the \emph{Fermi}/LAT spectrum of V906 Car (the highest S/N $\gamma$-ray detection to date; \citealt{Aydi_etal_2020}) applies to all novae, consistent with the findings of \citet{Franckowiak_etal_2018}. For the $>$100 MeV band, the conversion factor is then 1 photon cm$^{-2}$ s$^{-1} = 1.2794 \times 10^{-3}$ erg cm$^{-2}$ s$^{-1}$. References to the $\gamma$-ray analysis for these novae are also listed in Table \ref{table:characteristics1}.

\subsubsection[]{$\gamma$-Ray Non-Detections from 2008--2015}
As a comparison sample, we also considered all Galactic novae that were discovered during the date range Aug 2008 until the end of 2015 and were \emph{not} detected by \emph{Fermi}/LAT \citep{Franckowiak_etal_2018}. This sample is listed in Table \ref{table:characteristics2}.  Again, we exclude novae with known or suspected red giant companions, removing V1703 Sco \citep{Munari_etal_2013}, V5590 Sgr \citep{Mroz_etal_2014}, V745 Sco \citep{Schaefer_2010}, V1534 Sco \citep{Munari_Banerjee18}, V1535~Sco \citep{Linford_etal_2017}, V1708 Sco \citep{Mroz_Udalski20}, and V3664 Oph \footnote{(\citealt{Aydi_etal_2018b}, J.\ Mikolajewska, private communication; although the ATel does not specifically classify the event as a nova embedded in dense CSM, the narrow strong H lines coupled with deep absorption at v $\approx$ 0 km s$^{-1}$ are telltale signs of a dense red giant wind)}. Additionally, we excluded V1309 Sco, which is not a classical nova but a luminous red nova borne of a stellar merger \citep{Tylenda_etal_2011}.
We also exclude novae that do not have optical photometry available during the typical $\gamma$-ray emitting phase, between peak and $t_2$. We primarily rely on AAVSO data \citep{Kloppenborg23}, and supplement with ASAS-SN photometry (\citealt{Shappee_etal_2014}; because ASAS-SN only came online in 2013 and did not start observing the Galactic plane at high cadence until 2017, several novae in the earlier years of our study do not have good light curve coverage). The total number of novae not detected in $\gamma$-rays by \citet{Franckowiak_etal_2018} that we consider in this paper is 28.

95 per cent upper limits on the $\gamma$-ray flux for the non-detections in \citet{Franckowiak_etal_2018} were calculated over a 15-day window, and are listed in Table \ref{table:characteristics2}, assuming the same conversion factor from photon flux to energy flux as in \S \ref{sec:gamma}.

\subsubsection[Nondetections from 2016--2021]{$\gamma$-ray Non-Detections from 2016--2021}\label{sec:ray}
To enrich our samples, we also included optically bright novae ($V_{\rm peak} <$ 9 mag) that have erupted too recently to be in the \citet{Franckowiak_etal_2018} sample and do not have a reported $\gamma$-ray detection, covering 2016--2021. This added nine systems to our catalogue (also listed in Table \ref{table:characteristics2}), including FM Cir and V407 Lup, which have claims of marginal \emph{Fermi} detection reported \citep{Gordon_etal_2021, Wang_etal_2024}, but do not meet our threshold for significance. For the novae that lack published upper limits on their $\gamma$-ray fluxes, we analyse the \emph{Fermi} data to derive suitable upper limits (or check for \emph{Fermi} detections).

While V407 Lup has been claimed as a \emph{Fermi}/LAT detection by \citet{Gordon_etal_2021}, it tops out at $4 \sigma$ significance and so does not meet our $\gamma$-ray detection criteria (\S \ref{sec:gamma}). The nova was detected at $> 2 \sigma$ significance for just three days, during which the average $>$100 MeV flux was $(1.6 \pm 0.7) \times 10^{-7}$ phot s$^{-1}$ cm$^{-2}$. We therefore use a 95 per cent upper limit on its $\gamma$-ray flux as $2.8 \times 10^{-7}$ phot s$^{-1}$ cm$^{-2}$ ($3.6 \times 10^{-10}$ erg s$^{-1}$ cm$^{-2}$).

For the systems other than V407 Lup, we analysed their \emph{Fermi}/LAT data to constrain the GeV $\gamma$-ray flux, downloading the $\gamma$-ray data of the eight novae from the Fermi-LAT Data Server. The observations contain all the Pass 8 (\texttt{P8R3\_V3}) $\gamma$-ray events from 100~MeV to 300~GeV within 5 degrees from the novae. \texttt{Fermitools} (version 2.2.0), developed by the LAT science team, was employed for the data analyses. The so-called binned likelihood analysis was adopted by following the instructions in the LAT online data analysis manual\footnote{\url{https://fermi.gsfc.nasa.gov/ssc/data/analysis/scitools/binned_likelihood_tutorial.html}}. We first selected the events based on the class and type (i.e. \texttt{evclass}=128 and \texttt{evtype}=3), and obtained the good time intervals using the logical expression \texttt{(DATA\_QUAL>0)\&\&(LAT\_CONFIG==1)}. Source models for the nova fields were constructed based on the information from the 4FGL-DR3 catalogue \citep{Abdollahi_etal_2022}.
In the models, all the catalogued sources within 20 degrees from the nova were included with two additional background emission components, known as the Galactic diffuse emission and the isotropic emission (i.e. \texttt{gll\_iem\_v07} and \texttt{iso\_P8R3\_SOURCE\_V3\_v1}). The spectral models of all the novae were assumed to be a simple power law with a fixed photon index of 2.0. Except for the normalization factors for the nova $\gamma$-ray emission, all the spectral parameters in the model files are fixed  to the catalogued values to avoid overfitting, given the short time periods of the obtained observations (on daily time-scales). For each nova, we set the epoch of the optical peak as the start time, and performed a binned likelihood analysis with an end time at $t_2$. It turns out that none of the novae result in a significant $\gamma$-ray detection (i.e. TS $< 25$), and thus 95 per cent upper limits were computed for them, quoted in Table \ref{table:nondet}. We also give the upper limits on $\gamma$-ray flux in units of erg s$^{-1}$ cm$^{-2}$ in Table \ref{table:characteristics2}, assuming the same conversion factor from photon flux to energy flux as in \S \ref{sec:gamma}.

\begin{table}
\centering
\caption{New \emph{Fermi}/LAT upper limits on the 0.1--300 GeV $\gamma$-ray flux for eight recent novae, with the derived test statistic values.}
\begin{tabular}{lccc}
\hline
Name & 95 per cent upper limit on F$_{\gamma}$ (avg) & TS\\
& ($10^{-10}$ erg s$^{-1}$ cm$^{-2}$) & &\\
\hline
\hline
V1710 Sco & 2.83 & 5.8\\
V6595 Sgr & 0.73 & 0.6\\
V1112 Per & 0.18 & 0.0\\
V659 Sct & 1.65 & 5.4\\
V3666 Oph & 0.9 & 0.9\\
V408 Lup & 0.12 & $-0.1$\\
FM Cir & 0.1 & 0.8\\
V612 Sct & 0.14 & 0.0\\
\hline
\end{tabular}
\label{table:nondet}
\end{table}

\subsection{Optical Light Curve Properties}

We downloaded optical light curves for the novae in our samples from the AAVSO database \citep{Kloppenborg23}, focusing on data in the $V$-band and 
a clear band anchored to the $V$-band scale (AAVSO 
\kvs{filter name} 
`CV'). AAVSO light curves often begin a few days after discovery, so to constrain the earliest phases of the eruption, we also include photometry reported in circulars and telegrams. We also query the ASAS-SN survey \citep{Shappee_etal_2014}, which provides high-cadence $g$-band coverage for many novae. The plotted light curves are shown in the Appendix as Figures~\ref{Fig:V1674_Her_LC} to \ref{Fig:QY_Mus_LC}, with the figure captions including references to supplementary photometry. 

From the optical light curves, we measured the maximum brightness $V_{\rm peak}$ of the nova, presented in Tables \ref{table:characteristics1} and \ref{table:characteristics2}. We measured $V_{\rm peak}$ preferentially in the $V$-band whenever possible, but in some cases, the light curve maximum is only covered by an unfiltered or visual band; these are noted in Tables \ref{table:characteristics1} and \ref{table:characteristics2}. An accurate uncertainty on $V_{\rm peak}$ is not possible to derive for many of the novae in our sample with the current data in hand, because it depends not only on the cadence of the light curve, but also the shape of the underlying light curve, and nova light curve shapes are diverse and unpredictable \citep{Strope_etal_2010}. Additionally, for most novae the limited coverage of light curve peak dominates over magnitude measurement uncertainty. We therefore made the decision to not quantify uncertainty on $V_{\rm peak}$ in this study. In a few cases, the cadence of the optical light curve was poor, and it seems likely that light curve maximum fell in a data gap; we only place an upper limit on $V_{\rm peak}$ (lower limit on the peak optical flux) in cases where this gap around light curve maximum is greater than 8 days or greater than $\frac{1}{3} t_2$ (see below for discussion of $t_2$). Three of our $\gamma$-ray detected novae, YZ Ret, V392 Per and V357 Mus, along with some of the $\gamma$-ray non-detected novae, only have upper limits on the $V$-band magnitudes.

We also report the date of optical maximum ($t_{\rm max}$) in Tables \ref{table:characteristics1} and \ref{table:characteristics2}, with uncertainty on this time determined by the cadence of the light curve around maximum. \hl{This uncertainty is defined as the time between the brightest magnitude in our light curve and the preceding and following points, providing an uncertainty on each side of the maximum. For the majority of our light curves, this provides a natural constraint on the observed peak times.} In some cases, novae show multiple maxima in their optical light curves of comparable brightness (e.g. V612 Sct; Figure~\ref{Fig:V612_Sct_LC}); we generally took the first maximum in these cases, but also used optical spectral evolution (\S \ref{sec:specprop}) to inform the choice of maximum \citep{Aydi_etal_2020b}. We use the presence of the broad emission features produced by the fast outflow as an indicator that the nova is passed the optical peak.

We also measure the decline time-scale of the optical light curve, quantified as $t_2$ or the time for the light curve to decline from peak by two magnitudes. We list $t_2$ measurements in Tables \ref{table:characteristics1} and \ref{table:characteristics2}. In order to find $t_2$, we linearly interpolated between photometric points (preferring $V$-band measurements wherever possible). Some novae, like V5668 Sgr, show `jitters' around maximum, and the light curve may reach two magnitudes from maximum multiple times; e.g. Figure~\ref{Fig:V5668_Sgr_LC}. In these cases, we took the last time the light curve dipped two magnitudes below maximum. As with $t_{\rm max}$, we estimate an uncertainty on the timing of $t_2$ from the cadence of the light curve around this time, \hl{with uncertainties defined as the time between the point when the magnitude falls to 2 below peak and the preceding and following points in the light curve.} The uncertainty on the value of $t_2$ (also listed in Tables \ref{table:characteristics1} and \ref{table:characteristics2}) was calculated by adding this in quadrature with the uncertainty on $t_{\rm max}$, treating the upper and lower bounds separately. In the case of V392 Per, where the uncertainty on the peak time is large, the uncertainty on $t_2$ formally extends to values below zero; we truncate this at zero as negative $t_2$ values are unphysical. V392 Per is the only nova in our sample for which this is required.

\subsection{Spectroscopic Properties}\label{sec:specprop}

We use optical spectroscopy to measure the dust extinction along the line of sight to the nova, and to estimate the ejecta velocities. 
Some novae in the sample have medium to high-resolution spectra (R $\ge$ 5,000 to 67,000) available, obtained through the database of the Astronomical Ring for Amateur Spectroscopy (ARAS; \citealt{Teyssier_2019}), or with the Goodman spectrograph \citep{Clemens_etal_2004} on the 4.1\,m Southern Astrophysical Research (SOAR) telescope, or obtained with the CHIRON spectrograph on the Small and Medium Aperture Telescope System (SMARTS; \citealt{Walter_etal_2012}). In addition, for a small number of sources, we make use of suitable published high-resolution spectra taken with the High Resolution Spectrograph (HRS; \citealt{Barnes_etal_2008,Bramall_etal_2010,Bramall_etal_2012,Crause_etal_2014})  on the Southern African Large Telescope (SALT; \citealt{Buckley_etal_2006,Odonoghue_etal_2006}). 
The logs of the spectral observations are in Tables~\ref{table:spec_log_1} and~\ref{table:spec_log_2}.

\begin{figure}
\begin{center}
  \includegraphics[width=\columnwidth]{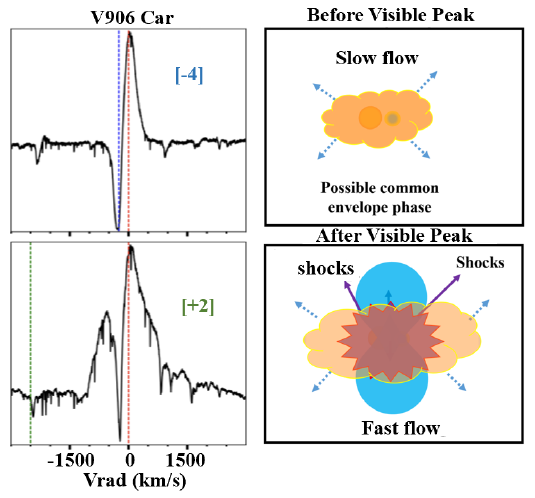}
\caption{\textit{Left}: the H$\alpha$ line profiles before (\textit{top}) and after (\textit{bottom}) optical peak for nova V906~Car. The red dashed lines represent rest velocity ($v_{\mathrm{rad}}$ = 0\,km\,s$^{-1}$). The blue and green dashed lines represent the velocities of the slow flow ($v_1 = 250$ km s$^{-1}$) and fast flow ($v_2 = 2500$ km s$^{-1}$), respectively. The numbers in brackets are the day of observation relative to the optical peak. The spectrum on day +2 shows absorption from the slow component superimposed on emission from the fast component, suggesting that the slow flow is external to and distinct from the fast flow. \hl{We note that the fast outflow velocity shown here is faster than the one shown in \citealt{Aydi_etal_2020b}; this is an effect of the fast outflow velocity varying in this nova, and we adopt a velocity measured during the measured $\gamma$-ray emission for this nova.} \textit{Right}: Schematic illustrations showing the suggested evolution of the nova ejecta near visible peak. \textit{Top}: Before visible peak, the accreted envelope puffs up due to the thermonuclear runaway, engulfing the system in a common envelope phase and becoming concentrated in the orbital plane (e.g. \citealt{Livio_etal_1990,Chomiuk_etal_2014,Sokoloski_etal_2017}). This slow flow manifests as a P Cygni profile in the spectrum, characterized by slow (a few 100s km\,s$^{-1}$) velocities. \textit{Bottom}: Near peak, a continuous fast wind starts, driven by radiation from the ongoing nuclear burning on the surface of the white dwarf (e.g. \citealt{Bath_Shaviv_1976,Shaviv_2001}). The fast flow can propagate more freely in the polar directions due to the concentration of the pre-existing slower ejecta in the orbital plane, and it manifests as broad emission in the spectral lines. When the two outflows collide, they produce shocks -- the likely origin of the $\gamma$-ray emission. Figure~adapted from \citealt{Aydi_etal_2020b}.}
\label{Fig:spec_scheme}
\end{center}
\end{figure}

\subsubsection{Velocity Measurements}
We make use of these spectra to measure the velocities of two 
\kvs{velocity} 
components that are typically observed in the early spectra of novae before and after the visible peak, as outlined by \citet{Aydi_etal_2020b}. Focusing on H Balmer lines, these authors found consistent evidence for a low velocity (a few hundred km\,s$^{-1}$) spectral component which is present during the rise to peak with a P Cygni profile (top left of Figure~\ref{Fig:spec_scheme}). After the nova reaches visible peak, a faster component emerges, exhibiting velocities of a few thousand km\,s$^{-1}$ (bottom left panel of Figure~\ref{Fig:spec_scheme}). Both components co-exist for a period of time, and absorption from the slow component is superimposed on the fast component's emission profile. This led \citet{Aydi_etal_2020b} to associate these components with two distinct outflows: an initial slow flow followed by a faster flow, which interact leading to shocks and $\gamma$-ray emission \citep{Li_etal_2017_nature,Aydi_etal_2020}. One of the goals of this study is to test the dependence of the energetics of the shocks on the properties of the interacting outflows, especially their velocities. Therefore, we use the radial velocities measured from the absorption/emission lines in the spectra as a proxy for the speeds of these colliding outflows. For novae with dedicated spectroscopic follow-up, we measure the velocities of the slow component ($v_1$), the fast component ($v_2$), and the difference between the two velocities ($v_2 - v_1 = \Delta v$). The spectral plots used to measure these parameters are presented in Figures~\ref{Fig:line_profiles_spec_gamma_slow}, \ref{Fig:line_profiles_spec_gamma_fast}, \ref{Fig:line_profiles_spec_non_gamma_slow}, \ref{Fig:line_profiles_spec_non_gamma_mod}, and~\ref{Fig:line_profiles_spec_non_gamma_fast}. For novae where spectroscopic observations before the visible peak are lacking, we measure the velocity of the fast component only. The spectral plots used to measure the velocity of the fast flow for these novae are presented in Figures~\ref{Fig:line_profiles_non_gamma_cyan_1} and~\ref{Fig:line_profiles_non_gamma_cyan_2}. For two $\gamma$-ray emitting novae, namely V357~Mus and YZ~Ret, pre-maximum spectra are lacking. In these cases, we use post-maximum spectra to measure a lower limit on the slow velocity component and increase our sample of $\gamma$-ray emitting novae with $\Delta v$ values. The spectra of these two novae are presented in Figure~\ref{Fig:line_profiles_Mus_Ret}. Our measurements of $v_1$ and $v_2$ are presented in Tables \ref{table:characteristics1} and \ref{table:characteristics2}.

\subsubsection{Extinction Measurements}
For novae with medium or high-resolution spectra ($R \gtrsim$ 9,000), we measure the foreground dust extinctions using the equivalent widths of the Diffuse Interstellar Bands (DIBs). The equivalent widths of DIBs are correlated with $E(B-V)$ and can be used to estimate the line of sight extinction \citep{Friedman_2011,Vos_2011,Lan_2015,Krelowski_2019}. We follow the correlations provided in \cite{Friedman_2011}, and utilize all eight of the DIBs that are considered in that paper. 
Once $E(B-V)$ measurements are in hand, we 
\kvs{compute $A_V = R_V \times E(B-V)$, assuming $R_V = 3.1$ \citep[e.g.][]{1989ApJ...345..245C,2006LNEA....2..189M}.}

\begin{figure*}
\begin{center}
  \includegraphics[width=\textwidth]{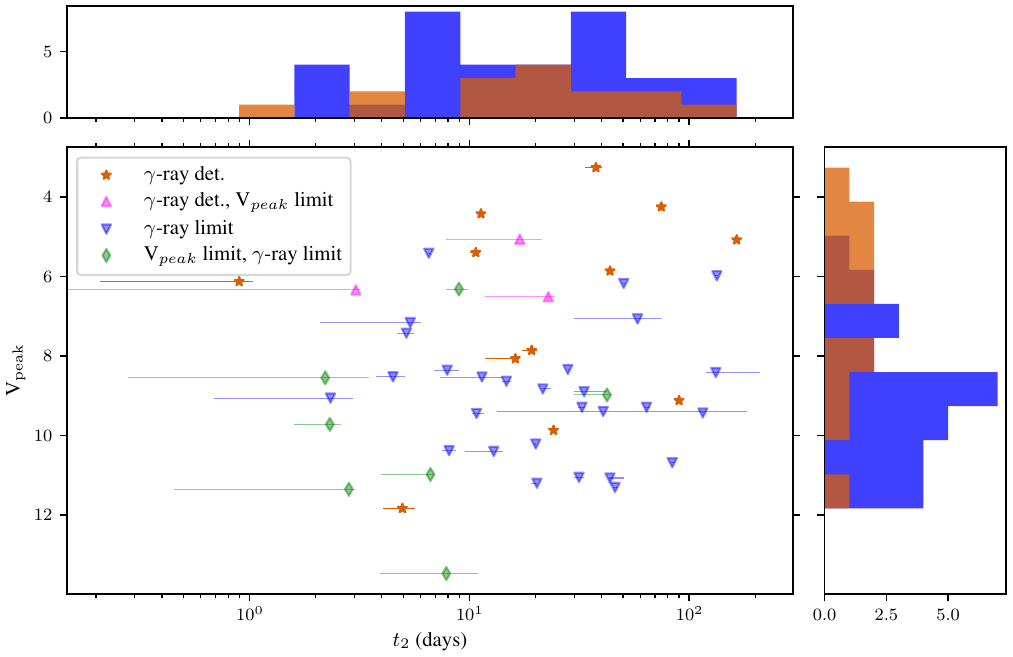}
\caption{The $V$-band magnitude of optical peak ($V_{\mathrm{peak}}$), plotted against the time to decline from peak by 2 magnitudes ($t_2$). The blue triangular data points represent the novae that have not been detected by \emph{Fermi}. The orange stars represent the systems that have been detected in $\gamma$-rays, and the magenta triangles show novae that are $\gamma$-ray detected, but only have a limit on the peak $V$-band magnitude. Green diamonds represent points that are $\gamma$-ray non-detections and only have limits on $V_{peak}$. Histograms in the top and right panels show the distribution of $t_2$ and $V_{\mathrm{peak}}$, respectively. In each histogram the orange bars display the $\gamma$-ray detected sample, while the blue bars represent the $\gamma$-ray limit sample. Novae with limits on $V_{\mathrm{peak}}$ are excluded from both histograms. Typically, optically bright novae are $\gamma$-ray detected, while fainter novae are not. There are several clear exceptions to this, however, indicating that there are other factors at play here beyond the optical flux.}
\label{Fig:vmax_vs_t2}
\end{center}
\end{figure*}

\begin{figure*}
\begin{center}
  \includegraphics[width=\textwidth]{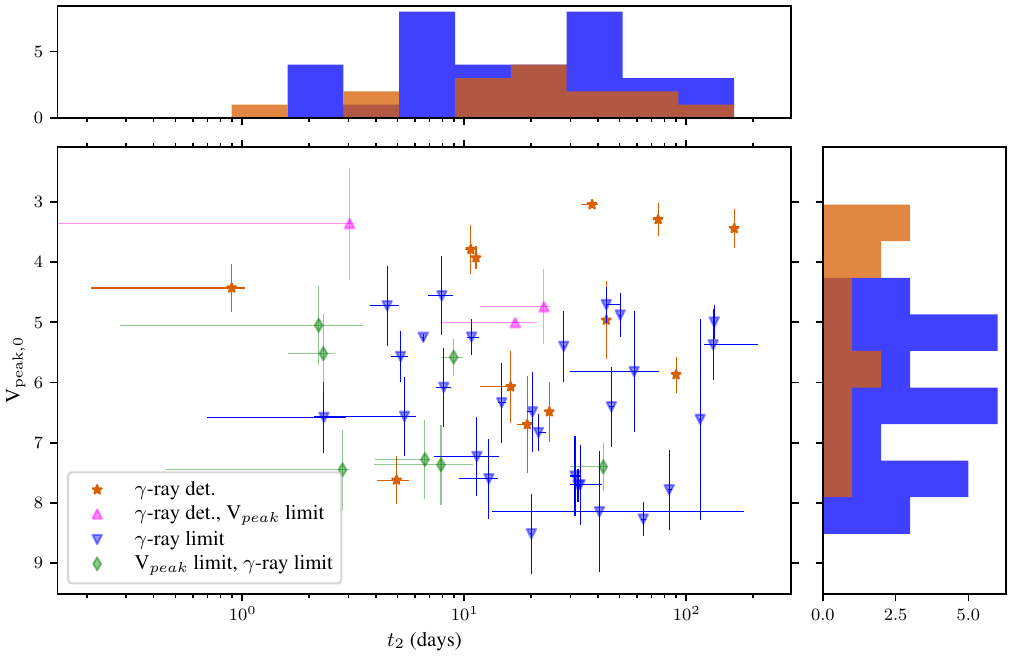}
\caption{\hl{Same as Figure \ref{Fig:vmax_vs_t2}, except now showing the extinction corrected peak $V$ band magnitudes ($V_{\mathrm{peak},0}$).} Sources with $V_{\mathrm{peak},0} < 5$ are frequently detected, and on occasion fainter optical novae are also \emph{Fermi} detected.}
\label{Fig:vpeakint_vs_t2}
\end{center}
\end{figure*}

The equivalent width measurements are derived following \cite{Craig2025}
\kvs{using} 
the publicly available \texttt{Python} code
\footnote{\url{https://github.com/AdkPete/reddening-dibs}}. Each equivalent width is measured by directly integrating across the line, with a linear continuum model connecting points in the continuum on either side of the line. The end points for the continuum are selected based on a model fit to the local continuum and are intended to provide a good estimate of the local continuum level. Typically, multiple DIBs are available for a given nova, and we calculate a weighted average $E(B-V)$ across all available DIBs based on the equivalent width uncertainties. When multiple spectra are used for a nova, we compute a weighted average of the equivalent width measurements for each DIB first, then compute $E(B-V)$ for each DIB. The final $E(B-V)$ value is then the weighted average of the  individual $E(B-V)$ values. 

In addition to the measurement uncertainties, there is scatter in the DIB equivalent width--$E(B-V)$ relations of a physical origin \citep{Krelowski_2019}. This scatter is thought to be due to varying physical properties in the absorbing clouds along the line of sight. Our uncertainty estimates include an estimate of the scatter derived from the data in \cite{Friedman_2011}, which is combined in quadrature with the uncertainties from the equivalent width measurements. The uncertainty derived from the relationship scatter is typically larger than the equivalent width uncertainty.

Not all of the novae in our sample have adequate spectroscopic coverage to allow for DIB measurements. In this case, we use 3D dust maps \citep{Green_etal_2019, Marshall_etal_2006, Drimmel_etal_2003} to derive $E(B-V)$ in novae that have reliable \emph{Gaia} DR3 parallaxes \citep{GaiaDR3,GaiaMission}. When parallax distances are not available, we use the colour calibrations from \citet{Craig2025}, combined with $(B-V)$ measurements from the light curves, to constrain the reddening. We prefer measurements of the colour at $t_2$ when possible due to the lower scatter, but will use the colour at peak if a colour measurement near $t_2$ is not available. In one case, V5855 Sgr, none of these methods are viable, so we adopt an estimated extinction based on 2D dust maps \citep{Schlafly_Finkbeiner_2011}. The integrated Galactic extinction in that direction is estimated as $A_V = 1.6$, and we adopt 0.8 $\pm$ 0.8 as a rough extinction estimate. The resulting $A_V$ values across our sample can be seen in Tables \ref{table:characteristics1} and \ref{table:characteristics2}.

\subsection{Distances}

Some of the novae in our sample have accurate parallax measurements available through \emph{Gaia} DR3 \citep{GaiaDR3,GaiaMission}. For these sources, distances derived from the parallaxes are the most direct distance measurements available. For this reason, when novae have $\geq 3\sigma$ parallaxes, we use the distances from \citet{Schaefer2022} that are derived using these parallaxes. To systematically constrain the distances to the remaining novae in Tables \ref{table:characteristics1} and \ref{table:characteristics2}, we follow the methods of \citet{Craig2025}. To start, we compare the measured foreground reddening with three-dimensional dust maps. Our distances use the \citet{Zucker2024} maps when possible, followed by the \cite{Chen2019} map, \cite{Green_etal_2019} map, and finally the \cite{Marshall_etal_2006} map. In addition, we used the Galactic mass model described in \citet[][see also \citealt{Kawash_etal_2021b}]{Kawash_etal_2022} to further constrain the distances. We generated a population of $10^7$ simulated novae in this model, distributed in proportion to stellar mass as described by a realistic  three-dimensional model of the Milky Way \citep{Robin_etal_2003}. It is possible to further constrain the distances using the peak absolute magnitude, but we do not use this information here. This is because we intend to use the distances for the purposes of constraining the $\gamma$-ray luminosity (and M$_V$), so including assumptions about the luminosity in the distance measurements will bias our resulting luminosity measurements. This is especially relevant here since the optical and $\gamma$ ray luminosities may be related, and therefore we don't want to use the optical luminosities in constraining the $\gamma$-ray luminosities. As shown in \cite{Craig2025}, the \emph{Gaia} parallax distances typically provide good agreement with the distances derived based on Galactic mass models and 3D dust maps.

\begin{table*}
\caption{Correlation coefficients and corresponding p-values for the $\gamma$-ray properties of novae. The last two columns give the size of the data set, with N$_{det}$ showing the number of $\gamma$-ray detected novae and N$_{limit}$ giving the number of upper limits for novae not detected in the $\gamma$-rays. The \emph{Fermi} non-detected novae are only used for the censored Spearman column; the other cases only use the $\gamma$-ray detected sample. All of the correlations in this table are computed using the average $\gamma$-ray luminosity or flux for each case. P-values below 0.05, which match our significance threshold, are marked in bold.}\label{tab:corrstats}
\begin{tabular}{cccccccccc}
\hline
x & y & \multicolumn{2}{|c|}{Pearson} & \multicolumn{2}{|c|}{Spearman} & \multicolumn{2}{|c|}{Censored Spearman} & N$_{det}$ & N$_{limit}$\\
\hline
  &  &  \multicolumn{1}{|c}{r} & \multicolumn{1}{c|}{p-value} & \multicolumn{1}{|c}{r$_s$} & \multicolumn{1}{c|}{p-value} & \multicolumn{1}{|c}{r$_s$} & \multicolumn{1}{c|}{p-value} & & \\
\hline
\multicolumn{10}{|c|}{V5668 Sgr Distance = \hl{3.9} kpc}\\
\hline
 $\Delta v$ & $\log_{10}$(L$_{\gamma,avg}$) & 0.69 & \bf{0.006} & 0.67& \bf{0.012} & 0.70 & \bf{0.0003} & 14 & 14 \\ 
 $v_1$ & $\log_{10}$(L$_{\gamma,avg}$) & 0.14 & 0.657 & $-0.2$ & 0.492 & $-0.2$ & 0.486 & 14 & 14 \\
 $v_2$ & $\log_{10}$(L$_{\gamma,avg}$) & 0.46&0.088 & 0.61 & \bf{0.023} & 0.59 & \bf{0.002} & 14 & 14 \\ 
 M$_V$ & $\log_{10}$(L$_{\gamma,avg}$) & $-0.54$&0.073 & $-0.39$ & 0.208 & $-0.28$ & 0.076 & 12 & 40 \\ 
 t$_{2}$ & $\log_{10}$(L$_{\gamma,avg}$) & $-0.38$&0.166 & $-0.49$ & 0.067 & $-0.27$ & 0.054 & 15 & 38 \\
 $\gamma$-ray duration & $\log_{10}$(L$_{\gamma,avg}$) & $-0.4$ & 0.134 & $-0.29$ & 0.296 & -- & -- & 15 & 0 \\
\hline
\multicolumn{10}{|c|}{V5668 Sgr Distance = 1.2 kpc}\\
\hline
 $\Delta v$ & $\log_{10}$(L$_{\gamma,avg}$) & 0.79& \bf{0.001} & 0.72&\bf{0.004} & 0.74& \bf{0.0001} & 14 & 14 \\
 $v_1$ & $\log_{10}$(L$_{\gamma,avg}$) & 0.18&0.561 & $-0.13$&0.642 & $-0.02$&0.934 & 14 & 14 \\ 
 $v_2$ & $\log_{10}$(L$_{\gamma,avg}$) & 0.54&\bf{0.037} & 0.67&\bf{0.014} & 0.63&\bf{0.001} & 14 & 14 \\ 
 M$_V$ & $\log_{10}$(L$_{\gamma,avg}$) & $-0.61$&\bf{0.034} & $-0.57$&0.055 & $-0.72$&0.090 & 12 & 40 \\
 t$_{2}$ & $\log_{10}$(L$_{\gamma,avg}$) & $-0.42$&0.133 & $-0.5$&0.057 & $-0.28$&\bf{0.048} & 15 & 38 \\
 $\gamma$-ray duration & $\log_{10}$(L$_{\gamma,avg}$) & $-0.56$&\bf{0.033} &$ -0.38$ & 0.168 & -- & -- &15 & 0 \\ 
\hline
\multicolumn{10}{|c|}{Distance Independent Stats}\\
\hline
 $\gamma$-ray duration & $\Delta v$ & $-0.68$&\bf{0.010} & $-0.63$&\bf{0.021} & -- & -- &15 & 0 \\ 
 $\gamma$-ray duration & $v_1$ & $-0.68$&\bf{0.001} & $-0.6$&\bf{0.026} & -- &-- & 15 & 0 \\ 
 $\gamma$-ray duration & $v_2$ & $-0.82$&\bf{0.0002} & $-0.79$&\bf{0.001} & -- & -- &15 & 0 \\ 
 $v_{peak,0}$ &  $\log_{10}$($\gamma$-ray Flux) & $-0.11$&0.780 & 0.0&0.987 & $-0.34$&\bf{0.033}   & 12 & 28 \\ 
 $\gamma$-ray duration & t$_{2}$ & 0.22&0.417 & 0.54&\bf{0.043} & -- & -- & 15 & 0 \\ 
\hline
\end{tabular}
\end{table*}

There is one $\gamma$-ray detected system, V5668 Sgr, that has a significantly lower distance reported in the literature than the one derived using our method. Our adopted distance towards this nova is 3.9$^{+0.3}_{-0.7}$\,kpc based on DIB extinction measurements combined with 3D dust maps and our Galactic mass model. Similarly, using the \cite{Chen2019} dust maps and estimated extinction, \cite{Gordon_etal_2021} estimate a distance of 2.8 $\pm$ 0.5\,kpc. Expansion parallax measurements, however, indicate a significantly smaller distance, with estimates of 1.2 $\pm$ 0.4\,kpc \citep{Takeda2022} and $1.54$\,kpc from \citep{Banerjee2016}. At our adopted distance, the absolute magnitude of this nova is $-9.6^{+0.5}_{-0.3}$, which is brighter than the average -7.5 \citep{Shafter_2017} typically seen for novae, consistent with an overestimated distance. Assuming a distance of 1.2 kpc instead, $M_V$ for this nova becomes -7.1. It is possible that the inclusion of the Galactic mass model is tending towards larger distances for this source, since many of the stars in this direction will be at closer to the Galactic center. For consistency, we retain our estimate for the distance for the main results of this paper, but all figures showing correlations between $\gamma$-ray and optical properties show this nova at both 3.9 and 1.2 kpc.

\section[]{What nova properties do $\gamma$-ray luminosity and duration depend on?}
\label{sec:gamma-ray-dependence}

In this section, we explore which properties of a nova are related to its $\gamma$-ray emission, including the outflow velocities (\S \ref{sec:vels}), optical luminosity and $t_2$ (\S \ref{sec:opt}), and the dust formation properties of novae (\S \ref{sec:dust}). This analysis is based on detecting correlations between parameters, which is done using Pearson and Spearman correlation coefficients. The Pearson coefficient measures the linear correlation between two parameters, and works well when the relationship is approximately linear. It makes the assumption that the data is normally distributed, and can be rather sensitive to outliers. The Spearman rank coefficient measures the monotonic relationship between parameters, based on the ranks of the data. This makes no assumptions about the underlying distribution of the variables, is more robust against outliers than the Pearson coefficient, and works on non-linear relationships (as long as they are monotonic). Spearman coefficients can struggle on data sets that have many equal values (creating tied ranks). We consider both coefficients, but prefer the Spearman coefficient when the relationship between variables may be non-linear.

In each case, we compute p-values (used in addition to the coefficients, with p-values less than 0.05 indicating significant correlations (at the 95 per cent level)). P-values are computed in \textsc{scipy} following the documentation, and using permutation testing (with a maximum number of samples set to 100,000), which can provide more accurate p-values on small sample sizes. These quantities are presented in Table \ref{tab:corrstats}.  Some of our parameters exhibit non-linear relationships, where the Spearman coefficient is more appropriate. We convert the $\gamma$-ray luminosities and fluxes into log space for the correlations, and the magnitudes are already in log space, while $t_2$, velocities and $\gamma$-ray durations are not switched to log space. This is only relevant for the Pearson coefficients, because the Spearman coefficient is calculated based on the ranks of the data, which are invariant under the log transformation.

For some of the cases examined here, the correlation analysis benefits from the inclusion of novae that only have limits on the $\gamma$-ray luminosity or fluxes. We use the methods implemented in \textsc{asurv} \citep{Feigelson1985,Isobe1986,asurv} to apply censored statistics to our correlations where appropriate, calculating a Spearman coefficient and corresponding p-value. The results of this analysis can also be found in Table \ref{tab:corrstats}. In most cases when the censored statistics detect a correlation, we can detect it with just the $\gamma$-ray detected sample as well. The only exceptions are when comparing the peak extinction corrected $V$-band magnitude against the $\gamma$-ray flux, and comparing $t_2$ with the average luminosity (where it is just a small difference in the p-value close to the significance threshold). We note that for the correlations that require both velocities and $\gamma$-ray luminosities (or upper limits), the full sample consists of 28 novae, which is small enough that the p-values on the censored Spearman test are unreliable (these are only considered reliable with samples larger than 30). However, these statistics agree well with the uncensored version in all cases examined here, so there is no impact from the potentially unreliable p-values.

\begin{figure}
\begin{center}
\includegraphics[]{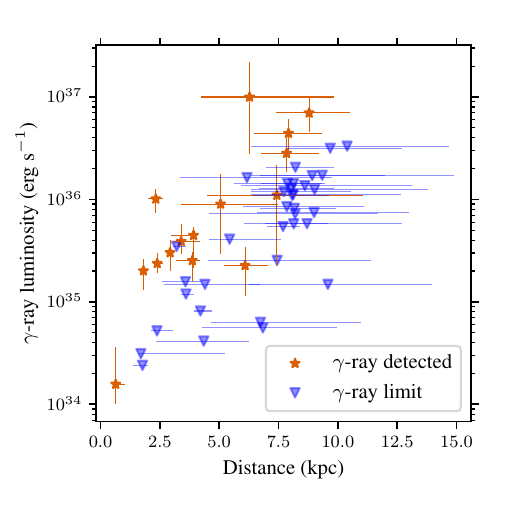}
\caption{$\gamma$-ray luminosity for the novae in our sample plotted against their distances. The $\gamma$-ray luminosities in our sample span three orders of magnitude, ranging from $10^{34}$ to 10$^{37}$ erg s$^{-1}$. At larger distances, such as around 8 kpc for novae close to the Galactic center, only the brighter end of this luminosity range is detectable.}
\label{fig:L_dist}
\end{center}
\end{figure}

\subsection{$\gamma$-ray Properties of Our Nova Sample}

\begin{figure*}
\begin{center}
  \includegraphics[]{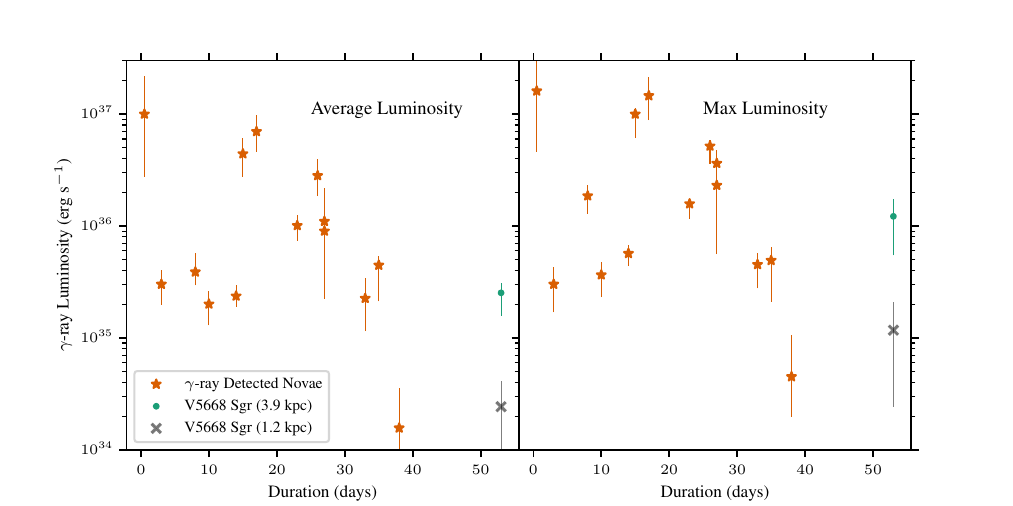}
\caption{The duration of the $\gamma$-ray emission plotted against the $\gamma$-ray luminosity. On the left panel, we have the average $\gamma$ ray luminosity, while the right panel displays the maximum $\gamma$-ray luminosity. In both cases our data indicates a significant negative correlation between these two parameters, with longer duration $\gamma$-ray emission corresponding to lower average and maximum luminosities. \hl{These correlations are driven by novae with $\gamma$-ray emission durations greater than 14 days, where the correlation is clear. This correlation does not translate down to the lower duration systems, where there are four novae at lower luminosities than would be expected from the observed correlation.} The translucent black x marks V5668 Sgr at a distance of 1.2 kpc, while the green round point shows V5668 Sgr assuming a distance of 3.9 kpc. The orange stars represent all of the other novae in our sample.}
\label{Fig:avgluminosity_vs_gamma_duration}
\end{center}
\end{figure*}

\begin{figure}
\begin{center}
  \includegraphics[]{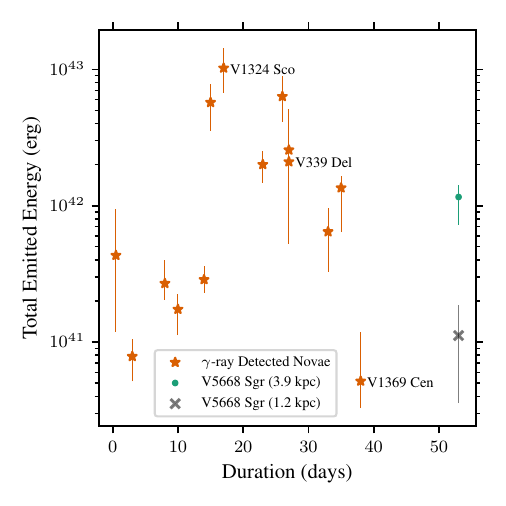}
\caption{Total emitted energy in the $\gamma$-rays as a function of the duration of the detected $\gamma$-ray emission. At short durations, the emitted energy does not seem to depend on the duration of the detectable emission, while at durations greater than 15 days there is a negative correlation. The novae that are labelled in this figure match the sample used for this analysis in \citet{Cheung_etal_2016}. The translucent black x marks V5668 Sgr at a distance of 1.2 kpc, while the green round point shows V5668 Sgr assuming a distance of 3.9 kpc. The orange stars represent all of the other novae in our sample.}
\label{fig:energy}
\end{center}
\end{figure}

\kvs{Novae that are brighter in the optical are more likely to be detected in $\gamma$-rays.}
Figure~\ref{Fig:vmax_vs_t2} displays the apparent peak $V$-band magnitudes of our sample plotted against $t_2$, 
broken up into the $\gamma$-ray detected and $\gamma$-ray non-detected samples. 
\kvs{The concentration of \emph{Fermi} detections among optically bright novae ($V_{\rm peak} \lesssim 8$ mag in most cases) is expected, 
since brighter novae are typically closer and thus more likely to produce detectable $\gamma$-ray flux for \emph{Fermi}/LAT.
Many optically faint novae likely also produced $\gamma$-rays, but below the \emph{Fermi}/LAT sensitivity threshold.}
Some relatively bright novae like KT Eri are not \emph{Fermi} detected \citep{Franckowiak_etal_2018}, 
while some fainter novae like V1707 Sco are \emph{Fermi} detected. 
Across our sample, 78 per cent of novae that peak brighter than 6th magnitude are \emph{Fermi} detected. 
Those brighter than 8th magnitude are 65 per cent detected, and novae brighter than 10th magnitude are 37 per cent detected. 
Only one nova fainter than 10th magnitude, V1707 Sco, has been \emph{Fermi} detected. 
Note that dust extinction in the Galactic plane can play a more significant role than distance 
in dimming a nova's optical flux \citep{Shafter_2017, Kawash_etal_2021b}. 
Therefore, Figure~\ref{Fig:vpeakint_vs_t2} is similar to Figure~\ref{Fig:vmax_vs_t2}, 
but we plot extinction-corrected peak apparent magnitude (see section \ref{sec:specprop} for details on the extinction measurements). 
The distribution of $V_{\rm peak,0}$ is narrower than $V_{\rm peak}$, but some of the \emph{Fermi}-detected novae, like V1707 Sco and V1324 Sco, 
still reside at relatively faint peak optical brightness.

The sample of novae detected by \emph{Fermi} is ultimately sensitivity limited. Shown in Figure~\ref{fig:L_dist} is the average $\gamma$-ray luminosity for the novae in our sample compared with their distances. At high distances, we lose out on detecting novae with lower $\gamma$-ray luminosities due to insufficient sensitivity to their reduced $\gamma$-ray flux. Only novae with relatively large $\gamma$-ray luminosities are detectable at the larger distance end.

From our sample of $\gamma$-ray detected novae, it is clear that novae have spread of several orders of magnitude in their $\gamma$-ray luminosity, ranging from $2 \times 10^{34}$ to $1.2 \times 10^{37}$ erg s$^{-1}$. This luminosity spread is too large to be explained by the uncertainties on derived \emph{Fermi} fluxes or the uncertainty in the distances to our sample. 
A substantial \kvs{intrinsic} spread in the $\gamma$-ray luminosities of 
\kvs{novae} 
is required to explain the measured fluxes.

We plot $\gamma$-ray duration against $\gamma$-ray luminosity in Figure~\ref{Fig:avgluminosity_vs_gamma_duration}. The duration of the $\gamma$-ray emission is negatively correlated with the $\gamma$-ray luminosity \hl{assuming a distance of 1.2 kpc for V5668 Sgr}. The Pearson coefficient is \hl{\nullstat{$-0.56$}}, with a corresponding p-value of \nullstat{$0.033$}, and the Spearman coefficient is \hl{\nullstat{$-0.38$}}, with a corresponding p-value of \hl{\nullstat{$0.168$}} (Table \ref{tab:corrstats}). This implies that more energetic shocks in the ejecta tend to be shorter lived, allowing for both large luminosities and short emission duration.

In Figure~\ref{fig:energy}, we display the total emitted energy in the $\gamma$-rays, compared with the duration of the $\gamma$-ray emission. Since longer duration systems tend to have lower average $\gamma$-ray luminosities, it is not generally the case that longer durations correspond to higher total energies. Other studies have found that novae with longer duration $\gamma$-ray emission tend to have lower overall emitted energies \citep{Cheung_etal_2016,Franckowiak_etal_2018}. Our sample includes the same systems considered in those papers, and these are labeled in Figure~\ref{fig:energy}. Across the systems with durations of at least 15 days, we recover the same trend with longer emission durations corresponding to lower total energies. Below 15 days, the novae in our sample show lower total emitted energies than what might be expected at these low durations. It could be that the short duration systems are relatively faint systems, where the flux drops below our detection threshold relatively quickly. There is some bias in these observations, in that the brighter systems will tend to be detected over a longer baseline. Despite this, long duration systems tend to have relatively small emitted energies.

The total emitted energies here are estimated based on the average $\gamma$-ray luminosities combined with the emission durations. Some of the total energies plotted here disagree significantly with those presented in \cite{Cheung_etal_2016}, with our estimates of the total energies tending to be larger. These discrepancies arise from a combination of three factors, the most significant of which being the assumed distance towards the nova. In some cases, such as V1324 Sco, there is a factor of $\sim 2$ difference in the estimated distances, in turn leading to factor of 4 differences in the energy. In addition, different assumptions are made for the spectral index of the $\gamma$-ray emission, which introduces an additional factor of up to 2 difference in the luminosity. Finally, we use fluxes estimated from different spectral models, which leads to similar, but slightly different, estimates for the $\gamma$-ray flux. These combined can amount to a difference of nearly an order of magnitude in the total measured energy. For the most part, different distance measurements and spectral indices are the primary factor for the different values in the total energy.

\subsection{$\gamma$-rays and Outflow Velocities}\label{sec:vels}

\begin{figure*}
\begin{center}
  \includegraphics[width=\textwidth]{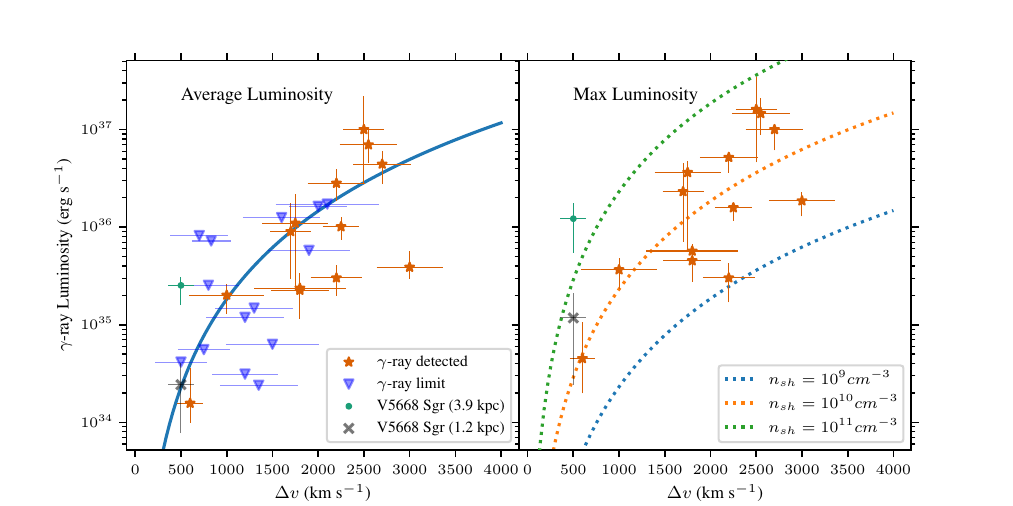}
\caption{The $\gamma$-ray luminosity of the novae in our sample plotted against the relative velocity between the slow and fast outflows ($v_2 - v_1 = \Delta v$). The left panel displays the average $\gamma$-ray luminosity over the duration of \emph{Fermi} detection, while the right panel displays the maximum $\gamma$-ray luminosity. The orange stars show novae with \emph{Fermi}-detected $\gamma$-rays, while the blue triangles represent upper limits for novae that lack a $\gamma$-ray detection. The translucent black x marks V5668 Sgr at a distance of 1.2 kpc, while the green round point shows V5668 Sgr assuming a distance of 3.9 kpc. There is evidence for a significant positive correlation in both cases, with the $\gamma$-ray luminosity increasing approximately with the cube of $\Delta v$. The blue solid line in the left panel shows the predicted cubic scaling for this relationship, fit to the \emph{Fermi} detected novae. The dotted lines in the right panel show the predicted $\gamma$-ray luminosity based on the observed velocities and an assumed ejecta density, assuming that 1 per cent of the shock luminosity is emitted in the $\gamma$-rays. Each line assumes a different, but possible ejecta density, with a one order of magnitude difference between lines and the lower density models having lower $\gamma$-ray luminosities.}
\label{Fig:avgluminosity_vs_dv}
\end{center}
\end{figure*}

\begin{figure*}
\begin{center}
\includegraphics[width=\textwidth,trim={0.25cm 0.25cm 0.25cm 0.25cm}, clip]{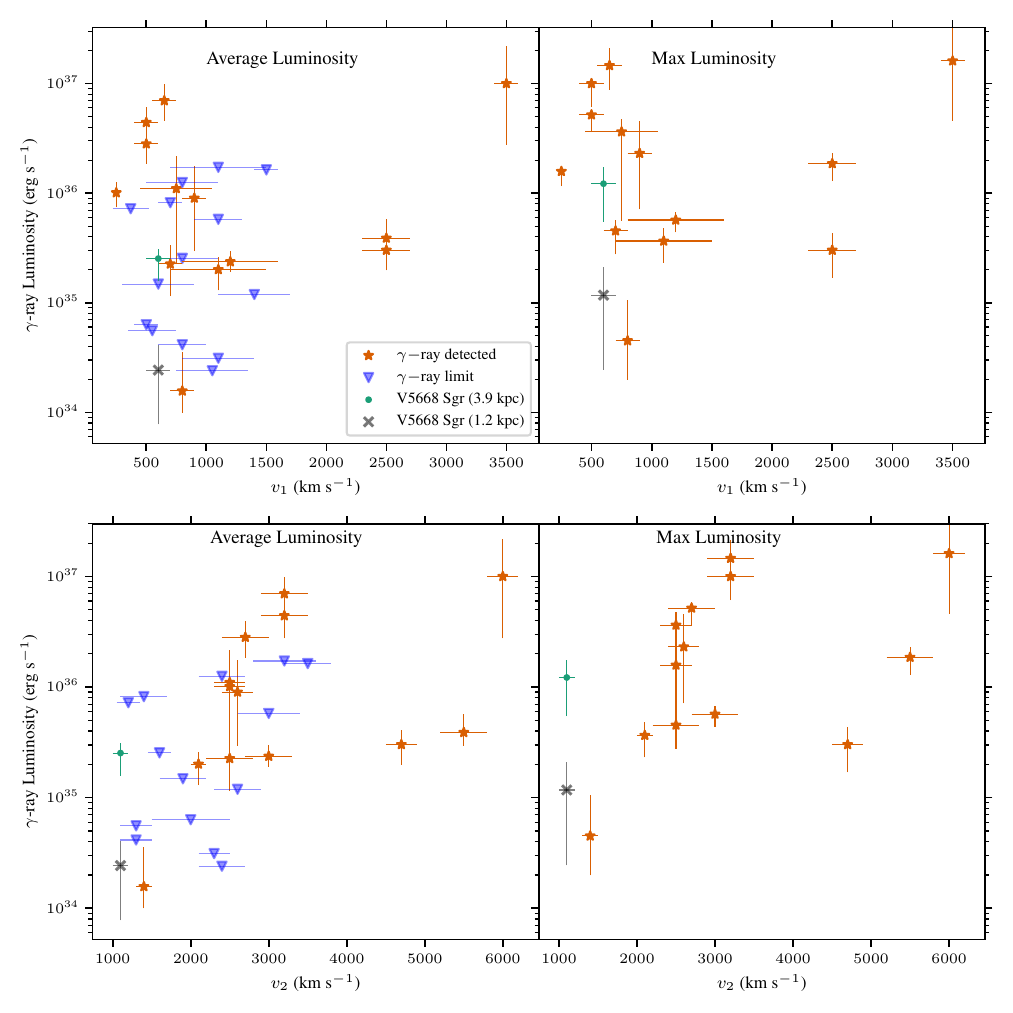}
\caption{Individual velocity components against the $\gamma$-ray luminosity. The left column displays the average $\gamma$-ray luminosity, while the right column displays the maximum $\gamma$-ray luminosity. The top row shows the slow outflow velocity ($v_1$), and the bottom row displays the fast outflow velocity ($v_2$). The orange stars show novae with \emph{Fermi}-detected $\gamma$-rays, while the blue triangles represent upper limits on $L_{\gamma}$ for novae that lack a $\gamma$-ray detection. The translucent black x marks V5668 Sgr at a distance of 1.2 kpc, while the green round point shows V5668 Sgr assuming a distance of 3.9 kpc. The fast outflow velocity is correlated with the luminosities, while the slow outflow velocity is not.}
\label{Fig:avgluminosity_vs_v1_v2}
\end{center}
\end{figure*}

If $\gamma$-rays are produced in shocks between interacting outflows, then the shock energy and luminosity are expected to scale with the cube of the shock velocity \citep{Metzger_etal_2015}. The expected shock luminosity, from \cite{Metzger_etal_2015}, is:
\begin{equation}\label{eqn:shockL}
L_{sh} = \frac{9\pi}{32}R_{ej}^2n_{sh}m_{p}v_{sh}^{3}
\end{equation}
Here, $n_{sh}$ is the number density of the gas being shocked, $R_{ej}$ is the radius of the ejecta, $m_p$ is the mass of a proton, and $v_{sh}$ is the shock velocity. 

The model that we test here is that the collision of the fast flow ($v_2$) with the slow flow ($v_1$) is the origin of the shocks and $\gamma$-rays (Figure~\ref{Fig:spec_scheme}). In this scenario, it is the relative velocity 
between the outflow components, $v_{sh} = \Delta v = v_2 - v_1$,  that is the velocity of the shock.
This scaling of $L_{sh} \propto v_{sh}^3$, combined with the observed $\Delta v$ measurements, leads to a predicted spread in the $\gamma$-ray luminosity of more than two orders of magnitude, in good agreement with observed luminosities. The shock properties also depend on $n_{sh}$, which depends on the ejecta mass, geometry, and the radius of the ejecta when the shocks are present. Ejecta masses in novae can vary by up to four orders of magnitude \citep{Yaron_etal_2005,Henze_etal_2014_Mar}, and this can lead to large differences in the ejecta density and therefore shock luminosity. 
\kvs{As only a small fraction of the shock's energy is spent on accelerating high-energy particles that emit $\gamma$-rays, most}
of the shock's energy output will not arise in the $\gamma$-rays, and instead is likely to appear in the optical or X-ray bands. In novae with main-sequence companions, the shocks are deeply embedded, and most of the shock luminosity appears to emerge in the optical band
 \citep{Li_etal_2017_nature,Aydi_etal_2020}. Only about 1 per cent of the shock energy is expected to result in GeV $\gamma$-ray emission \citep{Metzger_etal_2015}.

Across our sample, the observations are fully consistent with the predictions of shock models, in particular that novae with larger relative (and fast) outflow velocities show larger $\gamma$-ray luminosities compared to lower velocity eruptions. In Figure~\ref{Fig:avgluminosity_vs_dv}, we show the measured $\Delta v$ plotted against the GeV $\gamma$-ray luminosity measured by \emph{Fermi}/LAT. The Spearman correlation coefficient in this case is \hl{\nullstat{0.67}}, with a corresponding p-value of \hl{\nullstat{0.012}} (Table \ref{tab:corrstats}), indicative of a significant correlation. \hl{These values assume a distance of 3.9 kpc  for V5668 Sgr.} On the left panel, which displays the average $\gamma$-ray luminosity, we show a function fit to the data assuming that $L_{\gamma} = av_{sh}^3$, where $a$ is a constant selected to fit the data, intended to test the anticipated velocity scaling. In the right panel, we display the maximum observed $\gamma$-ray luminosity. The dotted lines are calculated using Equation \ref{eqn:shockL}, assuming an ejecta radius of 4\,AU and that 1 per cent of the shock luminosity is emitted in the $\emph{Fermi}$ band. This ejecta radius is calculated based on typical slow flow velocities and the time-scale for the $\gamma$-ray detection, but this can vary appreciably between novae. The number densities are estimated based on typical ejecta masses, choosing here masses of 10$^{-4}$, 10$^{-5}$ and 10$^{-6}$ M$_{\odot}$, and do not necessarily span the full range of plausible densities.

While the scaling overall appears to be correct, there remains a considerable amount of scatter in the average $\gamma$-ray luminosities across our sample. It is likely that this is derived from differences in the ejecta densities between novae, arising from differences in ejecta mass and geometry. Estimating ejecta densities in particular systems is beyond the scope of this paper, but may be included in future work. As seen in the right panel of Figure~\ref{Fig:avgluminosity_vs_dv}, all of the observations can be well described in this model by a suitable combination of ejecta radius and density, combined with observed velocities.

\begin{figure}
\begin{center}
  \includegraphics[]{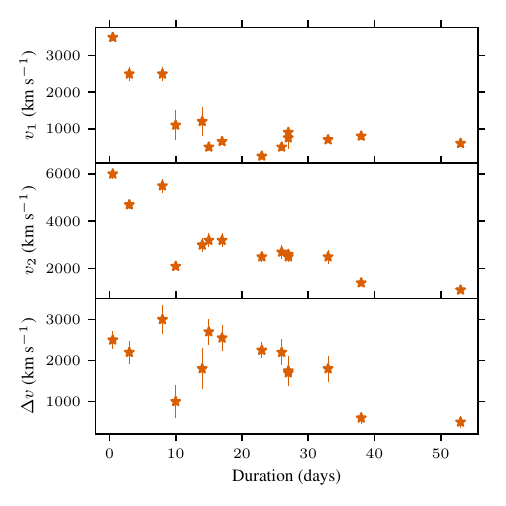}
\caption{Velocities for the two outflow components and their relative velocity ($v_1$, $v_2$, $\Delta v$) plotted against the duration of detectable $\gamma$-ray emission. There is a significant negative correlation between each of these outflow velocities and the $\gamma$-ray duration.}
\label{Fig:v_vs_gamma_duration}
\end{center}
\end{figure}

We also consider the location of V5668 Sgr with an assumed distance of $1.2$ kpc based on \cite{Takeda2022}. In the case of $\Delta v$ and the $\gamma$-ray luminosities, the point assuming the nearby distance to V5668 Sgr falls much closer to the evident relation between velocity and luminosity. Recomputing the correlation between $\Delta v$ and the average luminosity with this value for the distance, we recover an increased Spearman correlation coefficient of \hl{\nullstat{0.72}}, with a p-value of \hl{\nullstat{0.004}} (Table \ref{tab:corrstats}). This is an even stronger correlation than that obtained above, although the correlation is clearly present regardless of the distance used for V5668~Sgr.

In addition to correlations between $L_{\gamma}$ and $\Delta v$, we also examine how the luminosity scales with the individual velocity components, shown in Figure~\ref{Fig:avgluminosity_vs_v1_v2}. A positive correlation is detected between $v_2$ and the $\gamma$-ray luminosity, \hl{assuming a distance of 1.2 kpc towards V5668 Sgr}, of similar significance as the $L_{\gamma}-\Delta v$ relation (\hl{Spearman} p-value = \nullstat{0.014}). Meanwhile, there is no significant correlation between $v_1$ and the luminosity (\hl{Spearman} p-value of \nullstat{0.642}). The slow outflow velocity then appears to be less relevant in setting the strengths of the internal shocks than the fast outflow velocity. This is expected in our model where the shocks are driven by the collision of the slow and fast outflows, as $\Delta v$ is primarily driven by the fast outflow velocity.

The duration of the detectable $\gamma$-ray emission may also be expected to depend on the outflow velocities. In Figure~\ref{Fig:v_vs_gamma_duration} we show the duration of $\gamma$-ray emission plotted against $\Delta v$, $v_1$ and $v_2$. A significant negative correlation exists between the duration of the $\gamma$-ray emission and $\Delta v$. The correlation is seen with a Pearson coefficient of \nullstat{$-0.68$}, with a corresponding p-value of \hl{\nullstat{0.01}}, and with a Spearman coefficient of \nullstat{$-0.63$}, with a corresponding p-value of \hl{\nullstat{0.021}} (Table \ref{tab:corrstats}). We similarly find that both the slow and fast outflow components are correlated negatively with the $\gamma$-ray duration. In the context of our shock model, the relationship between duration and $\Delta v$ is likely responsible for the correlation between the emission duration and luminosity seen in the last section.

\subsection{$\gamma$-rays and Optical Light Curve Properties}\label{sec:opt}

There is strong evidence for a correlation between the optical and $\gamma$-ray light curves of novae \citep{Li_etal_2017_nature,Aydi_etal_2020}. The shock emission that does not arise in the $\gamma$-rays is expected to get reprocessed by the ejecta into optical emission, either through the reprocessing of thermal X-rays emitted by the shock-heated gas \citep{Metzger_etal_2015}, or possibly through turbulent mixing of shock-heated gas with cooler post-shock material \citep{Metzger2025}. The remaining shock energy is likely to be on a similar scale to the Eddington luminosity, and therefore will make a substantial contribution to the overall optical flux of the nova.

\begin{figure*}
\begin{center}
  \includegraphics[width=\textwidth]{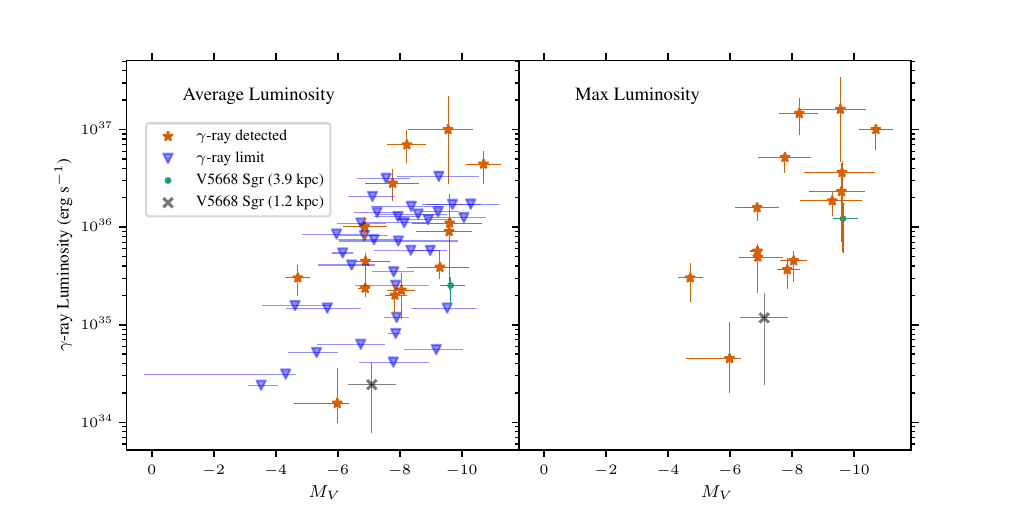}
\caption{Peak absolute $V$-band magnitudes plotted with the $\gamma$-ray luminosities for our sample. The left panel displays the average $\gamma$-ray luminosity, while the right panel displays the maximum $\gamma$-ray luminosity. The orange stars show novae with Fermi detected $\gamma$-rays, while the blue triangles represent novae that lack a $\gamma$-ray detection. The translucent black x marks V5668 Sgr at a distance of 1.2 kpc, while the green round point shows V5668 Sgr assuming a distance of 3.9 kpc. There is evidence for a correlation in both cases, where novae with smaller absolute magnitudes tending to have larger $\gamma$-ray luminosities. This is expected if the shocks are responsible for a substantial amount of optical flux, with stronger shocks then increasing both the optical and $\gamma$-ray luminosities.}
\label{Fig:Mv_vs_luminosity}
\end{center}
\end{figure*}

As a result, we should expect that the properties of the optical light curves of the novae in this sample might be correlated with the $\gamma$-ray luminosities and duration. For our sample, we plot the absolute $V$ band magnitude against the $\gamma$-ray luminosities in Figure~\ref{Fig:Mv_vs_luminosity}, correcting for interstellar extinction with $A_V$ values listed in Tables \ref{table:characteristics1} and \ref{table:characteristics2}. The $\gamma$-ray luminosities in our sample do show evidence for a correlation with the absolute $V$-band magnitude of the nova (Table \ref{tab:corrstats}), especially the maximum luminosity.

We note that the distance accuracy is of particular importance in this case. As can be seen with V5668 Sgr, inaccurate distances will tend to spread the systems out along the observed correlation. This naturally raises the concern that our observed effect could be due to inaccurate distances, as the distance errors could conceivably make the correlation appear stronger than it really is by spreading the systems out along this direction. However, aside from V5668 Sgr, the distances across the sample are in reasonable agreement with values from the literature derived using different techniques. In addition, inaccurate distances cannot explain the full range of observed $\gamma$-ray fluxes, and so some of the observed spread is due to intrinsic differences in the $\gamma$-ray luminosity.

\begin{figure*}
\begin{center}
  \includegraphics[width=\textwidth]{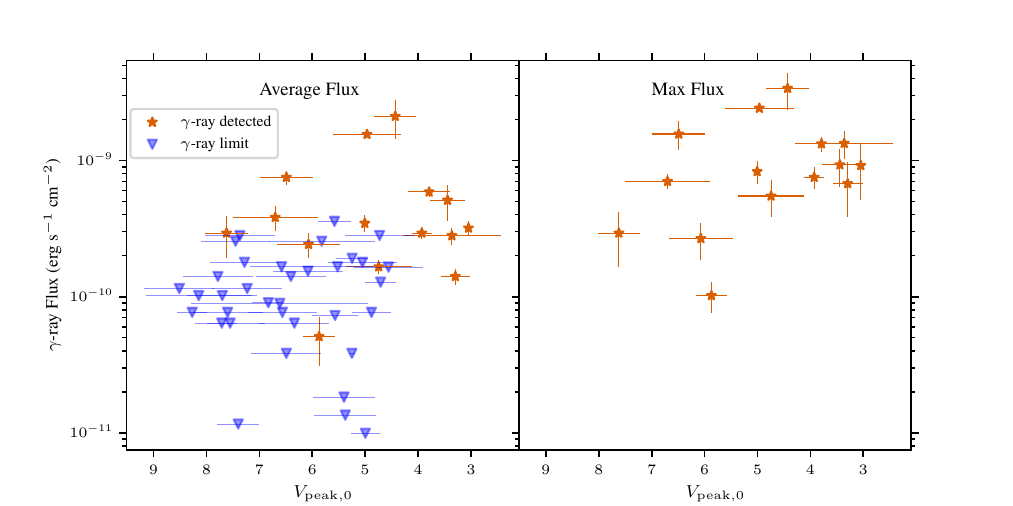}
\caption{The $\gamma$-ray flux plotted against the extinction corrected peak $V$-band magnitude ($V_{\mathrm{peak,0}}$). In the left panel, we show the average $\gamma$-ray flux, and in the right panel we plot the maximum observed $\gamma$-ray flux. Here the blue triangles represent $F_{\gamma}$ upper limits for novae that are not detected by \emph{Fermi}, while the orange stars do have $\gamma$-ray detections from \emph{Fermi}. The right panel instead displays the maximum $\gamma$-ray flux. A correlation is detected when utilizing censored statistics and including the upper limits on the average $\gamma$-ray flux, but not when only using the $\gamma$-ray detected novae. This is a relatively weak correlation even with the inclusion of the upper limits; there is a lot of scatter evident in this case.}
\label{Fig:maxflux_vs_vpeakintrinsic}
\end{center}
\end{figure*}

In Figure~\ref{Fig:maxflux_vs_vpeakintrinsic}, we show the relationship between the observed $\gamma$-ray and optical fluxes. Here there is no requirement for a distance measurement to the source, but this still allows a comparison between the optical and $\gamma$-ray emission. Here we consider the extinction-corrected $V$-band apparent magnitudes to compare with the $\gamma$-ray flux. Using only the novae with $\gamma$-ray detections, no significant correlation is recovered (Table \ref{tab:corrstats}). When using censored statistics and including the upper limits on the $\gamma$-ray luminosity, a correlation is detected. This case has a censored Spearman coefficient of \hl{\nullstat{$-0.34$}} and a corresponding p-value of \hl{\nullstat{0.033}} across a sample of 15 detected systems and 28 non-detections. This provides further evidence for a relationship between the optical and $\gamma$-ray luminosities. A comparison of peak apparent magnitudes with the $\gamma$-ray fluxes can also be found in \cite{Franckowiak_etal_2018}, who does not find a correlation. However, this analysis does not include the censored statistics, and does not apply an extinction correction to the $V$ band magnitudes.

\begin{figure*}
\begin{center}
  \includegraphics[width=\textwidth]{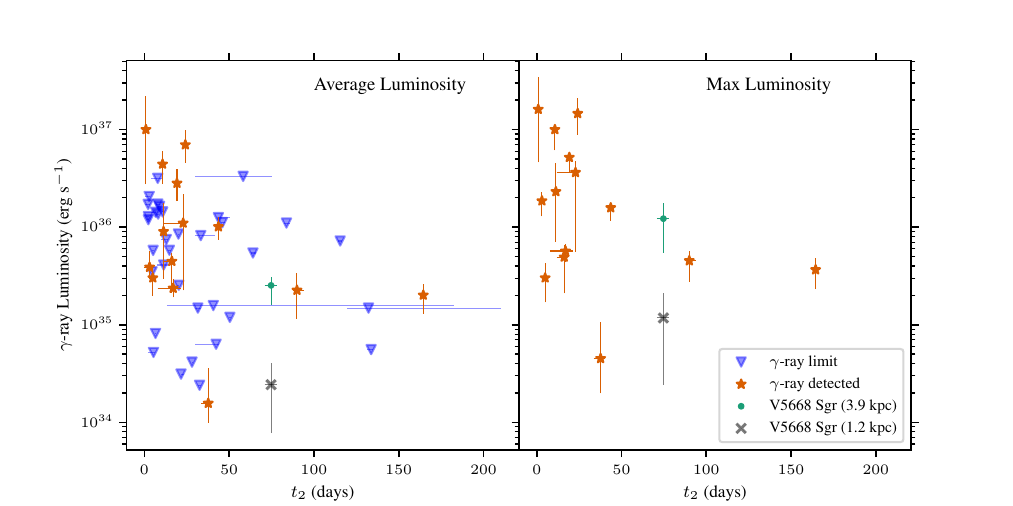}
\caption{The time for the $V$-band light curve to decline by two magnitudes ($t_2$) plotted against the $\gamma$-ray luminosity. The left panel displays the average $\gamma$-ray luminosity, while the right panel displays the maximum $\gamma$-ray luminosity. The orange stars show novae with \emph{Fermi}-detected $\gamma$-rays, while the blue triangles represent upper limits on $L_{\gamma}$ for novae that lack a $\gamma$-ray detection. The translucent black x marks V5668 Sgr at a distance of 1.2 kpc, while the green round point shows V5668 Sgr assuming a distance of 3.9 kpc. There is some evidence for a correlation between $t_2$ and the detected $\gamma$-ray luminosities in the $\gamma$-ray detected sample. A significant correlation is only found when V5668 Sgr has an assumed distance of 1.2 kpc, and when the upper limits are included.}
\label{Fig:avgluminosity_vs_t2}
\end{center}
\end{figure*}

\begin{figure}
\begin{center}
  \includegraphics[width=\columnwidth]{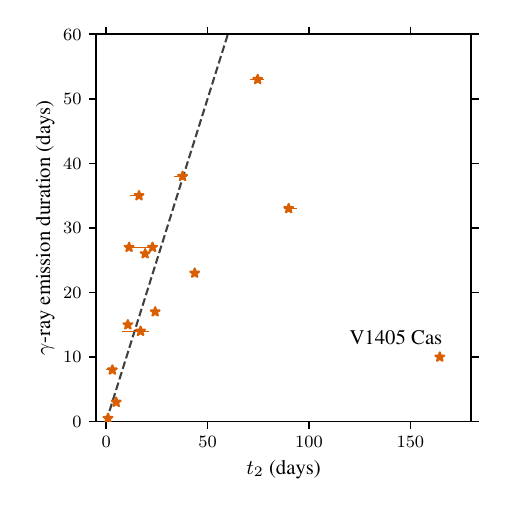}
\caption{The optical decline time-scale ($t_2$) of $\gamma$-ray detected novae, plotted against the duration of the $\gamma$-ray emission. There is generally a positive correlation between the two, with slower novae exhibiting larger duration $\gamma$-ray emission. V1405 Cas appears to be an exception to this, with a large $t_2$ of 164.5 days yet only 10 days of significant $\gamma$-ray emission. \hl{It is worth noting that this would not be true if we measured $t_2$ for this nova based on the first peak in the light curve alone, which falls by 2 magnitudes quickly (7 days) and would place this nova close to other novae with similar $\gamma$-ray durations. However, V1405 Cas has a complex light curve with multiple flares, which drives up our value for $t_2$. Formally the uncertainty in this case is small (0.2 days), and the error bar is smaller than the data point on this figure.}The dashed black line displays a 1:1 line for comparison. Typically the duration of the $\gamma$-ray emission is comparable to $t_2$, but not always.}
\label{Fig:t2_vs_gamma_duration}
\end{center}
\end{figure}

Continuing the comparison with the optical light curve properties, we look at the relationship between $t_2$ and $\gamma$-ray luminosity, shown in Figure~\ref{Fig:avgluminosity_vs_t2}. Such a correlation may be expected because novae with larger $\Delta v$ tend to have a shorter $t_2$. There is mild evidence in our data set for a negative correlation here, with slower novae tending to have lower $\gamma$-ray luminosities (Table \ref{tab:corrstats}). The correlation is of borderline significance, and only becomes significant when including the upper limits on the average $\gamma$-ray emission, and assuming a distance of 1.2 kpc for V5668 Sgr. Across our sample, we do not see any slow novae with large $\gamma$-ray luminosities, possibly indicating that slow novae are typically not very $\gamma$-ray luminous. The fast systems ($t_2 <$ 5), on the other hand, show a spread of two orders of magnitude in $\gamma$-ray luminosity. Other work has found that there is not a correlation between $t_2$ and the $\gamma$-ray flux \citep{Franckowiak_etal_2018}. It may be the case that $t_2$ and the $\gamma$-ray luminosity are weakly and indirectly related through relationships with $\Delta v$, and a larger sample is necessary to clearly identify it. 

There is a positive correlation between the duration of the $\gamma$-ray emission and $t_2$, shown in Figure~\ref{Fig:t2_vs_gamma_duration} (see also Table \ref{tab:corrstats}, such that slower novae tend to show longer durations of $\gamma$-ray emission. There is one nova that is a clear exception here, which is the very slow nova V1405 Cas. This system has a large $t_2$ of 164.5 days with a complex light curve, while the $\gamma$-ray emission only lasted for 10 days. In general, faster novae have larger values for $v_2$ and $\Delta v$ than slower novae, so it is likely that the differences in $\gamma$-ray properties between faster and slower novae is due to the different expansion/shock velocities. However, $t_2$ can also depend on other nova properties besides velocity, including ejecta mass \citep{Yaron_etal_2005}, white dwarf mass \citep{Kato_Hachisu_1994}, and dust formation \citep{Chong2025}. In \cite{Franckowiak_etal_2018}, no correlation is found between $t_2$ and the $\gamma$-ray durations. Our analysis includes a larger sample, allowing the correlation to be detected in spite of the substantial scatter apparent in this relationship.

\subsection{$\gamma$-rays and Dust Formation}\label{sec:dust}

Next we check if the $\gamma$-ray properties are related to dust formation in novae. Dust formation in novae is thought to be dependent on to cooling behind radiative shocks internal to the ejecta \citep{Metzger_etal_2015}, where dense regions can form, shielding the harsh radiation field of the white dwarf and allowing for dust formation \citep{Derdzinski_etal_2017}. There are also some indications that dust formation could be over-represented among the $\gamma$-ray bright novae \citep{Chong2025}. The signature dust dips signaling the presence of dust in nova eruptions have also been observed to occur concurrently with the radio synchrotron emission as seen in V1324 Sco \citep{Finzell_etal_2018}, V809 Cep \citep{Babul+22}, and V357 Mus \citep{Chomiuk_etal_2021radio}. Synchrotron emission is also driven by the shocks, so this temporal coincidence further indicates a connection between shock activity and dust formation.

We use the nova dust classifications of \citet{Chong2025} to examine the fraction of $\gamma$-ray detected novae that show evidence for dust formation. The novae classified in this catalogue are limited to those with IR light curves available, which restricts the sample of classifiable novae. Novae can be confirmed as dust forming based solely on optical photometry if a dust dip is present, but without information in the IR dust formation can not be ruled out. Systems without IR light curves could be IR excess novae, which display an increase in IR flux without any corresponding dust dip in the optical. We add a dust classification for V549 Vel to this sample, as this source has the required $V$ and $K$ band data available from SMARTS at around $t_2$ \citep{Walter_etal_2012}, when dust formation typically begins. This nova lacks $K$ band coverage at the peak, so we estimate the $(V-K)$ colour at peak using the estimated intrinsic colour of $(V-K)_0$ = 1.1 from \cite{Chong2025}. This is coupled with an estimate for $E(V-K)$ based on the measured $E(B-V) = 1.1$ (see Table \ref{table:characteristics1}) and using the extinction law from \cite{Wang2019}, which yields an expected observed colour at peak of $(V-K) = 4.1$. The maximum redward colour change from this peak value is $\Delta (V-K) = 1.6$ magnitudes, below the threshold of 2.35 found in \cite{Chong2025}. This, coupled with the lack of a clear dust dip or IR excess in the light curves at around $t_2$, leads us to conclude that V549 Vel most likely did not form dust.

As a result, our sample of $\gamma-$ray novae with dust classifications includes 8 sources, of which 6 show clear signs of dust formation. The percentage of $\gamma$-ray detected novae that show dust formation is then $75^{+9}_{-19}$, consistent with the overall population, where 50--70 per cent show evidence for dust formation \citep{Chong2025}. The uncertainties on our dust formation rates are calculated assuming a binomial distribution with Bayesian confidence intervals, using the calculations from \cite{Cameron2011}. In our sample of non-detected novae, there are 11 systems with dust classifications available. Four of these show clear evidence of dust formation, another three are uncertain, and the remaining four show no evidence of dust formation. We estimate from this that, 36--64 percent of the novae in our sample without $\gamma$-ray detections were dust forming, in agreement with the \cite{Chong2025} sample of novae without $\gamma$-ray detections. The sample from \cite{Chong2025} is a bit larger than our sample with 21 novae, and when including novae with uncertain dust classifications, they find that 38--62 percent of novae without \emph{Fermi} detections are dust forming. The fraction of dust forming novae appears to be higher in our $\gamma$-ray detected sample, as suggested in \citep{Chong2025}, but it is not a significant difference in our sample given the size of the associated uncertainties. It may be the case $\gamma$-ray detected novae tend to be more likely to form dust, but we would need a larger sample to determine this conclusively.
\begin{figure}
\begin{center}
  \includegraphics[width=\columnwidth]{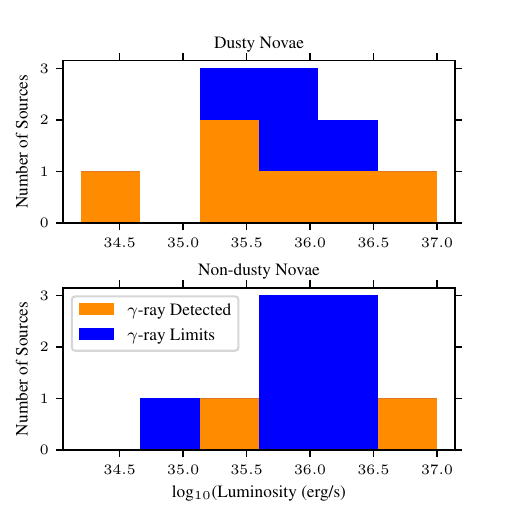}
\caption{Histograms displaying the distribution of $\gamma$-ray luminosities for the dust forming and not dust forming novae in our sample. The blue histogram shows the $\gamma$-ray upper limits and the orange one displays the $\gamma$-ray detected sample. The sample lacks sufficient size to determine if the distributions are significantly different from each other, but it is clear that both the dusty and non-dusty samples can span a large range of $\gamma$-ray luminosities.}
\label{Fig:dust_hist}
\end{center}
\end{figure}

We also investigate any dependence of the $\gamma$-ray luminosity on the dust forming status of a nova. Histograms showing the $\gamma$-ray luminosity distribution for both dusty and non-dusty novae can be seen in Figure~\ref{Fig:dust_hist}. From our limited sample, it is clear that both dusty and non-dusty novae can span a wide range of $\gamma$-ray luminosities, ranging anywhere from order $10^{34}$ erg s$^{-1}$ up to $10^{37}$ erg s$^{-1}$ in either case. Our sample does not contain any evidence for a relationship between the dust forming characteristics of a nova, and its $\gamma$-ray luminosity, but given our limited sample size this is not surprising.

\begin{figure}
\begin{center}
  \includegraphics[width=\columnwidth]{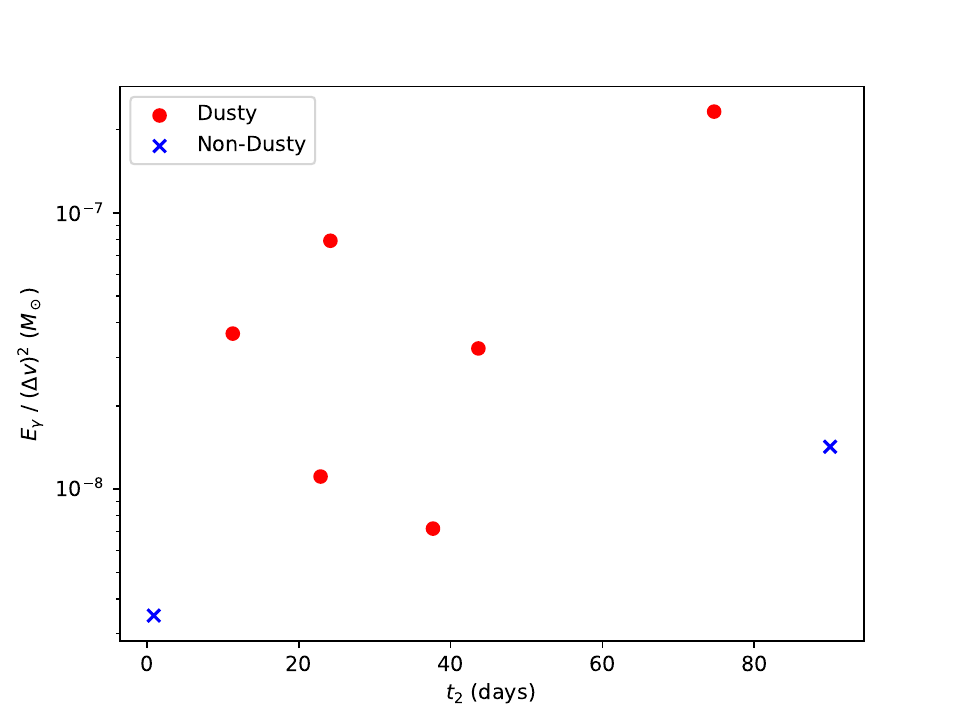}
\caption{Estimate of the mass processed through the shocks plotted against $t_2$ across our sample. The points are separated between dust forming and non-dust forming novae. The non-dusty nova in the lower left is V1674 Her, which has relatively high velocities and small $E_{\gamma}$, and likely has relatively little mass processed through the shocks.}
\label{Fig:dustmass}
\end{center}
\end{figure}

While $\gamma$-ray luminosity is expected to  correlate with both the velocity of the shock and the total mass passing through it \citep{Metzger_etal_2014}, the dust production is expected to scale more with the shocked mass, and might even anti-correlate with shock velocity \citep{Derdzinski_etal_2017}.
We therefore estimate the total mass that passes through the shock fronts, for comparison with dust properties. This mass is expected to scale roughly as M $\propto$ $E_{\gamma} / \Delta v^{2}$, where $E_{\gamma}$ is the total emitted energy from the $\gamma$-ray emission, used here as a tracer for the total shock energy. This mass estimate is plotted against $t_2$ in Figure~\ref{Fig:dustmass}. V1674 Her matches well with the expectation here, as it has a relatively small estimated mass and is not dust forming, while most of the $\gamma$-ray detected novae with more mass processed through the shocks are dust forming. The other non-dusty but $\gamma$-ray detected nova is V549 Vel, which is the slowest nova in our dust classified sample and shows a moderate mass estimate. Our sample size, especially for the $\gamma$-ray detected but not dust forming novae, remains too small to draw any substantive conclusions from this, but this plane appears to provide the best separation between dust-forming and not dust-forming novae in our sample.

\begin{figure*}
\begin{center}
  \includegraphics[width=\textwidth]{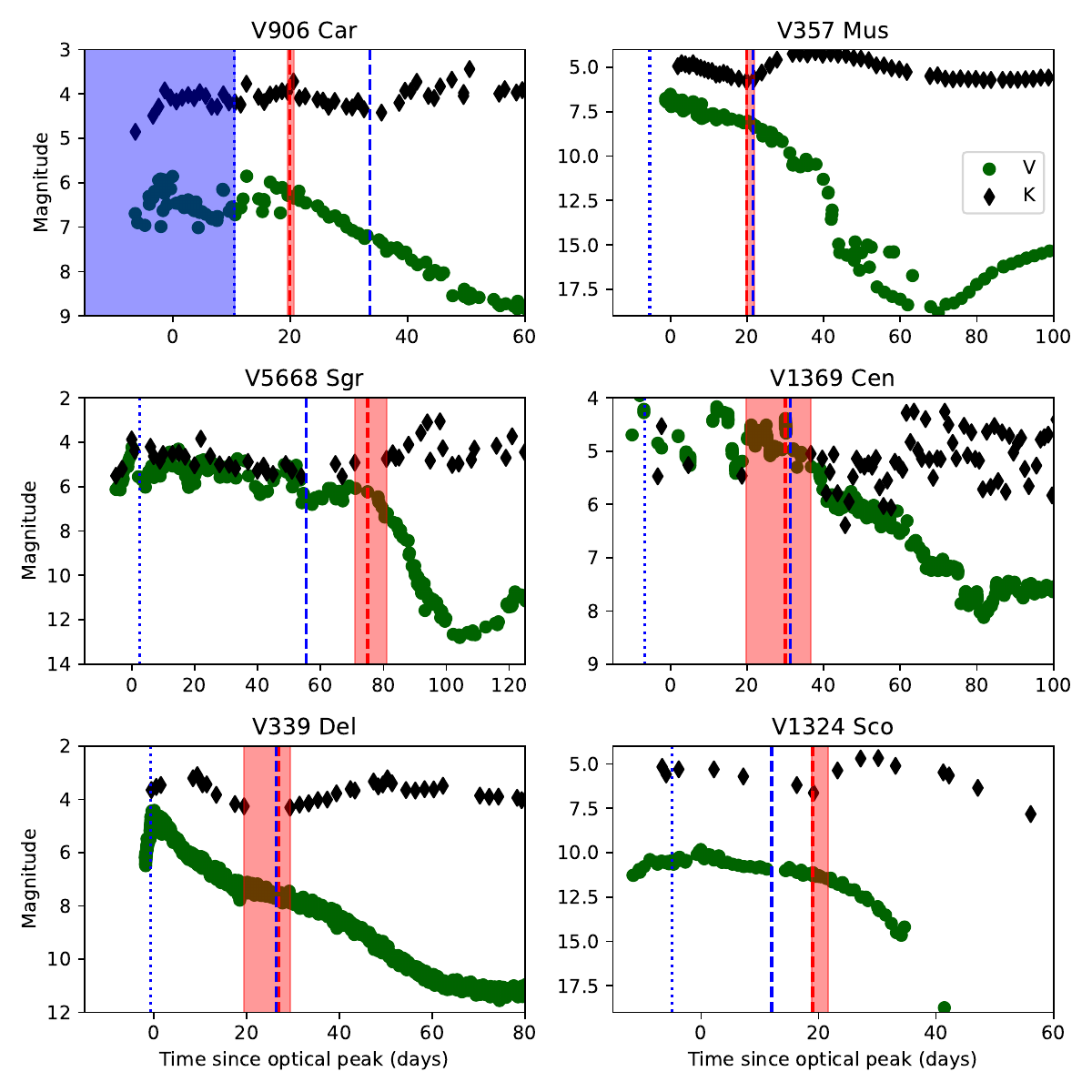}
\caption{$V$ and $K$ band light curves for six $\gamma$-ray detected novae. The dotted blue line displays the start time of the $\gamma$-ray emission, and the dashed blue line shows the end of the $\gamma$-ray emission. The dashed red line is the estimated time of onset of dust formation, and the shaded region shows the uncertainty. These values are from \citet{Chong2025}. In most cases, except for V906 Car, the dust formation begins near or after the end of the detectable $\gamma$-ray emission. In every case here, we see an increase in the $K$ band flux after the end of the $\gamma$-ray emission (even in the case of V906 Car, where the dust formation begins $\sim 15$ days before the end of the $\gamma$-ray detection, but the $K$ band flux still increases near the end of the $\gamma$-ray detection). The blue shaded region for V906 Car indicates that the start of the $\gamma$-ray emission is unknown, and could be somewhere in the blue region.}
\label{fig:dust-timing}
\end{center}
\end{figure*}

The timing of the onset of dust formation appears to be related to the end of the $\gamma$-ray emission across our sample. In Figure~\ref{fig:dust-timing} we show the $V$ and $K$ band light curves for all six dusty novae with $\gamma$-ray detections and available $K$ band data. The timing of the $\gamma$-ray emission and the apparent onset of dust formation are marked on these light curves. The dust formation times used here are from \cite{Chong2025}, and were determined based on the $(V-K)$ colour curves, which typically become redder as dust begins to form. There is consistent evidence that the dust forms soon after the end of the $\gamma$-ray emitting phase. All of the novae that have well sampled $K$ band light curves available display an increase in the $K$ band flux shortly after the end of the detectable $\gamma$-ray emission. In the case of V906 Car, there is dust formation during the $\gamma$-ray emitting phase, but the $K$ band flux still increases after the end of the $\gamma$-ray detection.

One possible model to explain the timing of the dust formation involves the emission from the shocks heating the dust forming regions enough to inhibit dust formation. Once the shock luminosity drops off, then the gas is able to cool effectively, and dust formation can begin. This requires that there is sufficient shielding between the dust forming region in the ejecta and the continuing supersoft emission from the white dwarf surface.  A related possibility is that the cosmic ray output of the shocks could serve to disrupt the dust forming regions. It is unlikely that a large fraction of the molecular formation within the dust forming region is inhibited by cosmic rays however, as only $\sim$0.01 per cent of the shock energy is expected to go into the dissociation of molecules in the ejecta \citep{Derdzinski_etal_2017}. 

An alternative possibility is that this effect is related to the optical depth of the ejecta. The fact that we don't see the dust until after the $\gamma$-ray emission turns off may be explained if the shock---and the dust-forming region behind it---are below the nova photosphere until the shock breaks out from the densest ejecta. At this time, the dust is unveiled and can have a significant effect on the optical/IR light curves, even as the shock luminosity plummets as it no longer has dense material to interact with.
This scenario might also explain the temporal coincidence between radio synchrotron flares in e.g. V1324 Sco and V357 Mus \citep{Finzell_etal_2018, Chomiuk_etal_2021radio} and dust formation, as synchrotron emission is expected to suffer significant free-free absorption from intervening ejecta, and would be suddenly unveiled were the shock to break out into lower density environs. The dust in this model could be present soon after the start of the $\gamma$-ray emission, but invisible until later times when the photosphere has receded sufficiently. In the case of either possibility, it is clear that dust formation only becomes apparent after the end of detectable $\gamma$-ray emission.

\section{Conclusions}
\label{sec:conclusions}

In this work, we have gathered a sample of novae with consistent measurements of their $\gamma$-ray emission properties, outflow velocities, distances, extinctions and optical light curve parameters. Our sample includes 15 $\gamma$-ray detected novae, and 37 novae that were not significantly detected by \emph{Fermi}, for which we have upper limits on their $\gamma$-ray luminosity. These eruptions are all between 2008 and \hl{2021}. Distances for this sample have been derived using a luminosity-independent technique, utilizing either Gaia DR3 parallaxes when available, or spectroscopic reddening measurements combined with 3D dust maps. Luminosity-independent distances allows for a fair comparison of the $\gamma$-ray luminosities of novae without any bias due to the relationship between optical and $\gamma$-ray luminosity.

This sample is then used to test for relationships between optical properties, such as the outflow velocities, and $\gamma$-ray properties, such as $\gamma$-ray luminosity and duration. The $\gamma$-ray luminosities of novae vary by at least three orders of magnitude, with the sources in our sample ranging from $2 \times 10^{34}$ to $1.2 \times 10^{37}$ erg/s. Most, but not all, of the $\gamma$-ray detected novae are optically bright ($V_{peak} < 6$), and they cover a wide range in $t_2$. The $\gamma$-ray emission can remain visible for less than one day in very fast systems, to more than 50 days in others.

We have shown that the $\gamma$-ray luminosities of novae are strongly correlated with the relative velocity between the slow and fast outflows, and the fast outflow itself. The predicted cubic scaling between this luminosity and $\Delta v$ from shocks internal to the ejecta agrees well with the available data, especially considering the anticipated differences in ejecta densities between novae. This result is consistent with shock-powered $\gamma$-ray production in novae arising from shocks forming from the collision between two mass outflows during the eruption. 

Simultaneously, the duration of the $\gamma$-ray emission is negatively correlated with both $\Delta v$ and the $\gamma$-ray luminosity. Novae with a higher $\Delta v$ have larger $\gamma$-ray luminosities over a shorter emission duration. The total emitted energy in the $\gamma$-rays does not cleanly depend on $\Delta v$, the $\gamma$-ray luminosity or the duration of the $\gamma$-ray emission.

We also report evidence for a relationship between the onset of dust formation in $\gamma$-ray detected systems and the turn off of the $\gamma$-ray emission. In all six of our dust forming $\gamma$-ray novae, the IR flux increases after the end of the $\gamma$-ray emission. This may be indicative of a relationship between  the $\gamma$-ray producing shocks and the production of dust forming regions within the nova ejecta. This dust formation may be inhibited by the flux of cosmic rays and radiation from the shock fronts disturbing this layer until after the shock energy drops and the $\gamma$-rays are no longer detected. Alternatively, this dust forming layer could be buried behind the photosphere, and therefore not visible, until after the $\gamma$-ray emission turns off. While this timing relation is apparent, there is no evidence in our sample for a difference in the $\gamma$-ray luminosities between dusty and non-dusty novae.

The \emph{Fermi} detected novae have varied properties, and our understanding of the $\gamma$-ray emission properties in novae is still limited by small number statistics. This is especially the case in regards to the relationship between the $\gamma$-ray emitting shocks and dust formation, for which the current sample is quite limiting. \kvs{The $\gamma$-ray properties of novae with evolved companions may differ from those observed here, warranting further investigation.}

\section*{Acknowledgements}

PC, AS, EA, LC, AC, AK, and KVS are grateful for support from NASA awards 80NSSC25K7334, 80NSSC23K1247, 80NSSC23K0497, and 80NSSC18K1746,  They also acknowledge NSF awards AST-1751874, AST-2107070, and AST-2205631, and a Cottrell fellowship of the Research Corporation. E.A. acknowledges support by NASA through the NASA Hubble Fellowship grant HST-HF2-51501.001-A awarded by the Space Telescope Science Institute, which is operated by the Association of Universities for Research in Astronomy, Inc., for NASA, under contract NAS5-26555.  JS was supported by the Packard Foundation. JLS was supported by NASA grant 80NSSC25K7068. LI was supported by grants from VILLUM FONDEN (project number 16599 and 25501). DAHB gratefully acknowledges the receipt of research grants from the National Research Foundation (NRF) of South Africa. FMW acknowledges support of the US taxpayers through NSF grant 1611443.  BDM acknowledges support through NASA (grants 80NSSC22K0807, 80NSSC24K0408), and the Simons Foundation (grant 727700).  The Flatiron Institute is supported by the Simons Foundation.

ASAS-SN thanks the Las Cumbres Observatory and its staff for its continuing support of the ASAS-SN project. ASAS-SN is supported by the Gordon and Betty Moore Foundation through grant GBMF5490 to the Ohio State University and NSF grant AST-1515927. Development of ASAS-SN has been supported by NSF grant AST-0908816, the Mt. Cuba Astronomical Foundation, the Center for Cosmology and AstroParticle Physics at the Ohio State University, the Chinese Academy of Sciences South America Center for Astronomy (CASSACA), the Villum Foundation, and George Skestos. 

We thank Robert E. Williams for useful comments and discussion. We thank Kristen Dage and Chelsea Harris for useful comments and support during this work.  
We thank the AAVSO observers from around the world who contributed their magnitude measurements to the AAVSO International Database used in this work.
We acknowledge all the ARAS observers for their optical spectroscopic observations which complement our data. This work is based on observations obtained at the Southern Astrophysical Research (SOAR) telescope, which is a joint project of the Minist\'{e}rio da Ci\^{e}ncia, Tecnologia, Inova\c{c}\~{o}es e Comunica\c{c}\~{o}es (MCTIC) do Brasil, the U.S. National Optical Astronomy Observatory (NOAO), the University of North Carolina at Chapel Hill (UNC), and Michigan State University (MSU). A part of this work is based on observations made with the Southern African Large Telescope (SALT), with the Large Science Programme on transients 2018-2-LSP-001 (PI: DAHB). 
This work is also partly based on observations collected at the European Organisation for Astronomical Research in the Southern Hemisphere. 
This paper includes data gathered with the 6.5 meter Magellan Telescopes located at Las Campanas Observatory, Chile. Polish participation in SALT is funded by grant no. MNiSW DIR/WK/2016/07.

\section*{Data Availability}
All photometry used in this work is publicly available, typically through the AAVSO or ASAS-SN websites. Many of the spectra utilized for measuring reddenings or velocities are publicly available from ARAS, although some SALT spectra are not public. The velocity measurements from these spectra are all listed in Tables \ref{table:characteristics1} or \ref{table:characteristics2}, and spectroscopic extinction measurements have been published in \citep{Craig2025}. Infrared light curves are obtained from the publicly available SMARTS atlas for (mostly) southern novae, and all $\gamma$-ray data is also publicly available through the LAT data server.

\clearpage
\clearpage 

\bibliographystyle{mnras}
\bibliography{biblio}
\clearpage

\appendix
 
\renewcommand\thetable{\thesection.\arabic{table}}    
\renewcommand\thefigure{\thesection.\arabic{figure}}   
\setcounter{figure}{0}

\renewcommand\thetable{\thesection.\arabic{table}}    
\renewcommand\thefigure{\thesection.\arabic{figure}}   
\setcounter{figure}{0}
\section{Nova Light Curves}

In this Appendix we present the light curves for all of the novae in this sample, shown with measurements of the peak time and $t_2$. The light curves in this section are presented in reverse chronological order.


\begin{figure*}
\begin{center}
  \includegraphics[width=0.9\textwidth]{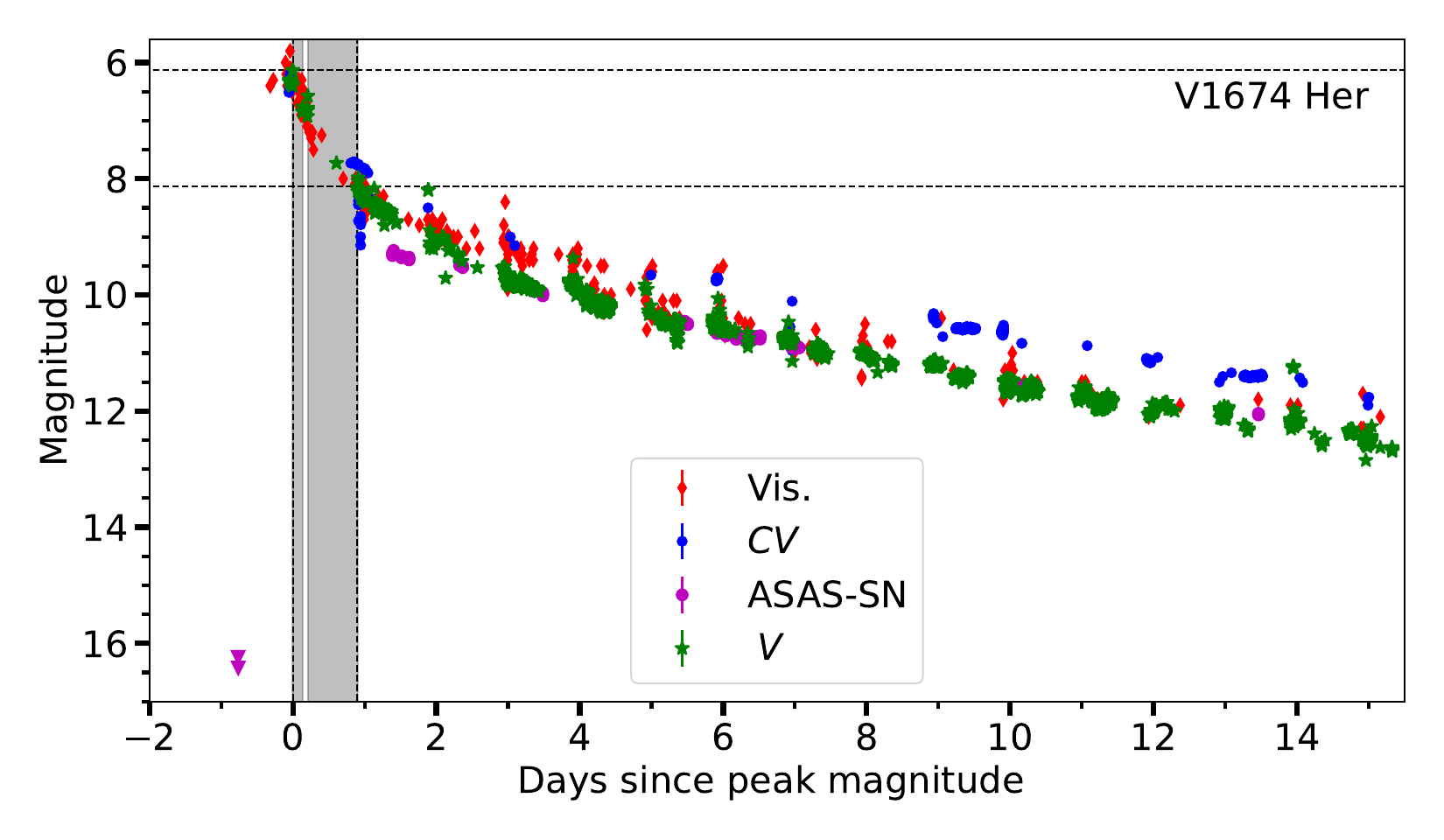}
  \caption{The AAVSO optical light curve of the $\gamma$-ray detected nova V1674 Her, plotted against time since peak magnitude ($t_{max} =$ 2021 Jun 13.0). The left and right vertical dashed lines represent the time of maximum brightness ($t_{max}$) and the time for the light curve to fade by two magnitudes ($t_2$), respectively. The top and bottom horizontal dashed lines represent the maximum magnitude ($V_{\mathrm{peak}}$) and two mags below peak ($V_{\mathrm{peak}} +2$), respectively. The two grey shaded regions indicate the uncertainties on $t_{max}$ and $t_2$. The light curve rise is well measured by photometry from AAVSO, and ASAS-SN which provided the non-detections shown as magenta triangles on the plot.} 
\label{Fig:V1674_Her_LC}
\end{center}
\end{figure*}

\begin{figure*}
\begin{center}
  \includegraphics[width=0.9\textwidth]{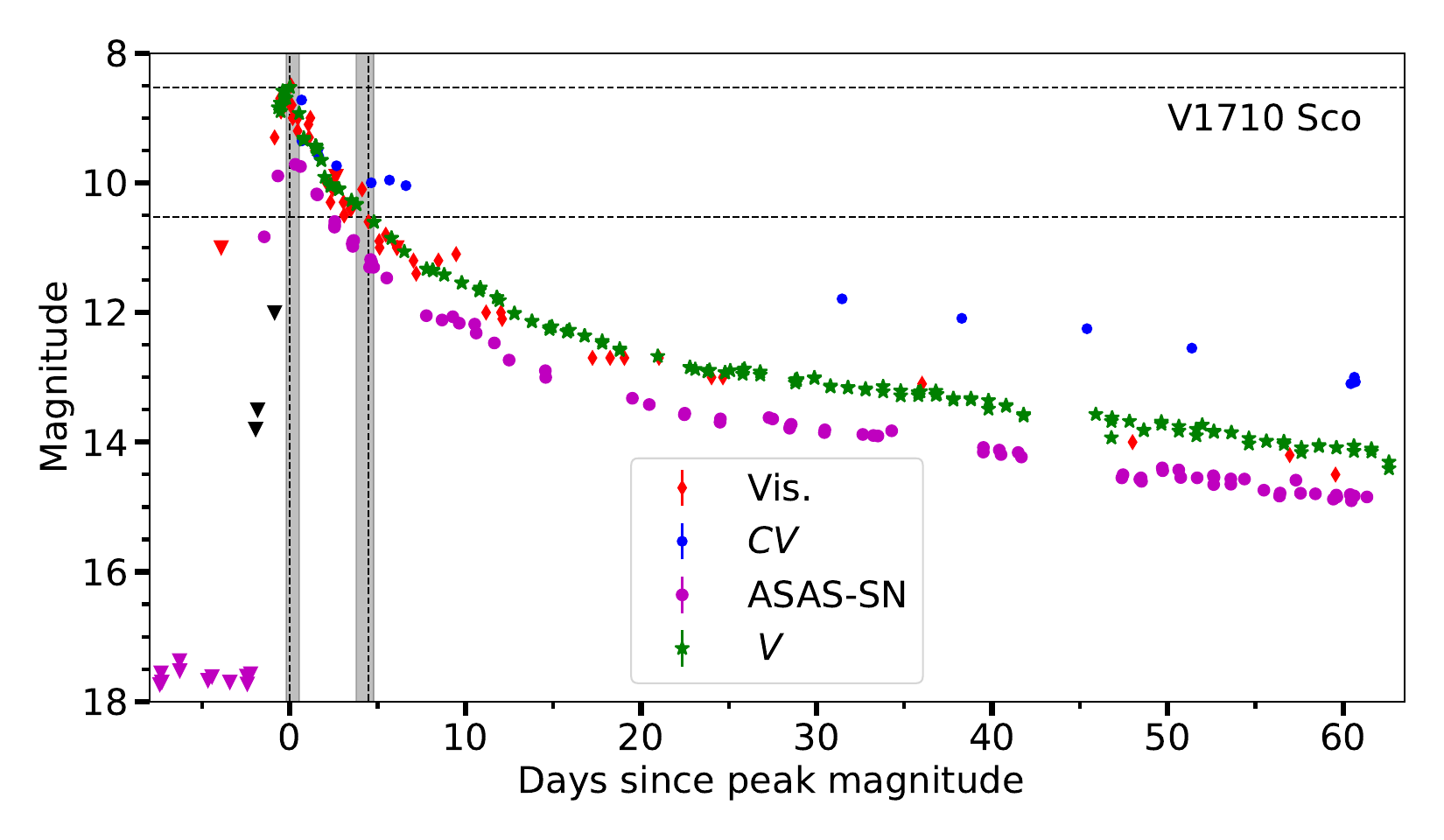}
\label{Fig:V1710_Sco_LC}
\caption{The optical light curve of V1710 Sco, plotted against time since peak magnitude  ($t_{max}=$ 2021 Apr 13.6). Horizontal and vertical dashed lines and the shaded regions have the same meaning as in Figure~\ref{Fig:V1674_Her_LC}. The rise to maximum is well captured by ASAS-SN and three other non-detections shown as black triangles (R.\ McNaught and P.\ Camilleri; CBAT "Transient Object Followup Reports").}
\end{center}
\end{figure*}

\begin{figure*}
\begin{center}
  \includegraphics[width=0.9\textwidth]{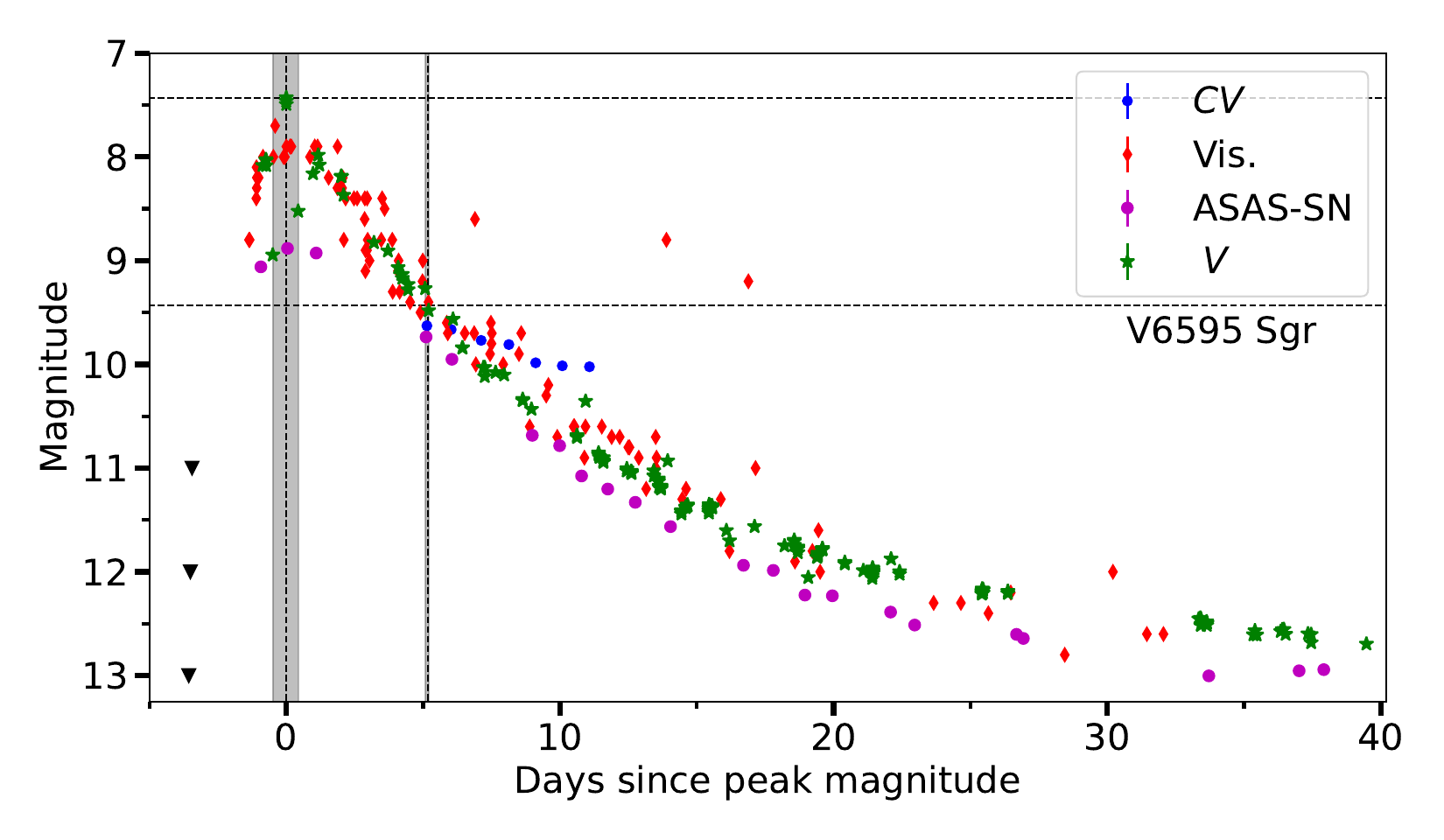}
\label{Fig:V6595_Sgr_LC}
\caption{The optical light curve of V6595 Sgr, plotted against time since peak magnitude  ($t_{max}=$ 2021 Apr 06.2). Horizontal and vertical dashed lines and the shaded regions have the same meaning as in Figure~\ref{Fig:V1674_Her_LC}. The rise to maximum is well-constrained by non-detections shown as black triangles (AAVSO Alert Notice 739).}
\end{center}
\end{figure*}


\begin{figure*}
\begin{center}
  \includegraphics[width=0.9\textwidth]{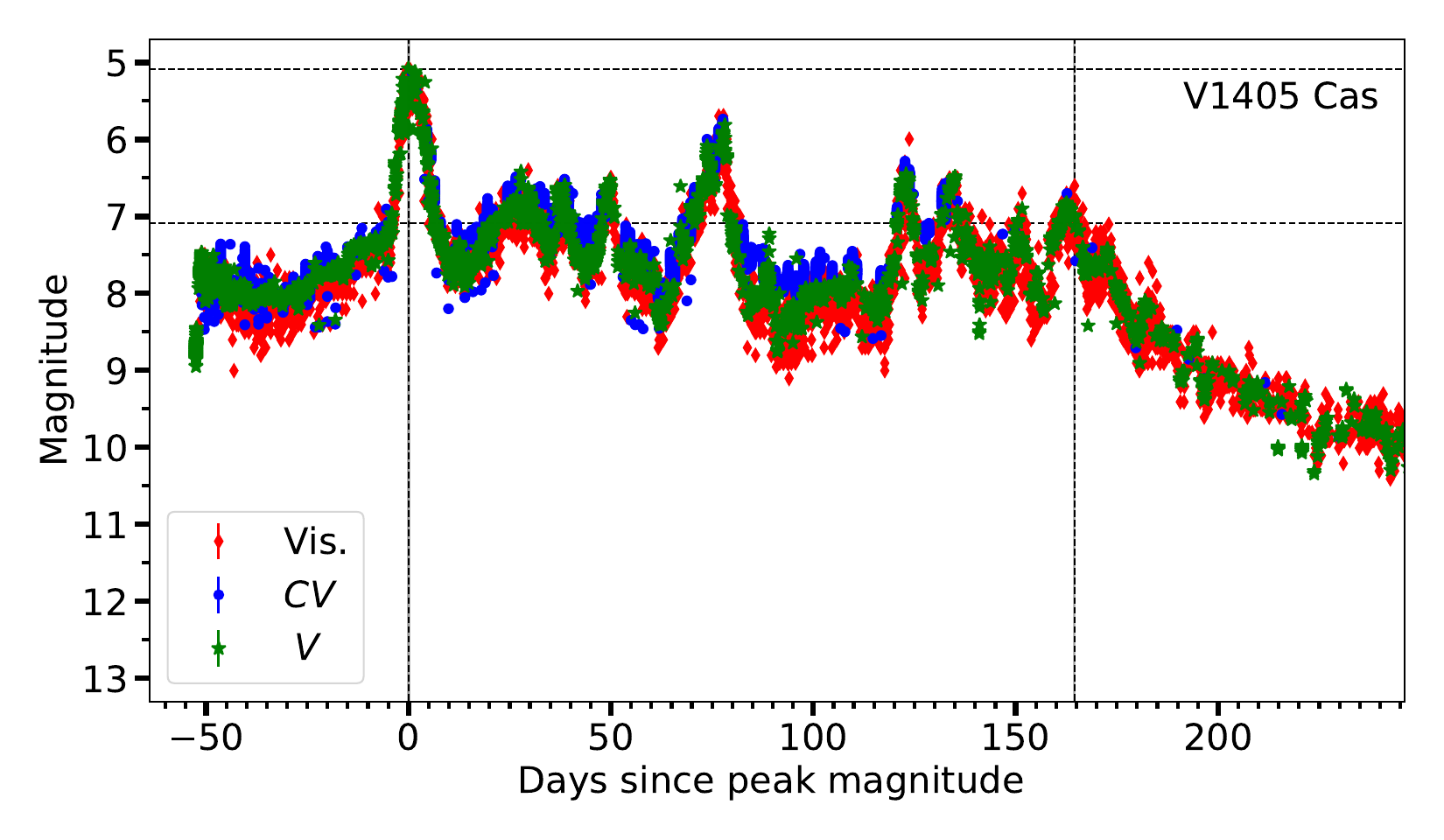}
  \caption{The optical light curve of the $\gamma$-ray detected nova V1405 Cas, plotted against time since peak magnitude  ($t_{max}=$ 2021 May 10.3). Horizontal and vertical dashed lines and the shaded regions have the same meaning as in Figure~\ref{Fig:V1674_Her_LC}. A non-detection from 2021 Mar 14.4 constrains the light curve rise well (Y.\ Nakamura; CBAT "Transient Object Followup Reports").}
\label{Fig:V1405_Cas_LC}
\end{center}
\end{figure*}


\begin{figure*}
\begin{center}
  \includegraphics[width=0.9\textwidth]{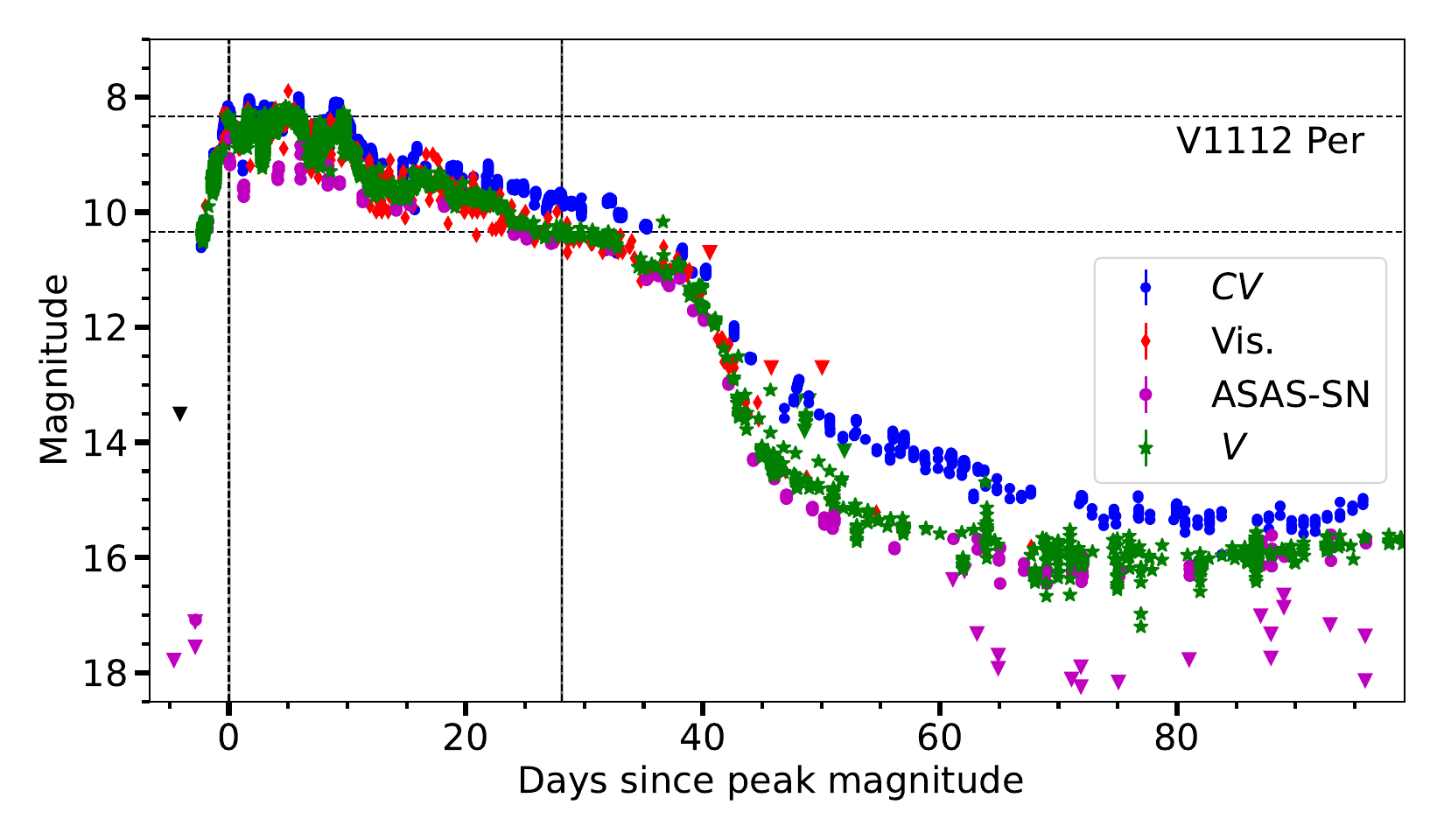}
\caption{The optical light-curve of V1112~Per, plotted against time since peak magnitude ($t_{max}=$ 2020 Nov 28.2). Horizontal and vertical dashed lines and the shaded regions have the same meaning as in Figure~\ref{Fig:V1674_Her_LC}. The rise to optical maximum is well observed, including pre-discovery non-detections (AAVSO Alert Notice 726).}
\label{Fig:V1112_Per_LC}
\end{center}
\end{figure*}


\begin{figure*}
\begin{center}
  \includegraphics[width=0.9\textwidth]{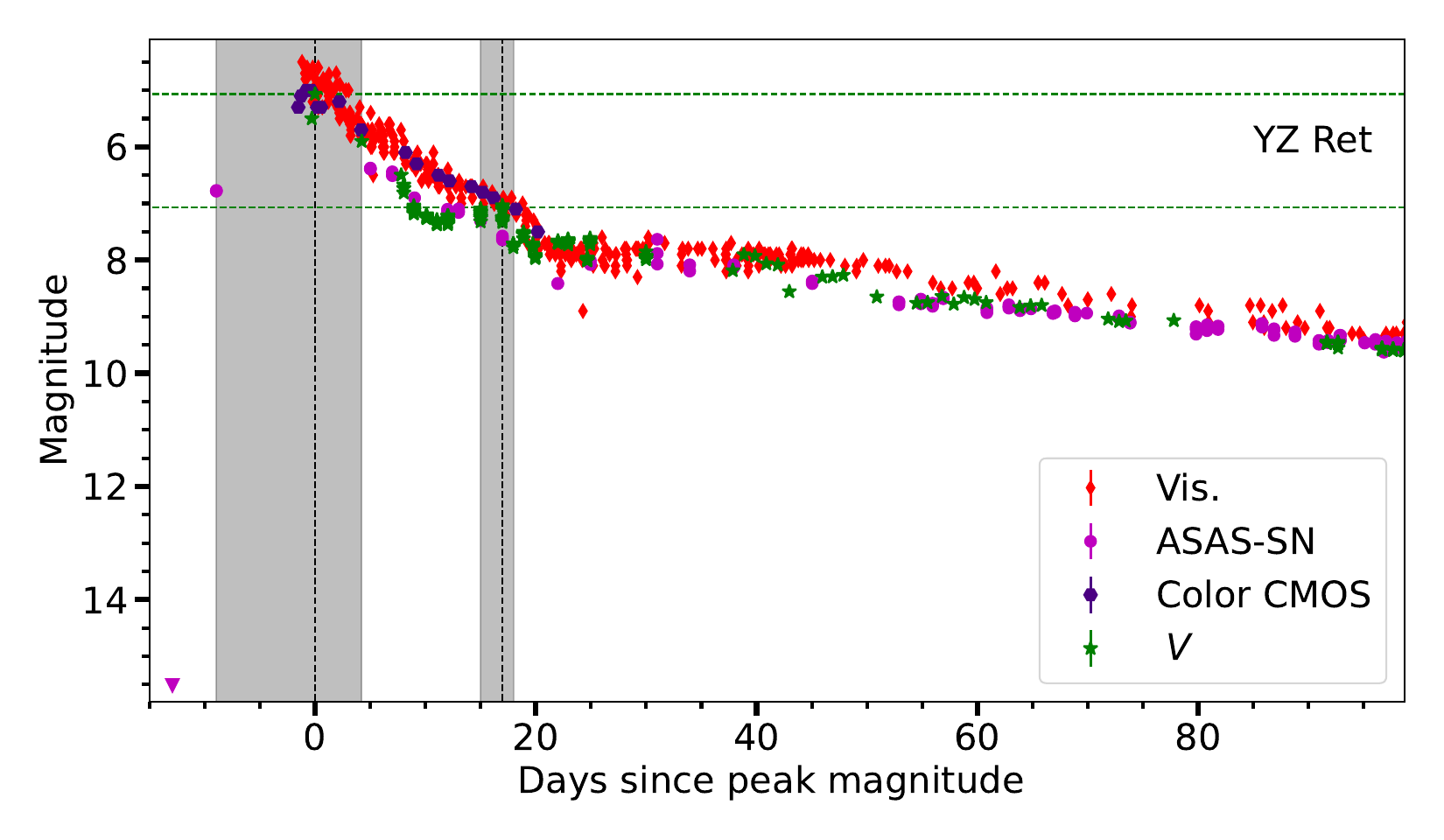}
\caption{The optical light-curve of the $\gamma$-ray detected nova YZ Ret, plotted against time since peak magnitude ($t_{max}=$ 2020 Jul 17.1). Horizontal and vertical dashed lines and the shaded regions have the same meaning as in Figure~\ref{Fig:V1674_Her_LC}. Here, the two horizontal dashed lines are green to indicate that the optical peak is an upper limit, meaning the peak may not have been observed. The time of optical maximum is constrained by an ASAS-SN observation on 2020 Jul 08.2. The Color CMOS data was provided by \protect\cite{Sokolovsky_etal_2022}.}
\label{Fig:YZ_Ret_LC}
\end{center}
\end{figure*}

\begin{figure*}
\begin{center}
  \includegraphics[width=0.9\textwidth]{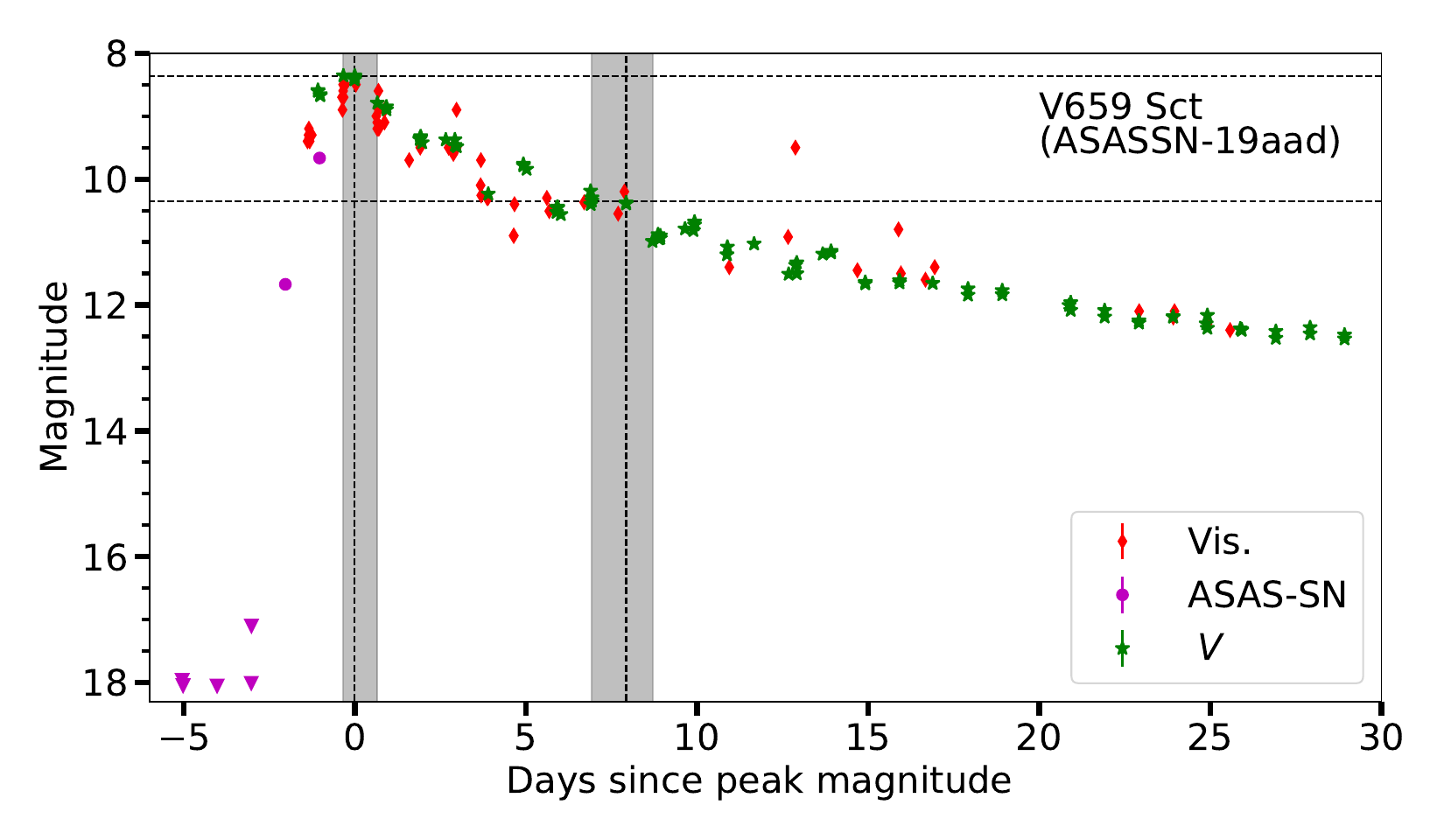}
\caption{The optical light-curve of V659~Sct, plotted against time since peak magnitude ($t_{max}=$ 2019 Oct 31.1). Horizontal and vertical dashed lines and the shaded regions have the same meaning as in Figure~\ref{Fig:V1674_Her_LC}. The rise to light curve maximum is captured by ASAS-SN.}
\label{Fig:V659_Sct_LC}
\end{center}
\end{figure*}


\begin{figure*}
\begin{center}
  \includegraphics[width=0.9\textwidth]{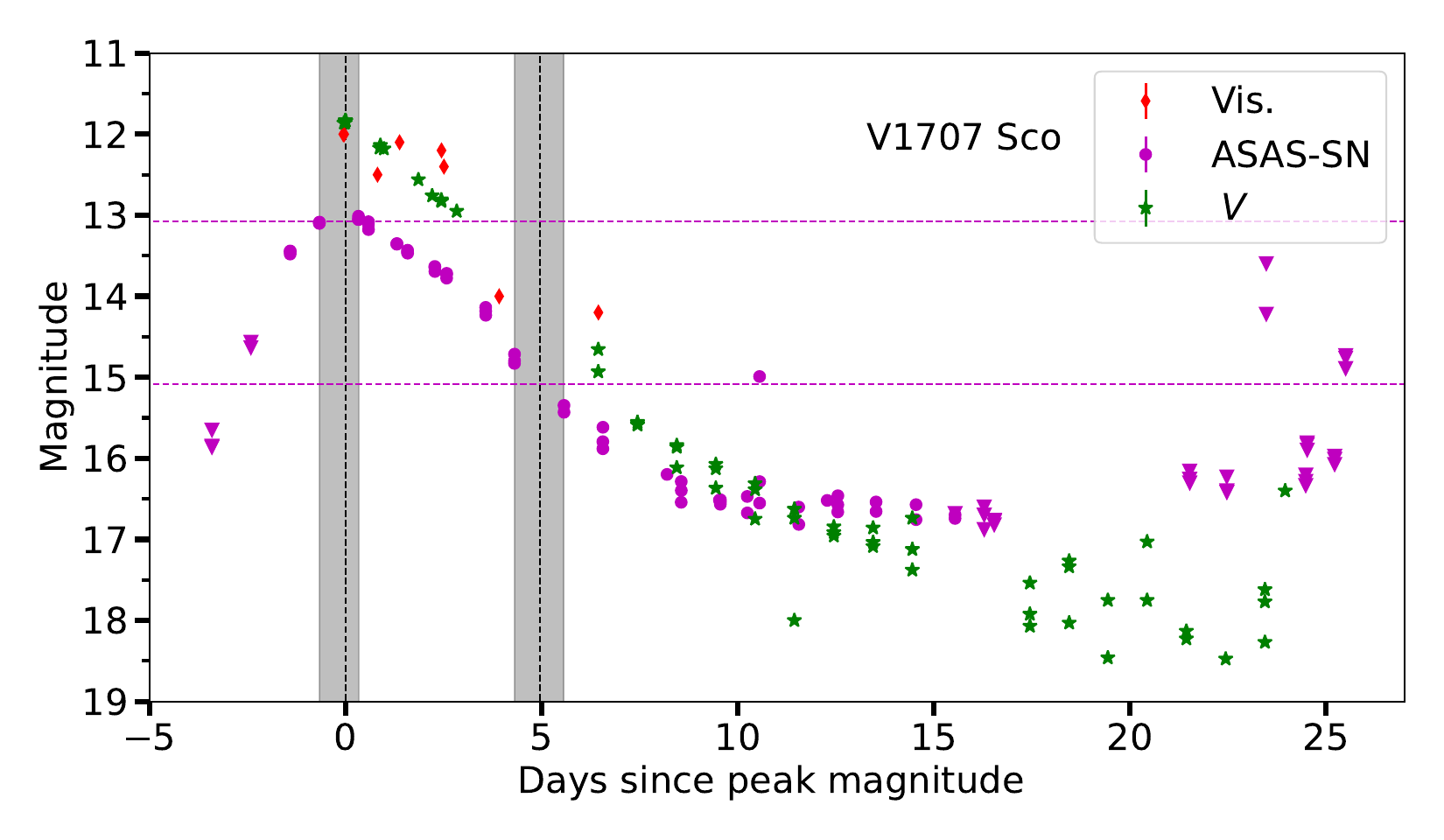}
\caption{The optical light-curve of the $\gamma$-ray detected nova V1707~Sco, plotted against time since peak magnitude ($t_{max}=$ 2019 Sep 15.6). Horizontal and vertical dashed lines and the shaded regions have the same meaning as in Figure~\ref{Fig:V1674_Her_LC}. Here, the two horizontal dashed lines are magenta to indicate that $V_{\mathrm{peak}}$ is being measured using V-band observations and $t_2$ is measured using the g-band ASAS-SN observations. The rise to light curve maximum is captured by ASAS-SN.}
\label{Fig:V1707_Sco_LC}
\end{center}
\end{figure*}


\begin{figure*}
\begin{center}
  \includegraphics[width=0.9\textwidth]{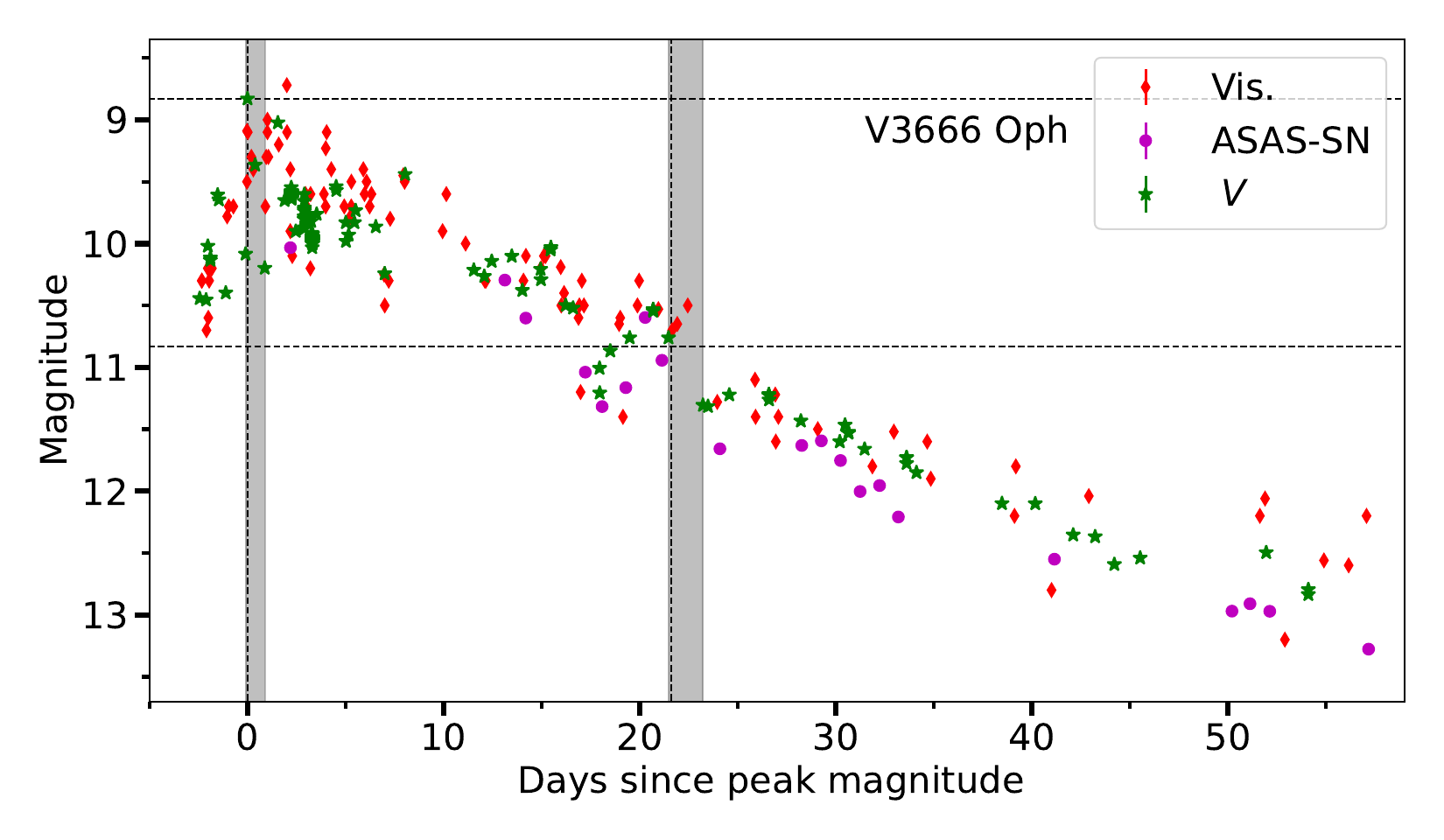}
\caption{The optical light-curve of V3666~Oph, plotted against time since peak magnitude ($t_{max}=$ 2018 Aug 11.9). Horizontal and vertical dashed lines and the shaded regions have the same meaning as in Figure~\ref{Fig:V1674_Her_LC}. The rise and maximum magnitude of this nova are well observed by AAVSO observers.}
\label{Fig:V3666_Oph_LC}
\end{center}
\end{figure*}


\begin{figure*}
\begin{center}
  \includegraphics[width=0.9\textwidth]{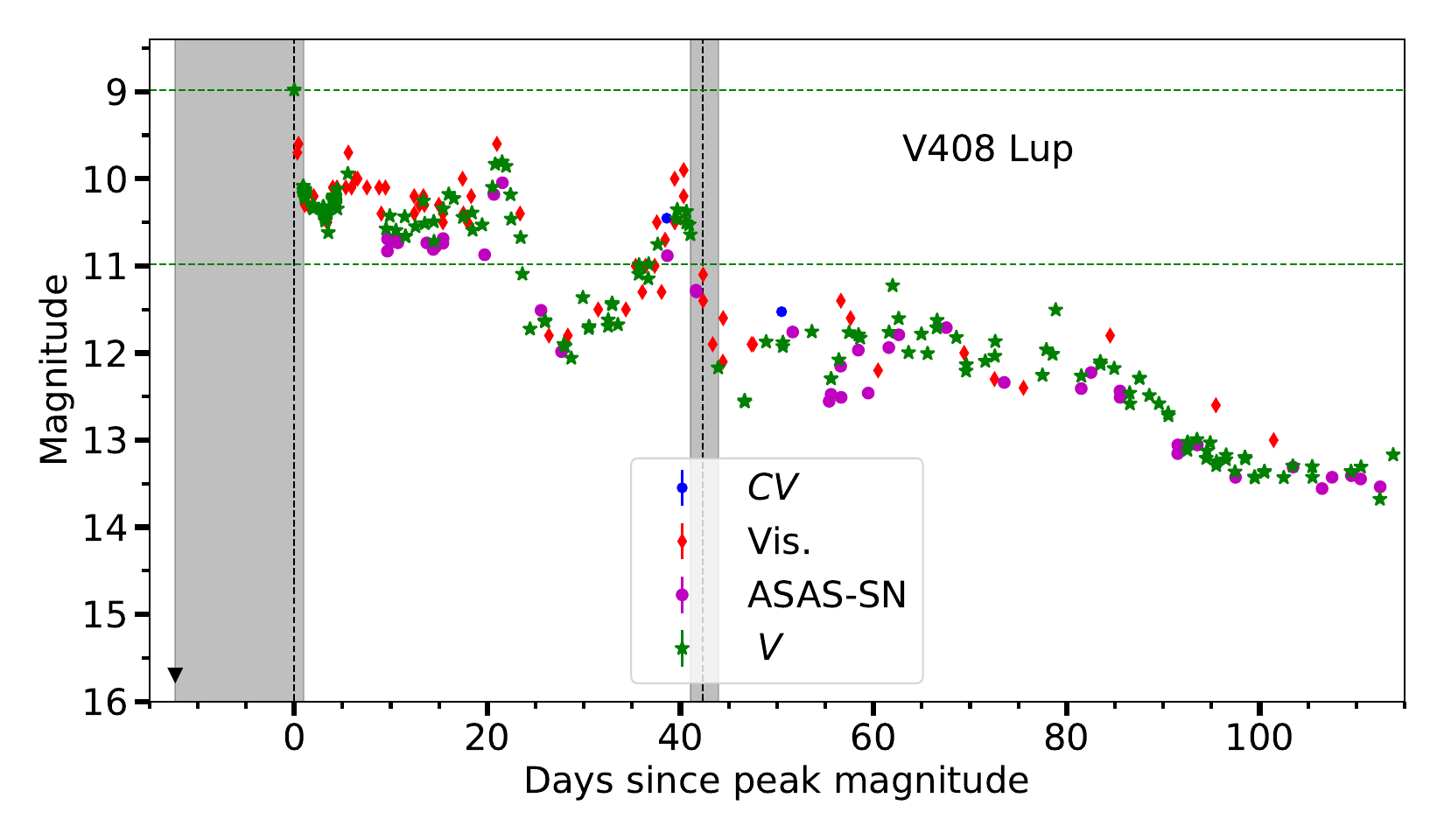}
\caption{The optical light-curve of V408~Lup, plotted against time since peak magnitude ($t_{max}=$ 2018 Jun 03.6). Horizontal and vertical dashed lines and the shaded regions have the same meaning as in Figure~\ref{Fig:V1674_Her_LC}. Here, the two horizontal dashed lines are green to indicate that the optical peak is an upper limit, meaning the peak may not have been observed. A non-detection from 2018 May 22.2 constrains the light curve rise well (R.\ Kaufman; CBAT "Transient Object Followup Reports"). The optical maximum of this nova around 2018 Jun 03.6 is supported by spectroscopy.}
\label{Fig:V408_Lup_LC}
\end{center}
\end{figure*}


\begin{figure*}
\begin{center}
  \includegraphics[width=0.9\textwidth]{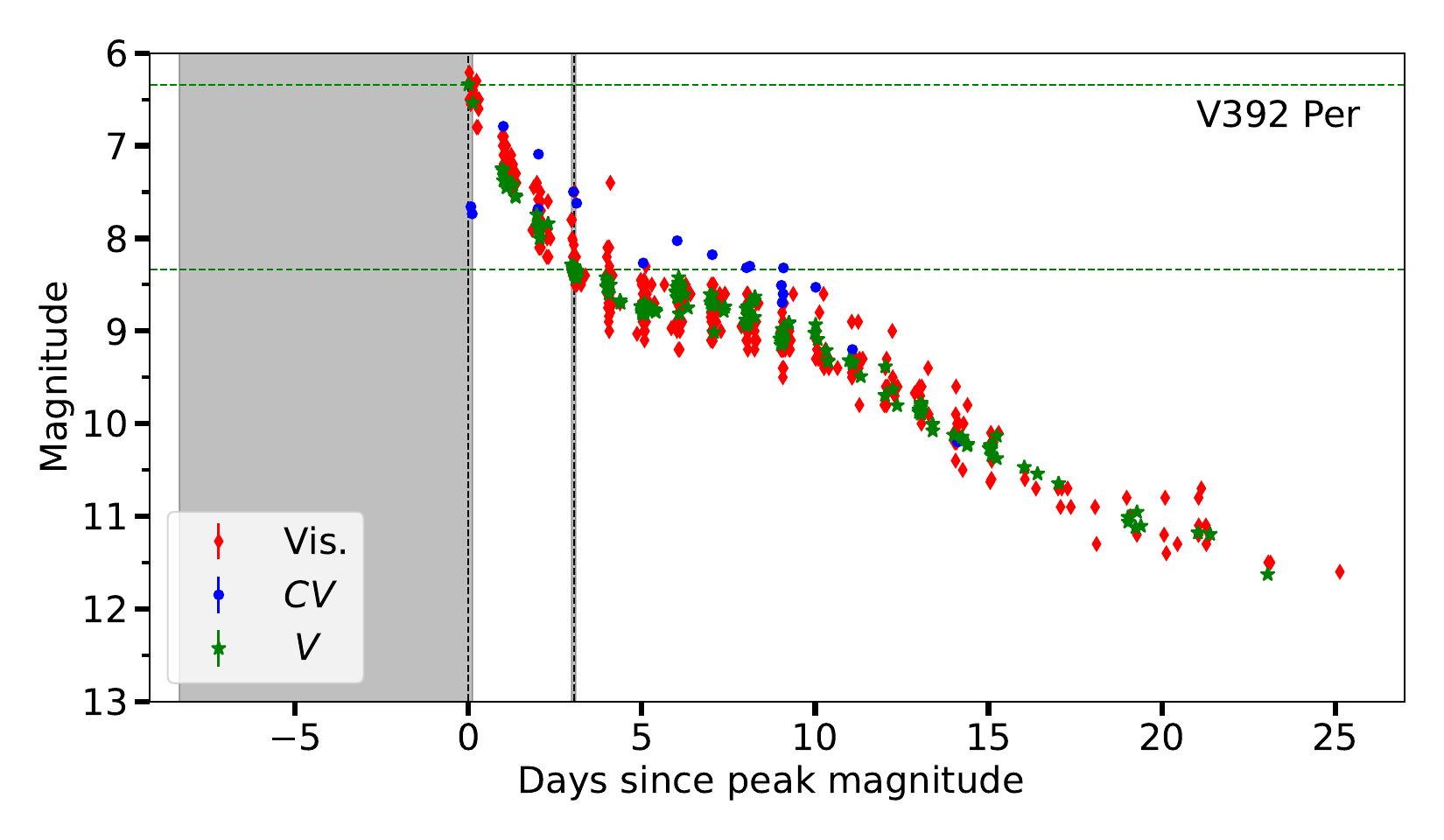}
\caption{The optical light-curve of the $\gamma$-ray detected nova V392~Per, plotted against time since peak magnitude ($t_{max}=$ 2018 Apr 29.8). Horizontal and vertical dashed lines and the shaded regions have the same meaning as in Figure~\ref{Fig:V1674_Her_LC}.  Here, the two horizontal dashed lines are green to indicate that the optical peak is an upper limit, meaning the peak may not have been observed. A non-detection from 2018 Apr 21.5 constrains the light curve rise (Y.\ Nakamura; CBAT "Transient Object Followup Reports").}
\label{Fig:V392_Per_LC}
\end{center}
\end{figure*}

\begin{figure*}
\begin{center}
  \includegraphics[width=0.9\textwidth]{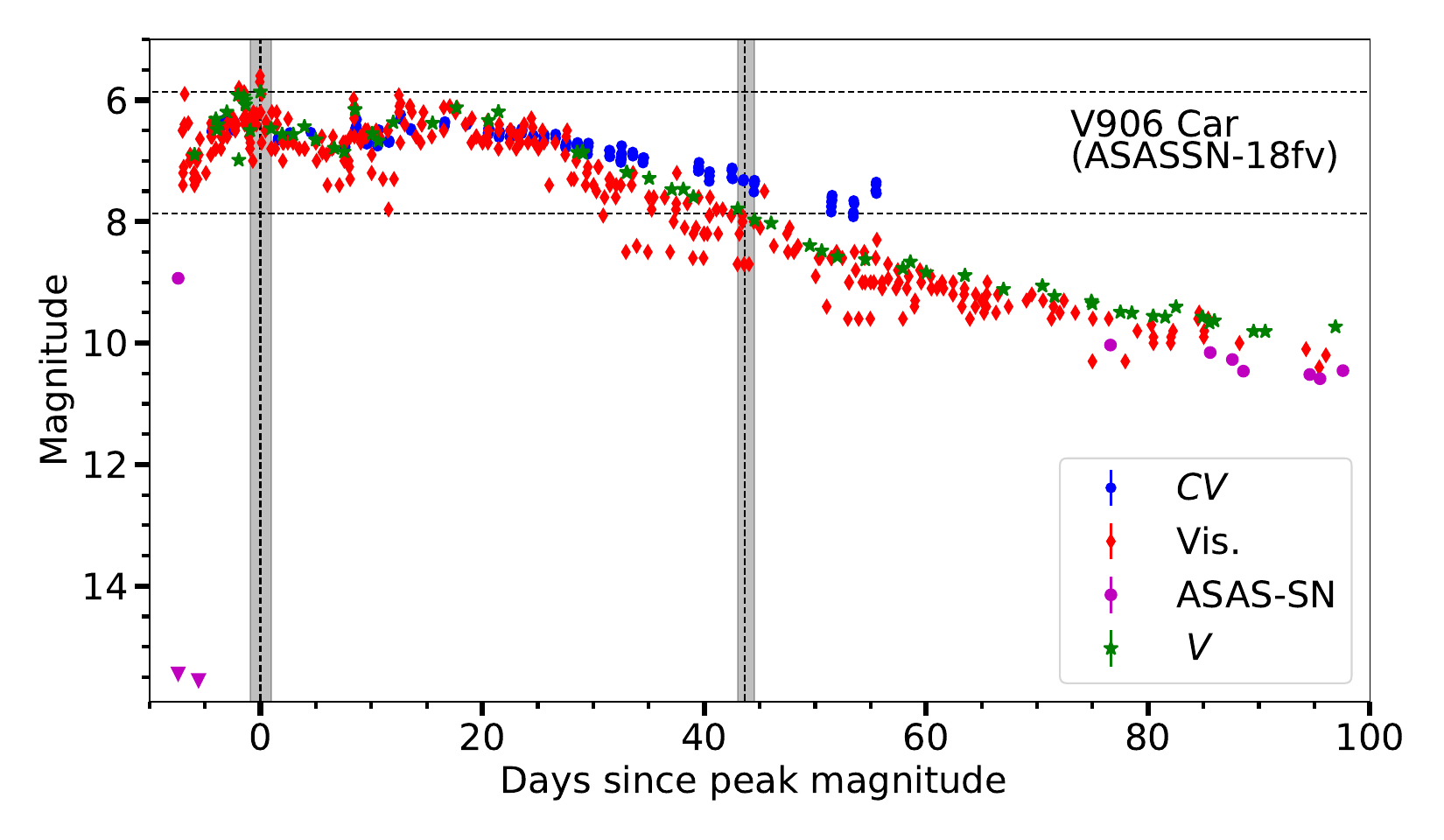}
\caption{The optical light-curve of the $\gamma$-ray detected nova V906~Car, plotted against time since peak magnitude ($t_{max}=$ 2018 Mar 28.5). Horizontal and vertical dashed lines and the shaded regions have the same meaning as in Figure~\ref{Fig:V1674_Her_LC}. The rise to light curve maximum is well captured by ASAS-SN and the BRITE constellation of nanosatellites. \citep{Aydi_etal_2020}}
\label{Fig:V906_Car_LC}
\end{center}
\end{figure*}


\begin{figure*}
\begin{center}
  \includegraphics[width=0.9\textwidth]{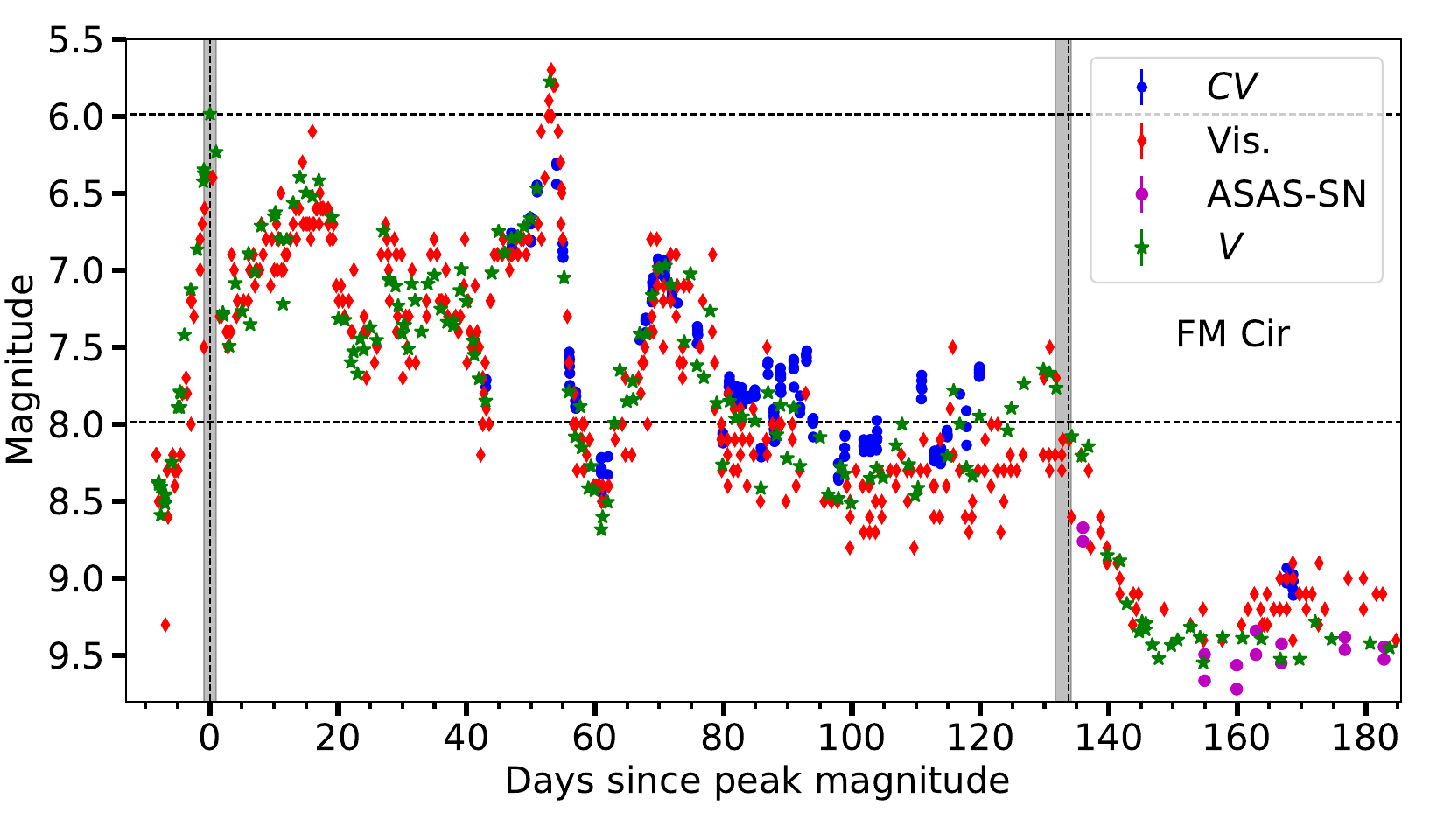}
\caption{The optical light-curve of FM~Cir, plotted against time since peak magnitude ($t_{max}=$ 2018 Jan 28.3). Horizontal and vertical dashed lines and the shaded regions have the same meaning as in Figure~\ref{Fig:V1674_Her_LC}. The rise to peak and the maximum magnitude are well-constrained by AAVSO observers.}
\label{Fig:FM_Cir_LC}
\end{center}
\end{figure*}


\begin{figure*}
\begin{center}
  \includegraphics[width=0.9\textwidth]{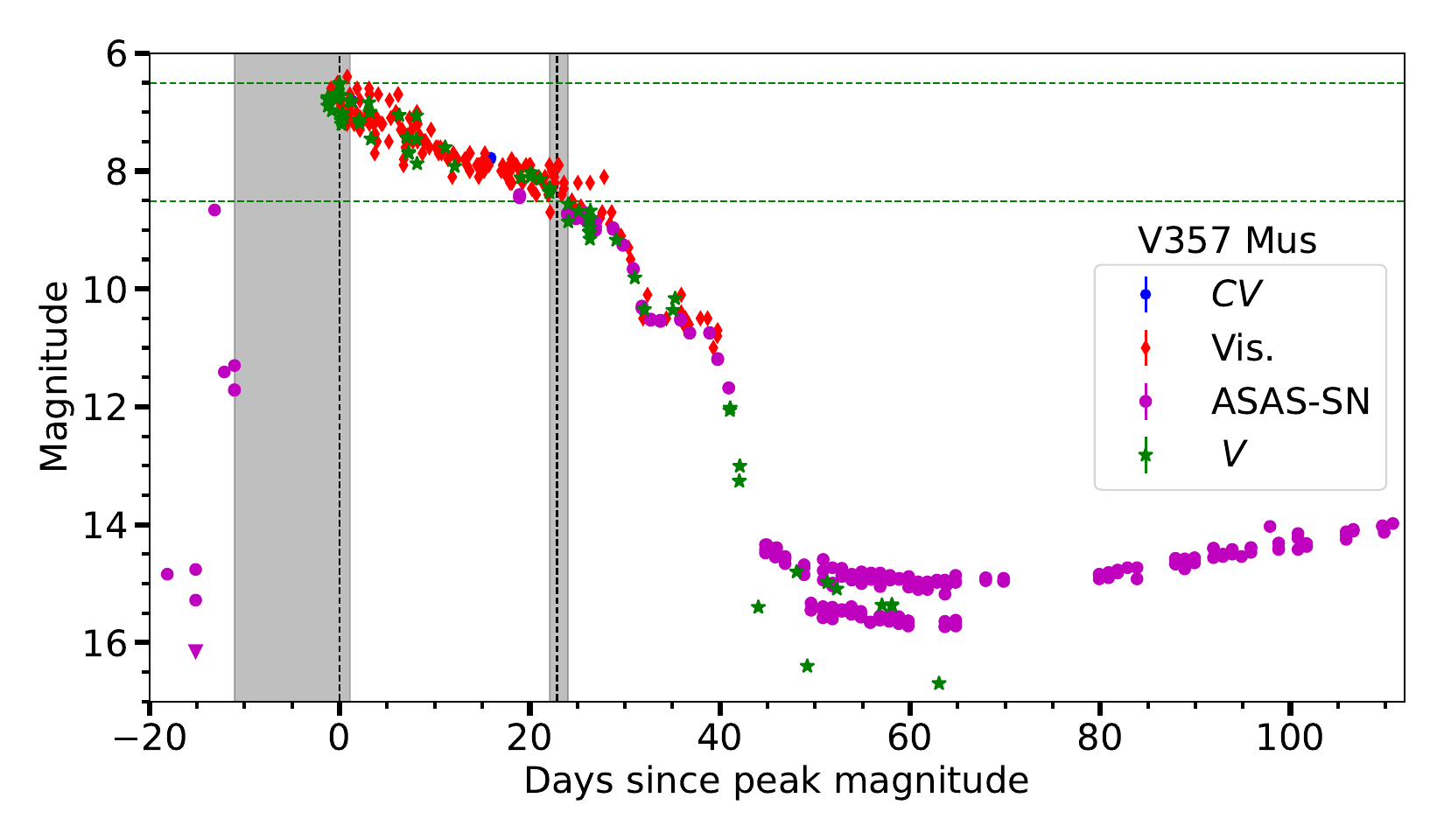}
\caption{The optical light-curve of the $\gamma$-ray detected nova V357~Mus, plotted against time since peak magnitude ($t_{max}=$ 2018 Jan 16.4). Horizontal and vertical dashed lines and the shaded regions have the same meaning as in Figure~\ref{Fig:V1674_Her_LC}. The rise and peak of the light curve is observed by ASAS-SN. However, the source was saturated for the peak, so we constrain $V_{peak}$ to be brighter than the saturation limit of ASAS-SN, represented on the plot with the top green horizontal line.}
\label{Fig:V357_Mus_LC}
\end{center}
\end{figure*}


\begin{figure*}
\begin{center}
  \includegraphics[width=0.9\textwidth]{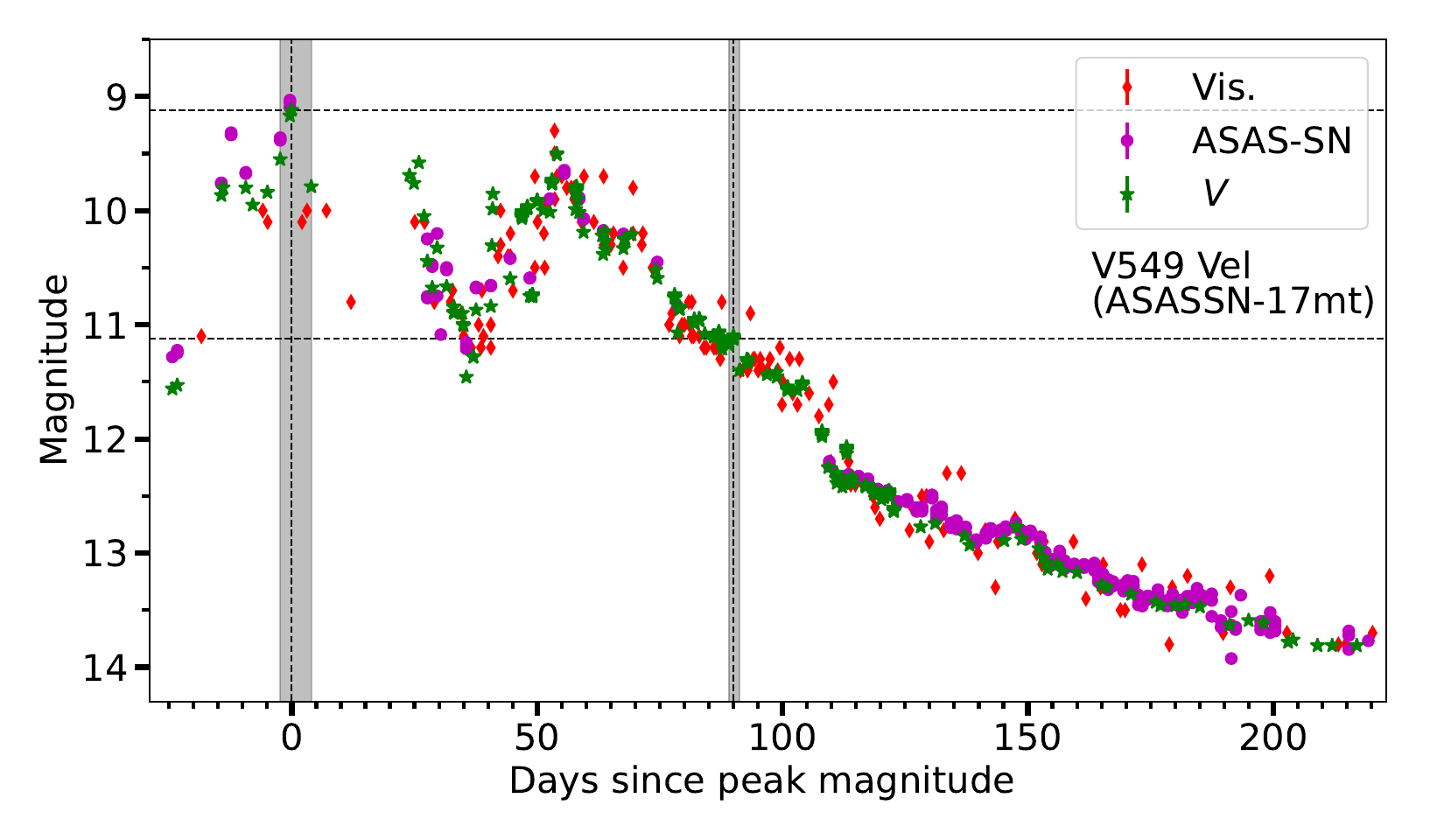}
\caption{The optical light-curve of the $\gamma$-ray detected nova V549~Vel, plotted against time since peak magnitude ($t_{max}=$ 2017 Oct 17.7). Horizontal and vertical dashed lines and the shaded regions have the same meaning as in Figure~\ref{Fig:V1674_Her_LC}. ANS collaboration photometry captures the rise to light curve maximum nicely (\protect\cite{Li_etal_2020}), and a light curve maximum around 2017 Oct 17.7 is also supported by optical spectroscopy.}
\label{Fig:V549_Vel_LC}
\end{center}
\end{figure*}


\begin{figure*}
\begin{center}
  \includegraphics[width=0.9\textwidth]{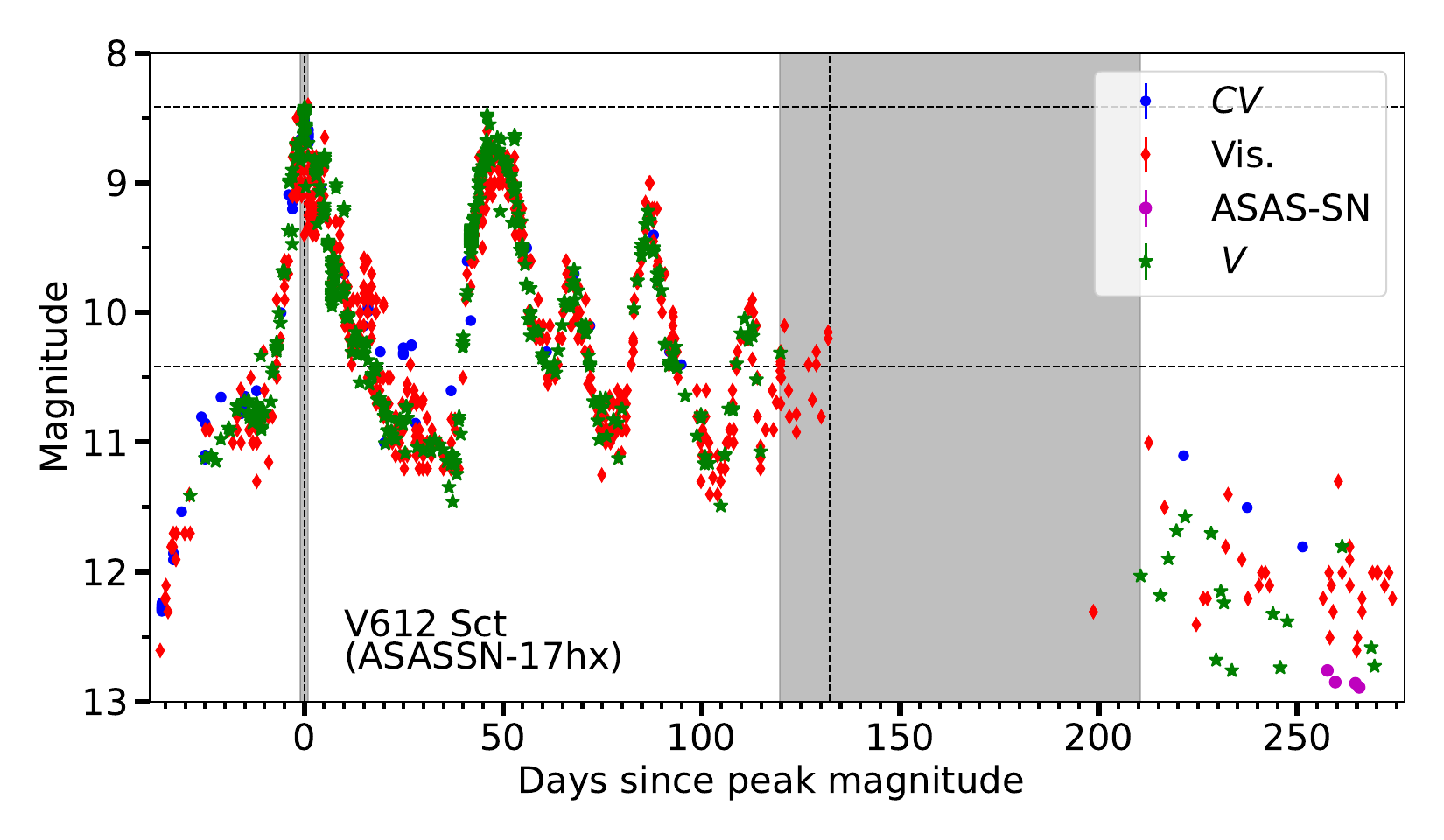}
\caption{The optical light-curve of V612~Sct, plotted against time since peak magnitude ($t_{max}=$ 2017 Jul 30.0). Horizontal and vertical dashed lines and the shaded regions have the same meaning as in Figure~\ref{Fig:V1674_Her_LC}. The nova was discovered by ASAS-SN, and the light curve rise is well observed by AAVSO observers. $t_2$ is less well constrained, because it occurred during solar conjunction (day 131--198).}
\label{Fig:V612_Sct_LC}
\end{center}
\end{figure*}


\begin{figure*}
\begin{center}
  \includegraphics[width=0.9\textwidth]{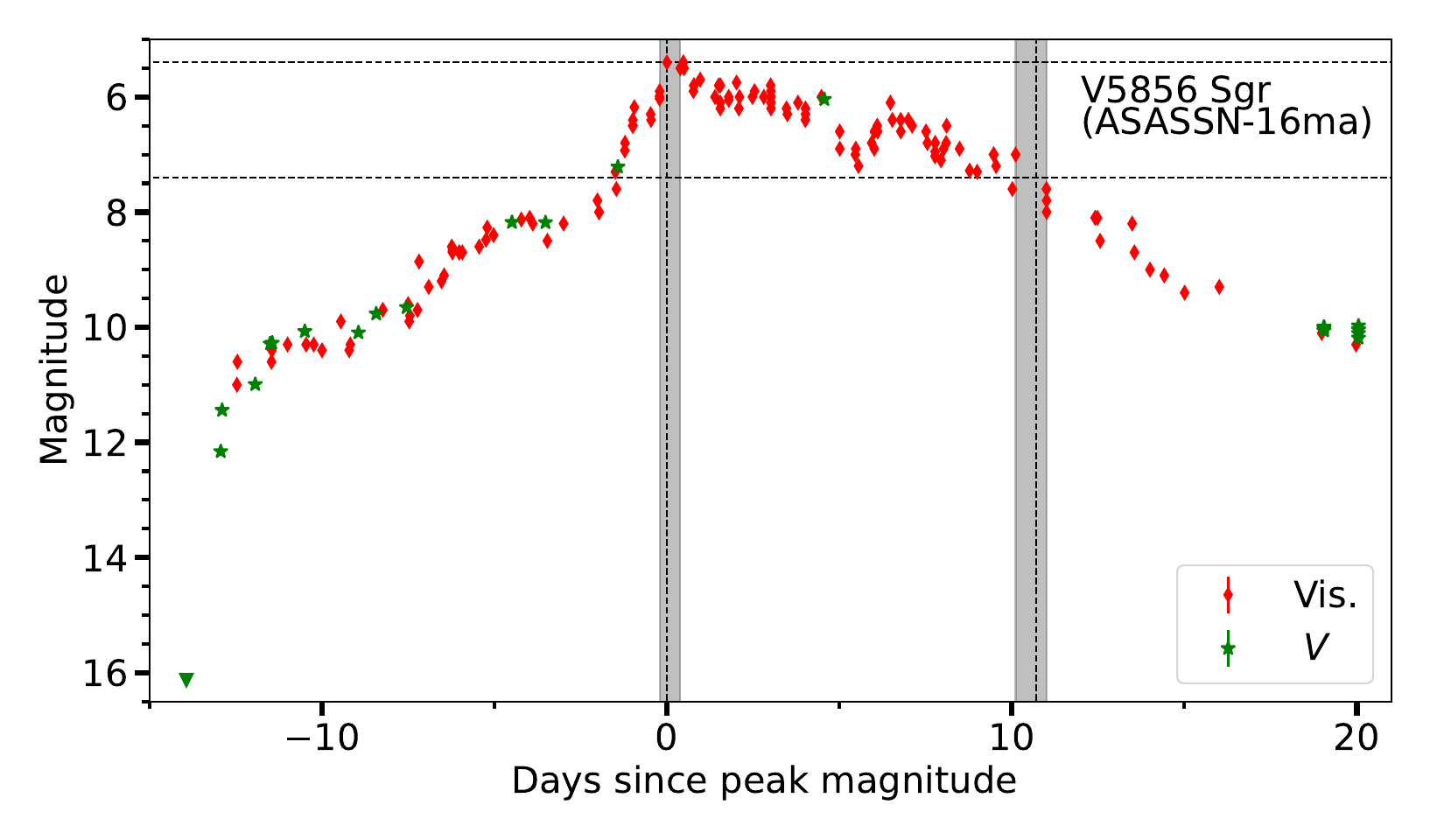}
\caption{The optical light-curve of the $\gamma$-ray detected nova V5856~Sgr, plotted against time since peak magnitude ($t_{max}=$ 2016 Nov 08.0). Horizontal and vertical dashed lines and the shaded regions have the same meaning as in Figure~\ref{Fig:V1674_Her_LC}. The green triangle represents an upper limit on the magnitude in the V-band. For this nova, the visual band of light was used to measure $V_{\mathrm{peak}}$ and $t_2$ because the rise to light curve maximum is well captured by AAVSO observations.}
\label{Fig:V5856_Sgr_LC}
\end{center}
\end{figure*}


\begin{figure*}
\begin{center}
  \includegraphics[width=0.9\textwidth]{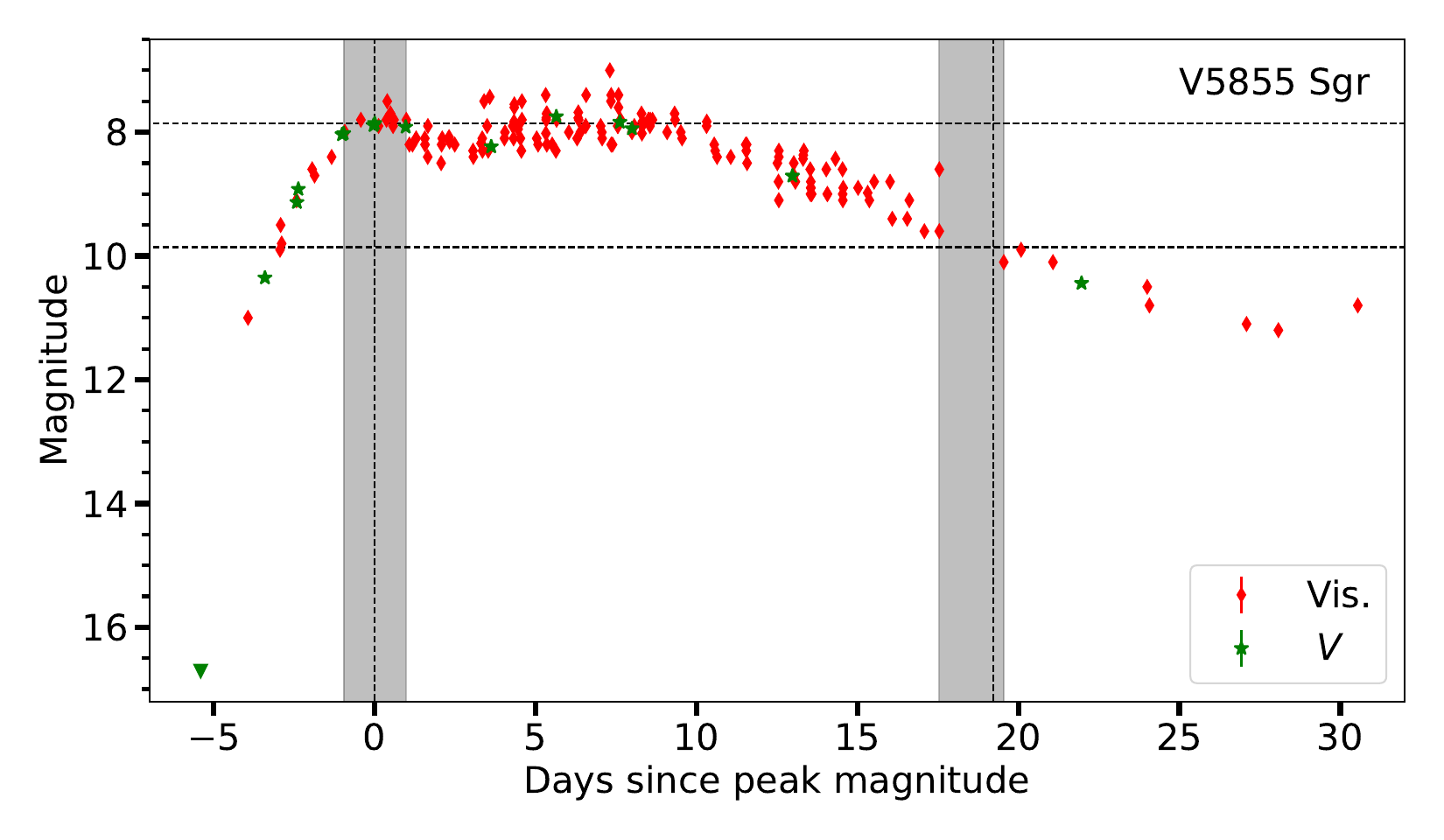}
\caption{The optical light-curve of the $\gamma$-ray detected nova V5855~Sgr, plotted against time since peak magnitude ($t_{max}=$ 2016 Oct 24.4). Horizontal and vertical dashed lines and the shaded regions have the same meaning as in Figure~\ref{Fig:V1674_Her_LC}. The green triangle represents an upper limit on the magnitude in the V-band.  The rise to light curve maximum is well captured by AAVSO observations and \protect\cite{Munari_etal_2017}. The time of optical maximum (which is  ambiguous from the light curve alone) is justified by the optical spectroscopic evolution.}
\label{Fig:V5855_Sgr_LC}
\end{center}
\end{figure*}


\begin{figure*}
\begin{center}
  \includegraphics[width=0.9\textwidth]{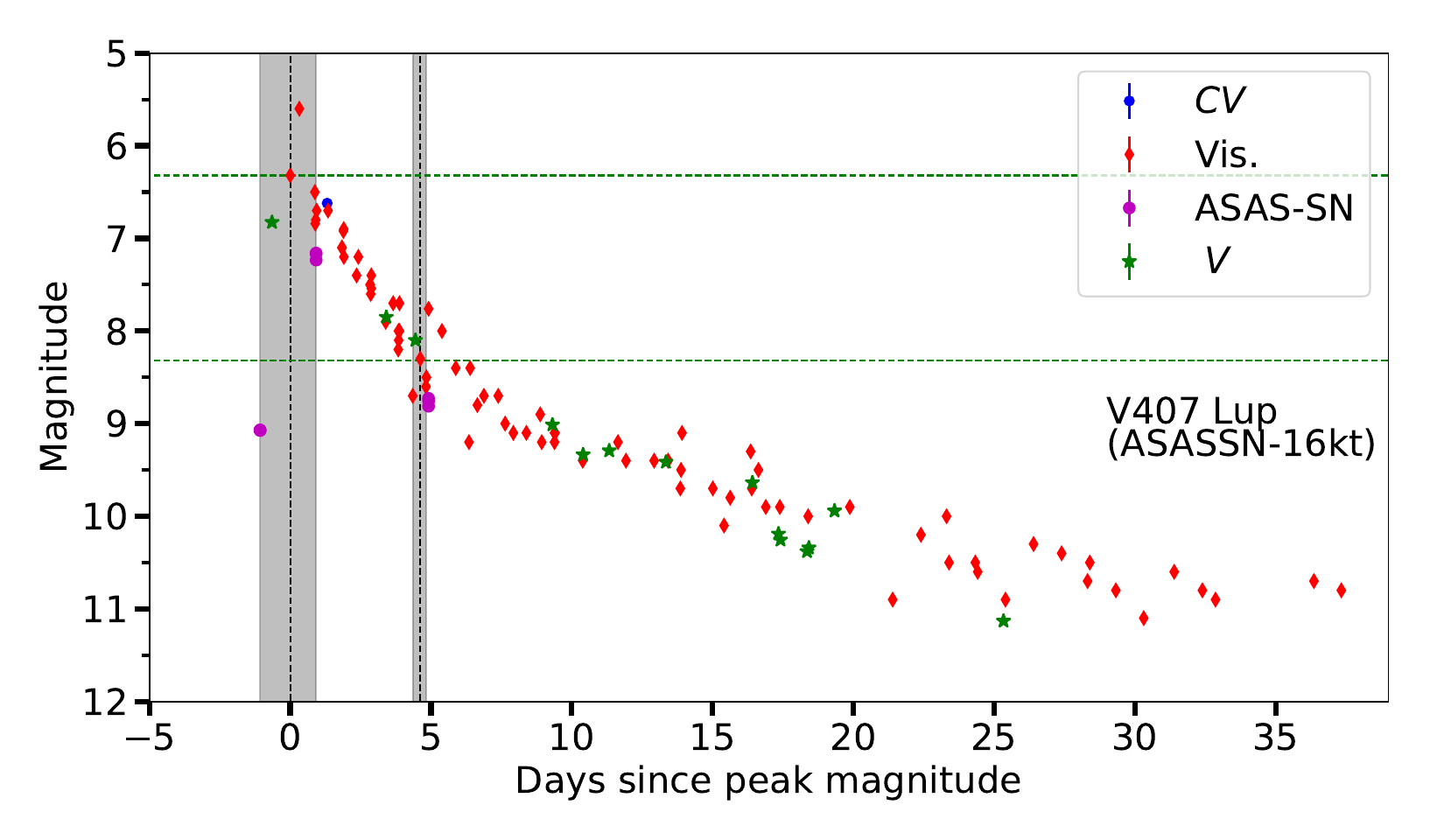}
\caption{The optical light-curve of V407~Lup, plotted against time since peak magnitude ($t_{max}=$ 2016 Sep 25.1). Horizontal and vertical dashed lines and the shaded regions have the same meaning as in Figure~\ref{Fig:V1674_Her_LC}. Here, the two horizontal dashed lines are green to indicate that the optical peak is an upper limit, meaning the peak may not have been observed. A pre-discovery observation from ASAS-SN on 2016 Sep 24 constrains the rise of the optical light curve well.}
\label{Fig:V407_Lup_LC}
\end{center}
\end{figure*}


\begin{figure*}
\begin{center}
  \includegraphics[width=0.9\textwidth]{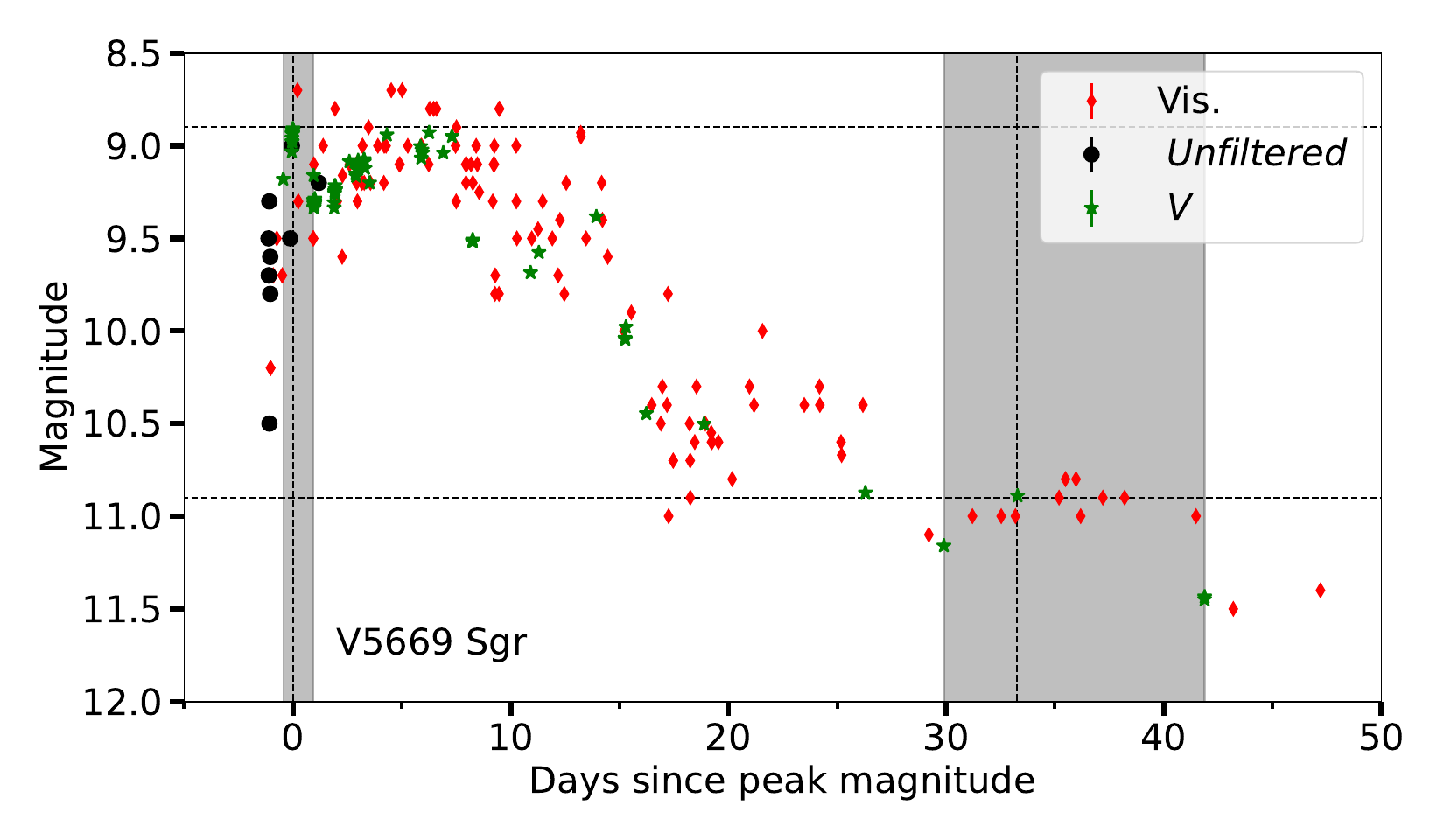}
\caption{The optical light-curve of V5669~Sgr, plotted against time since peak magnitude ($t_{max}=$ 2015 Sep 28.5). Horizontal and vertical dashed lines and the shaded regions have the same meaning as in Figure~\ref{Fig:V1674_Her_LC}. Pre-discovery non-detections from 2015 Sep 23.4 constrain the time of optical peak to be around 2015 Sep 28.5 \citep{2015CBET.4145....2N}.}
\label{Fig:V5669_Sgr_LC}
\end{center}
\end{figure*}


\begin{figure*}
\begin{center}
  \includegraphics[width=0.9\textwidth]{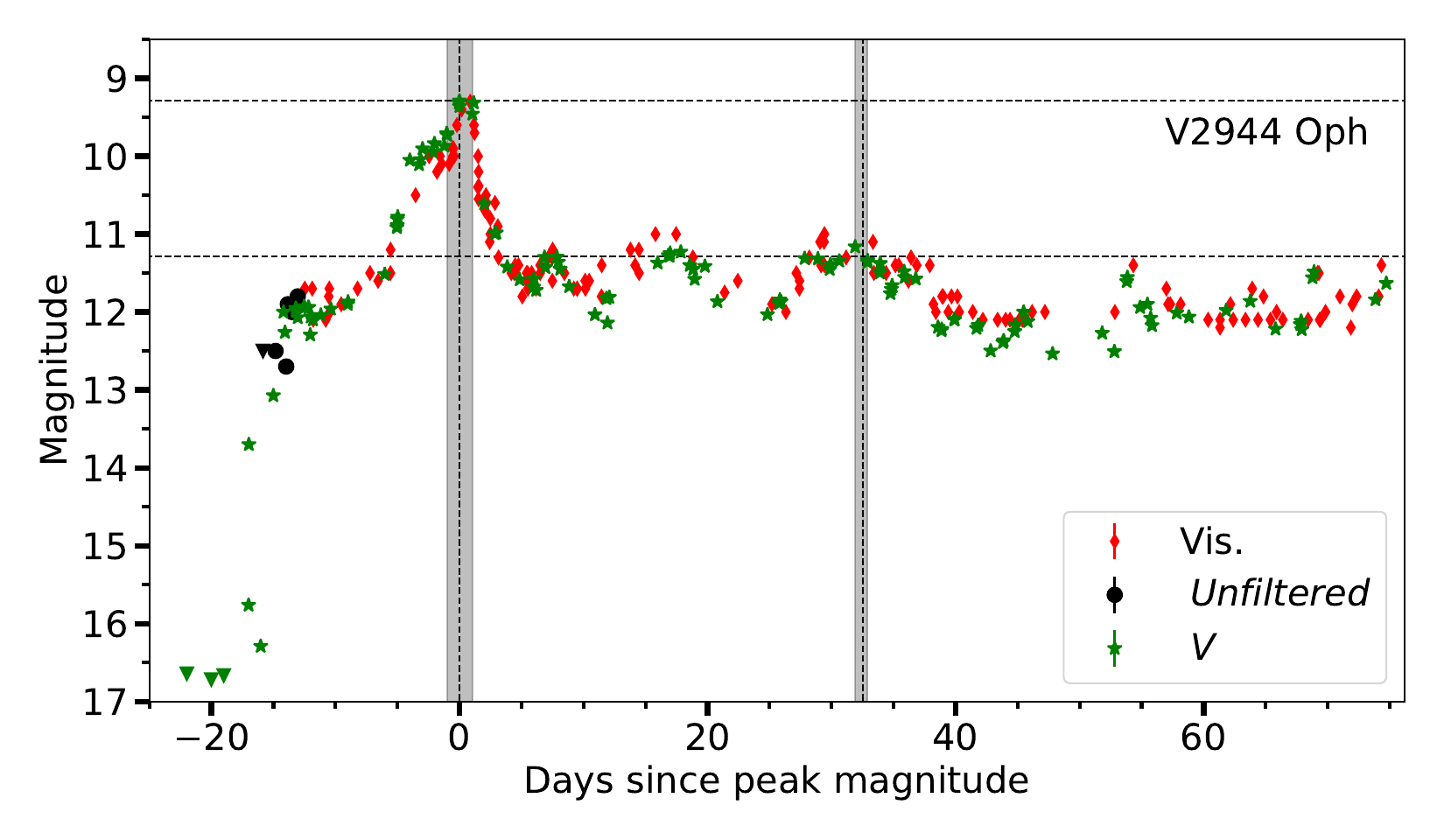}
\caption{The optical light-curve of V2944~Oph, plotted against time since peak magnitude ($t_{max}=$ 2015 Apr 13.6). Horizontal and vertical dashed lines and the shaded regions have the same meaning as in Figure~\ref{Fig:V1674_Her_LC}. The rise to light curve maximum is well captures by AAVSO observers and the non-detection on 2015 Mar 28.7 \citep{2015CBET.4086....2K}.}
\label{Fig:v2944_Oph_LC}
\end{center}
\end{figure*}


\begin{figure*}
\begin{center}
  \includegraphics[width=0.9\textwidth]{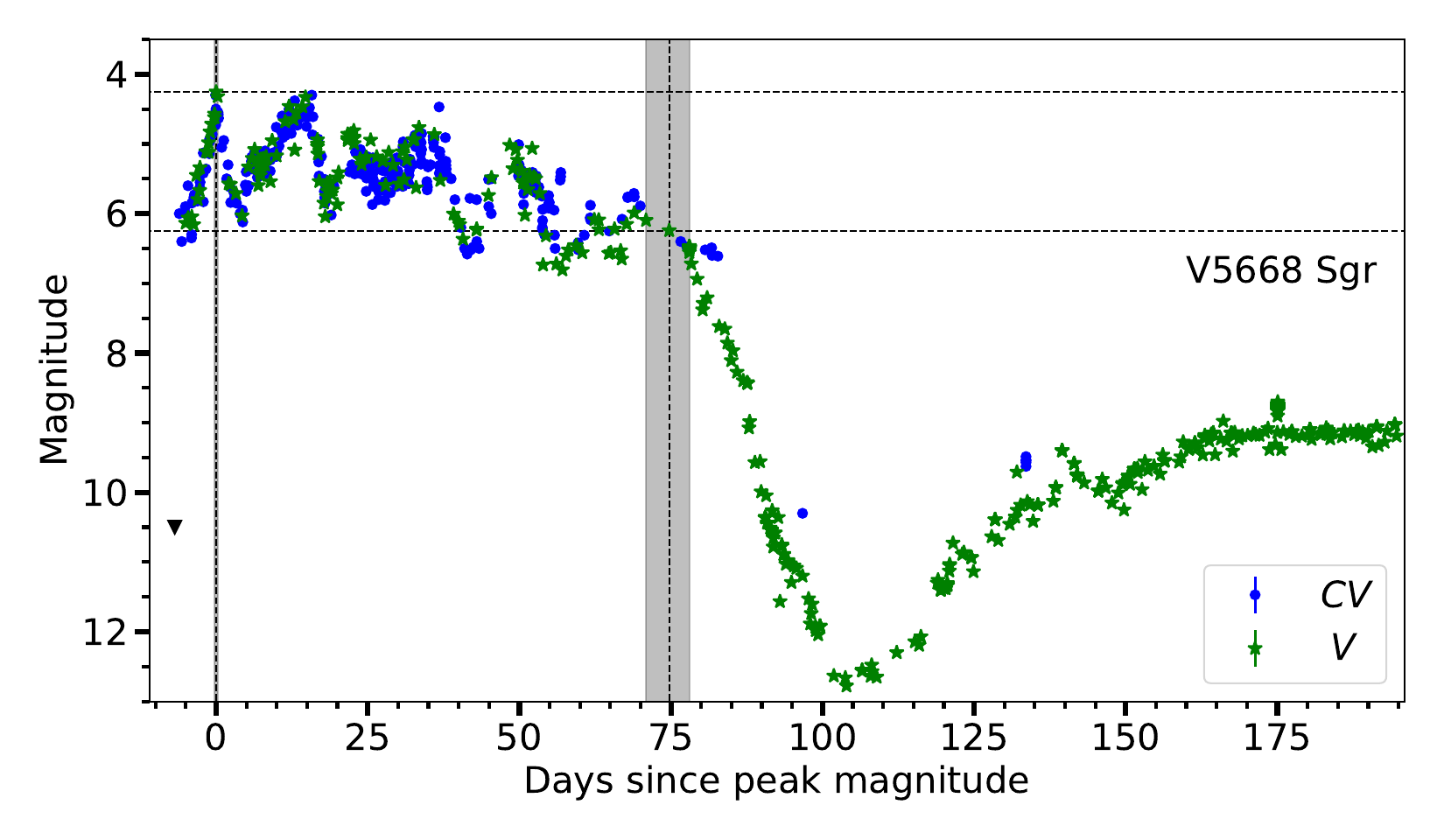}
\caption{The optical light-curve of the $\gamma$-ray detected nova V5668~Sgr, plotted against time since peak magnitude ($t_{max}=$ 2015 Mar 21.4). Horizontal and vertical dashed lines and the shaded regions have the same meaning as in Figure~\ref{Fig:V1674_Her_LC}. A pre-eruption non-detection on 2015 Mar 14.6 constrains well the rise to optical peak \citep{2015CBET.4080....2I}.}
\label{Fig:V5668_Sgr_LC}
\end{center}
\end{figure*}


\begin{figure*}
\begin{center}
  \includegraphics[width=0.9\textwidth]{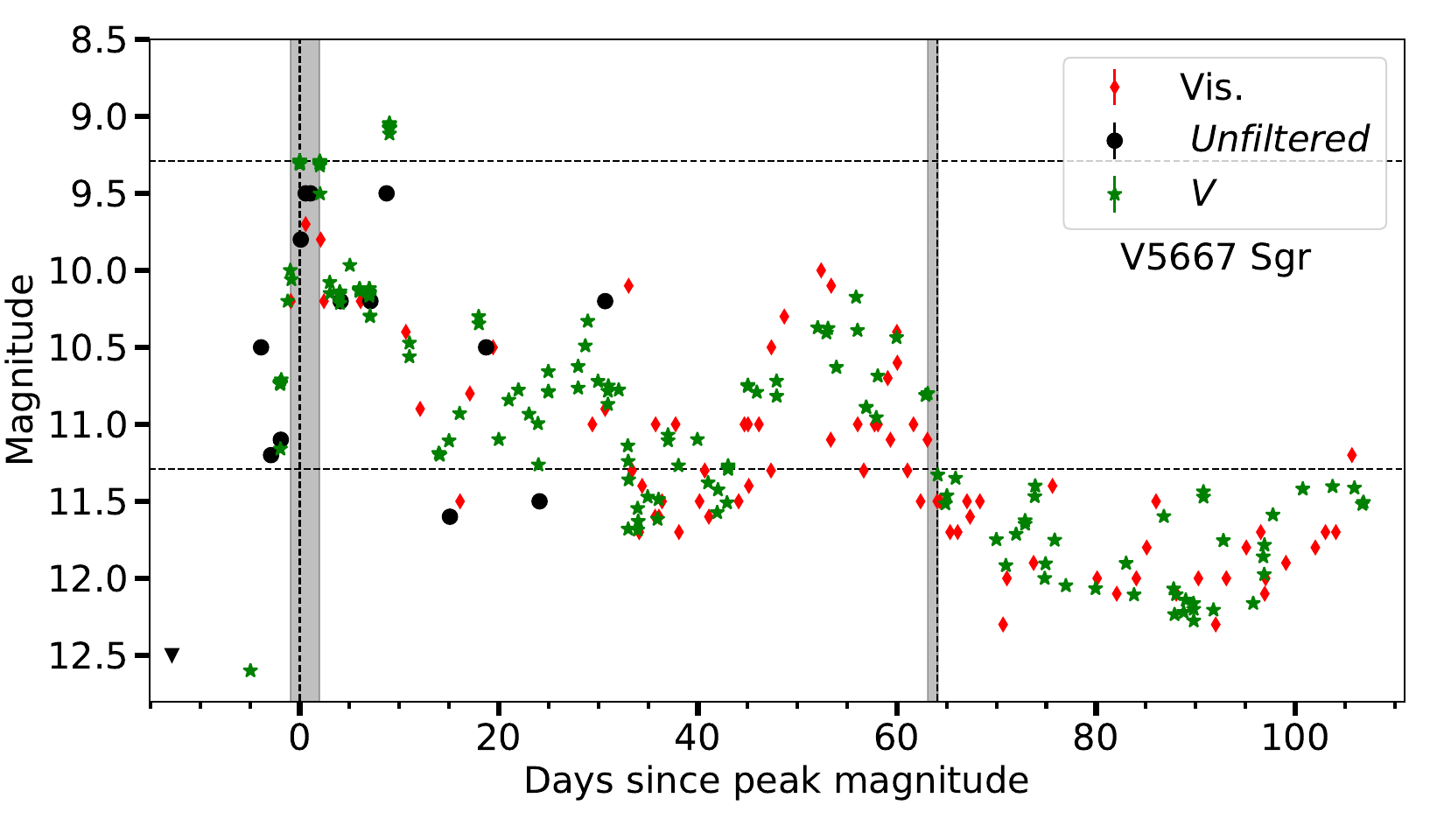}
\caption{The optical light-curve of V5667~Sgr, plotted against time since peak magnitude ($t_{max}=$ 2015 Feb 15.7). Horizontal and vertical dashed lines and the shaded regions have the same meaning as in Figure~\ref{Fig:V1674_Her_LC}. The rise to light curve maximum is captured well by \protect\cite{2015CBET.4079....2K} and AAVSO observations.}
\label{Fig:V5667_Sgr_LC}
\end{center}
\end{figure*}


\begin{figure*}
\begin{center}
  \includegraphics[width=0.9\textwidth]{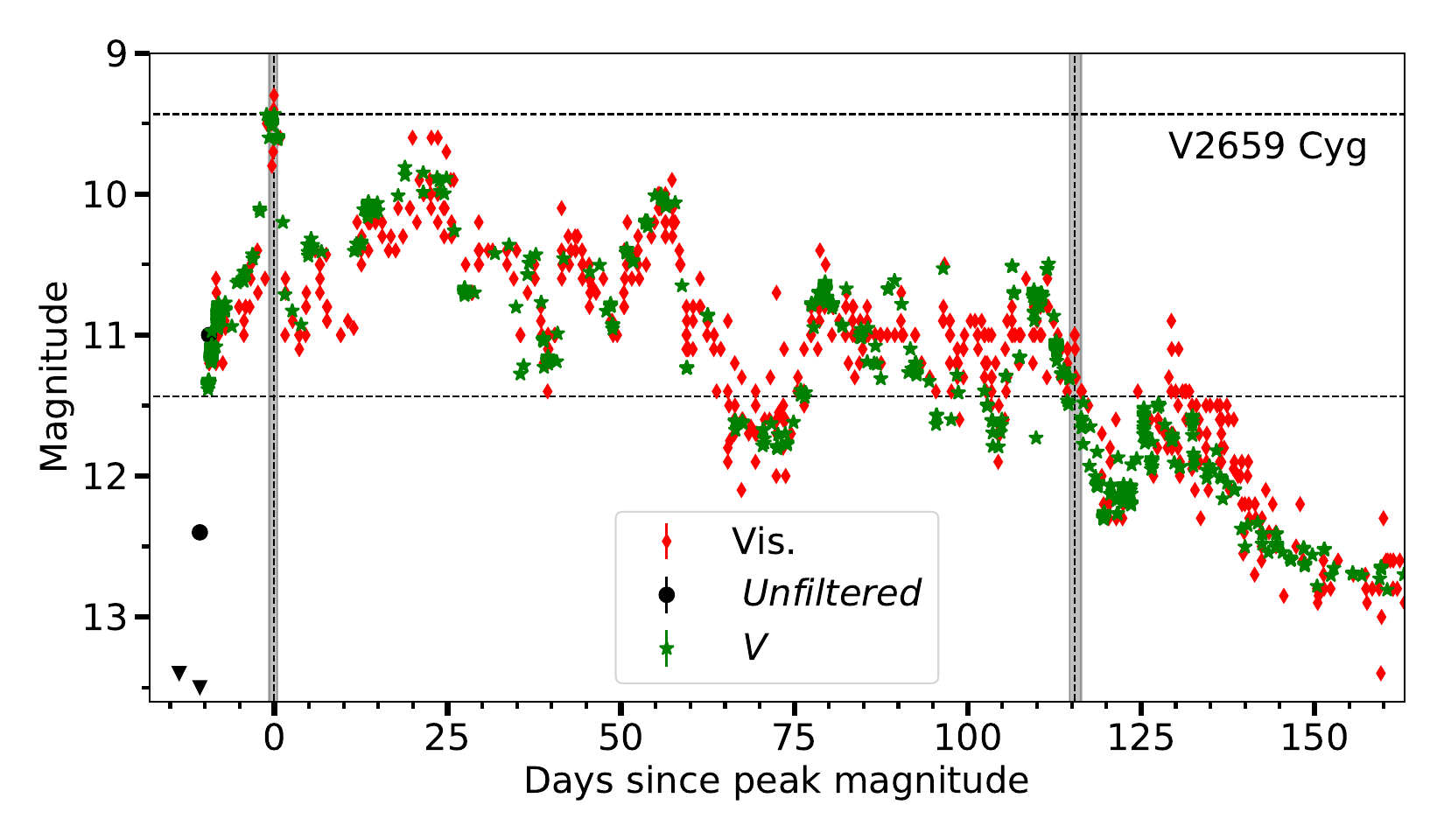}
\caption{The optical light-curve of V2659~Cyg, plotted against time since peak magnitude ($t_{max}=$ 2014 Apr 10.5). Horizontal and vertical dashed lines and the shaded regions have the same meaning as in Figure~\ref{Fig:V1674_Her_LC}. A non-detection on 2014 Mar 27.8 constrains the optical rise well \citep{2014CBET.3842....1N}.}
\label{Fig:V2659_Cyg_LC}
\end{center}
\end{figure*}


\begin{figure*}
\begin{center}
  \includegraphics[width=0.9\textwidth]{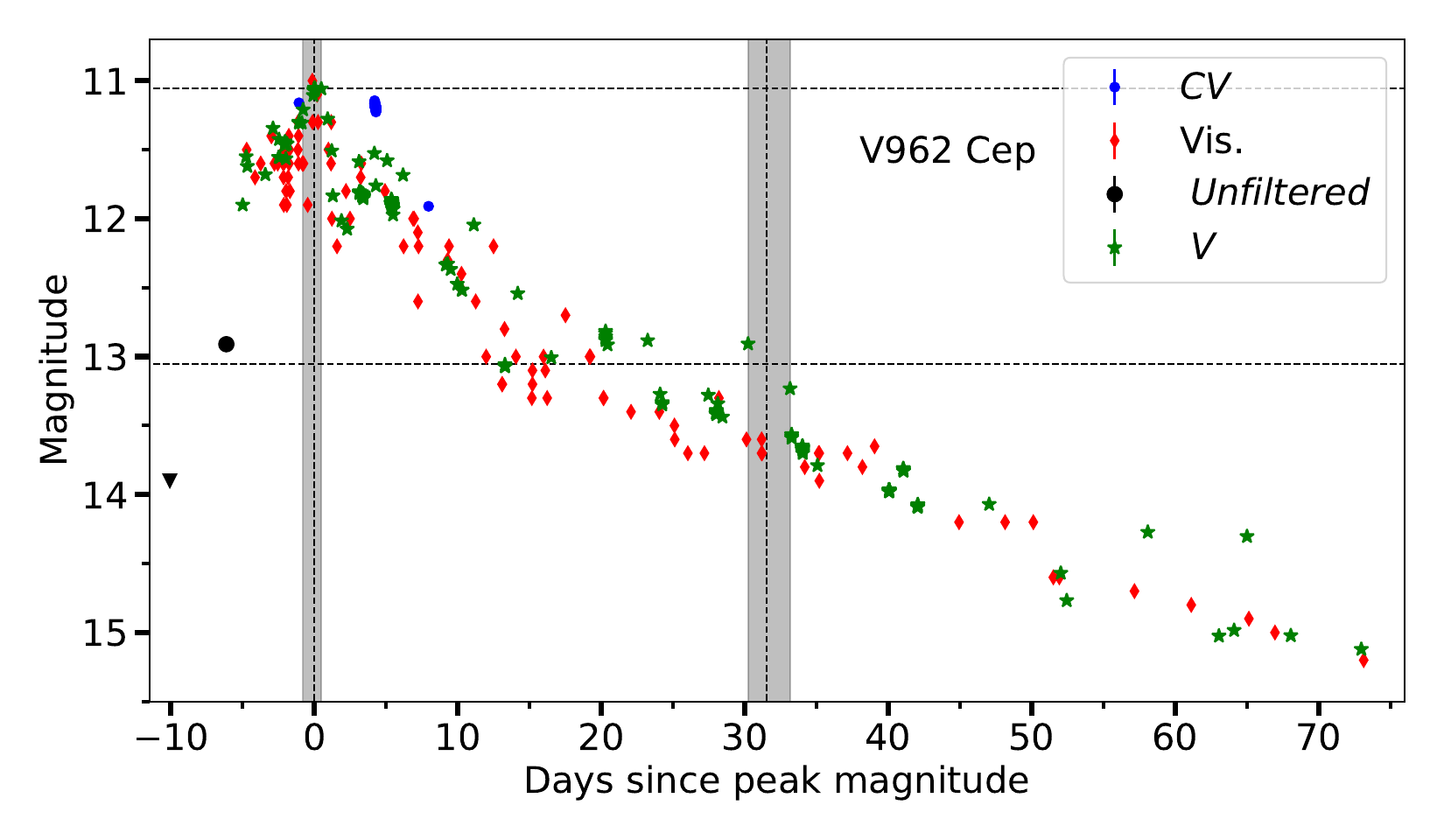}
\caption{The optical light-curve of V962~Cep, plotted against time since peak magnitude ($t_{max}=$ 2014 Mar 13.9). Horizontal and vertical dashed lines and the shaded regions have the same meaning as in Figure~\ref{Fig:V1674_Her_LC}. A pre-eruption non-detection on 2014 Mar 3.8 constrains the optical rise and time of optical maximum \citep{2014CBET.3825....2M}.}
\label{Fig:V962_Cep_LC}
\end{center}
\end{figure*}


\begin{figure*}
\begin{center}
  \includegraphics[width=0.9\textwidth]{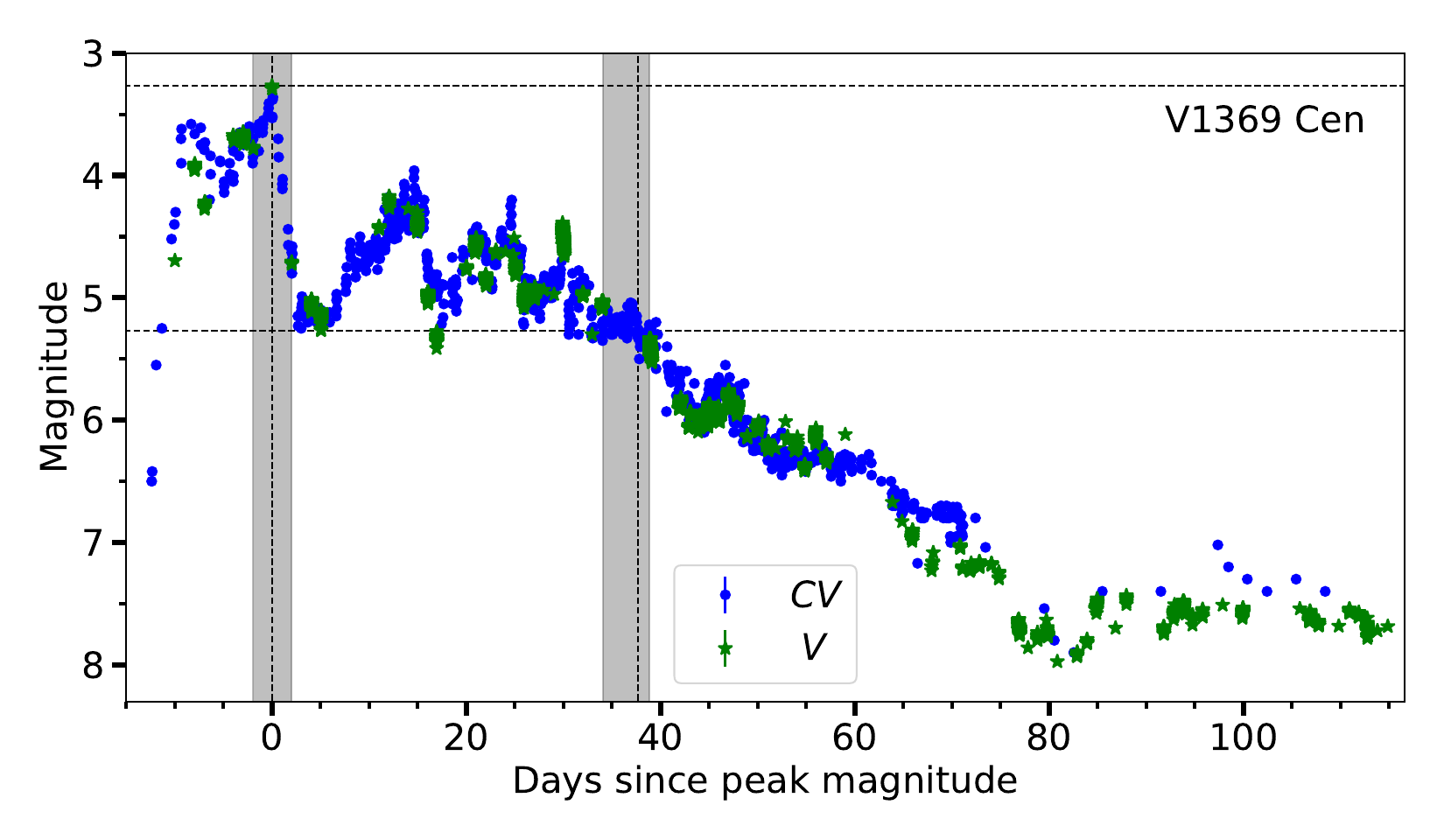}
\caption{The optical light-curve of the $\gamma$-ray detected nova V1369~Cen, plotted against time since peak magnitude ($t_{max}=$ 2013 Dec 14.7). Horizontal and vertical dashed lines and the shaded regions have the same meaning as in Figure~\ref{Fig:V1674_Her_LC}. The rise to light curve maximum is captured well by AAVSO observations.}
\label{Fig:V1369_Cen_LC}
\end{center}
\end{figure*}


\begin{figure*}
\begin{center}
  \includegraphics[width=0.9\textwidth]{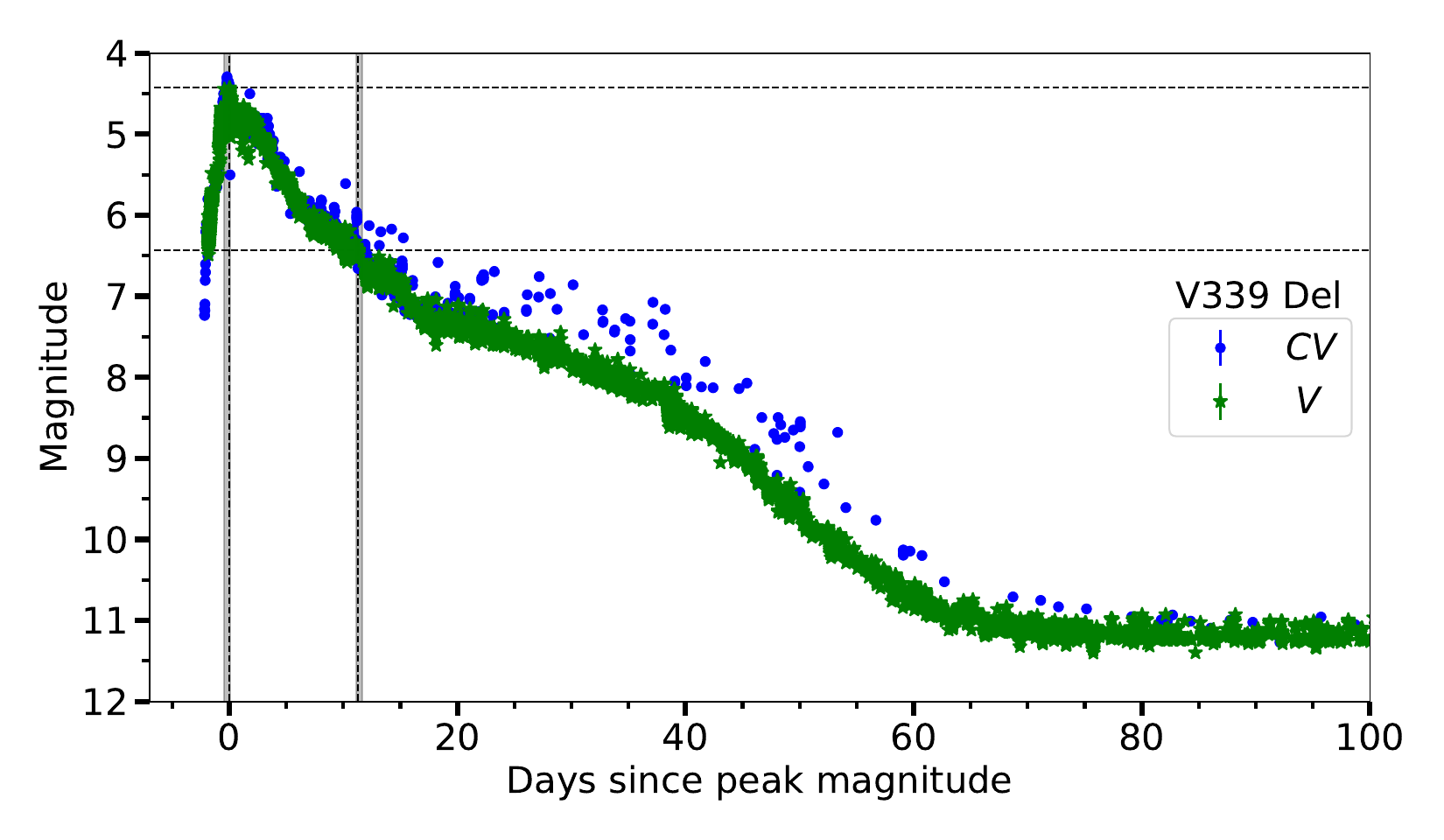}
\caption{The optical light-curve of the $\gamma$-ray detected nova V339~Del, plotted against time since peak magnitude ($t_{max}=$ 2013 Aug 16.7). Horizontal and vertical dashed lines and the shaded regions have the same meaning as in Figure~\ref{Fig:V1674_Her_LC}. The rise to light curve maximum is captured well by AAVSO observations.}
\label{Fig:V339_Del_LC}
\end{center}
\end{figure*}


\begin{figure*}
\begin{center}
  \includegraphics[width=0.9\textwidth]{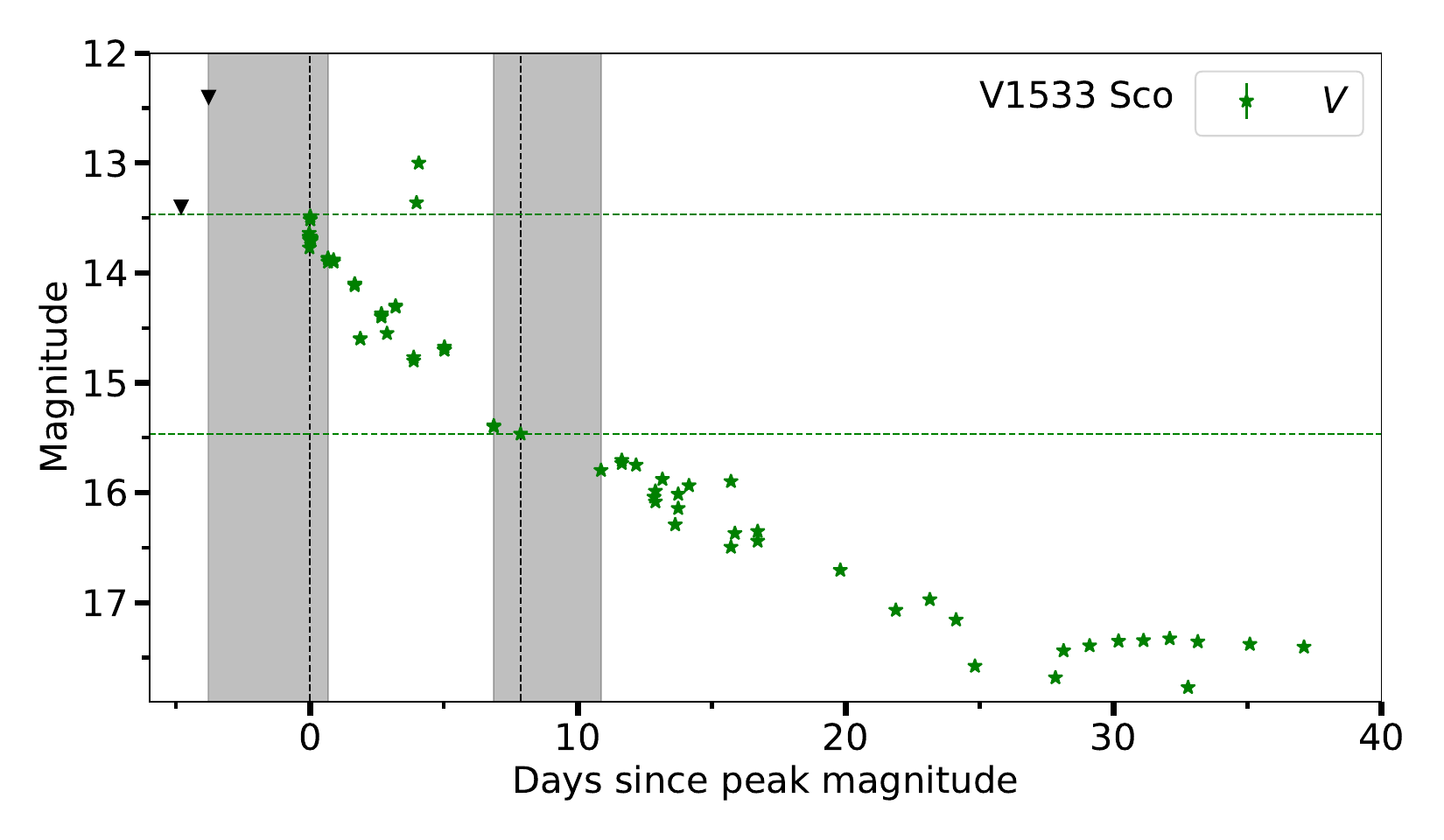}
\caption{The optical light-curve of V1533~Sco, plotted against time since peak magnitude ($t_{max}=$ 2013 Jun 04.5). Horizontal and vertical dashed lines and the shaded regions have the same meaning as in Figure~\ref{Fig:V1674_Her_LC}. Here, the two horizontal dashed lines are green to indicate that the optical peak is an upper limit, meaning the peak may not have been observed. A pre-eruption limit from 2013 May 30.6 \citep{2013CBET.3542....1Y} and from 2013 May 31.7 \citep{2013CBET.3556....1K} imply that light curve maximum occurred around 2013 Jun 04.5.}
\label{Fig:V1533_Sco_LC}
\end{center}
\end{figure*}


\begin{figure*}
\begin{center}
  \includegraphics[width=0.9\textwidth]{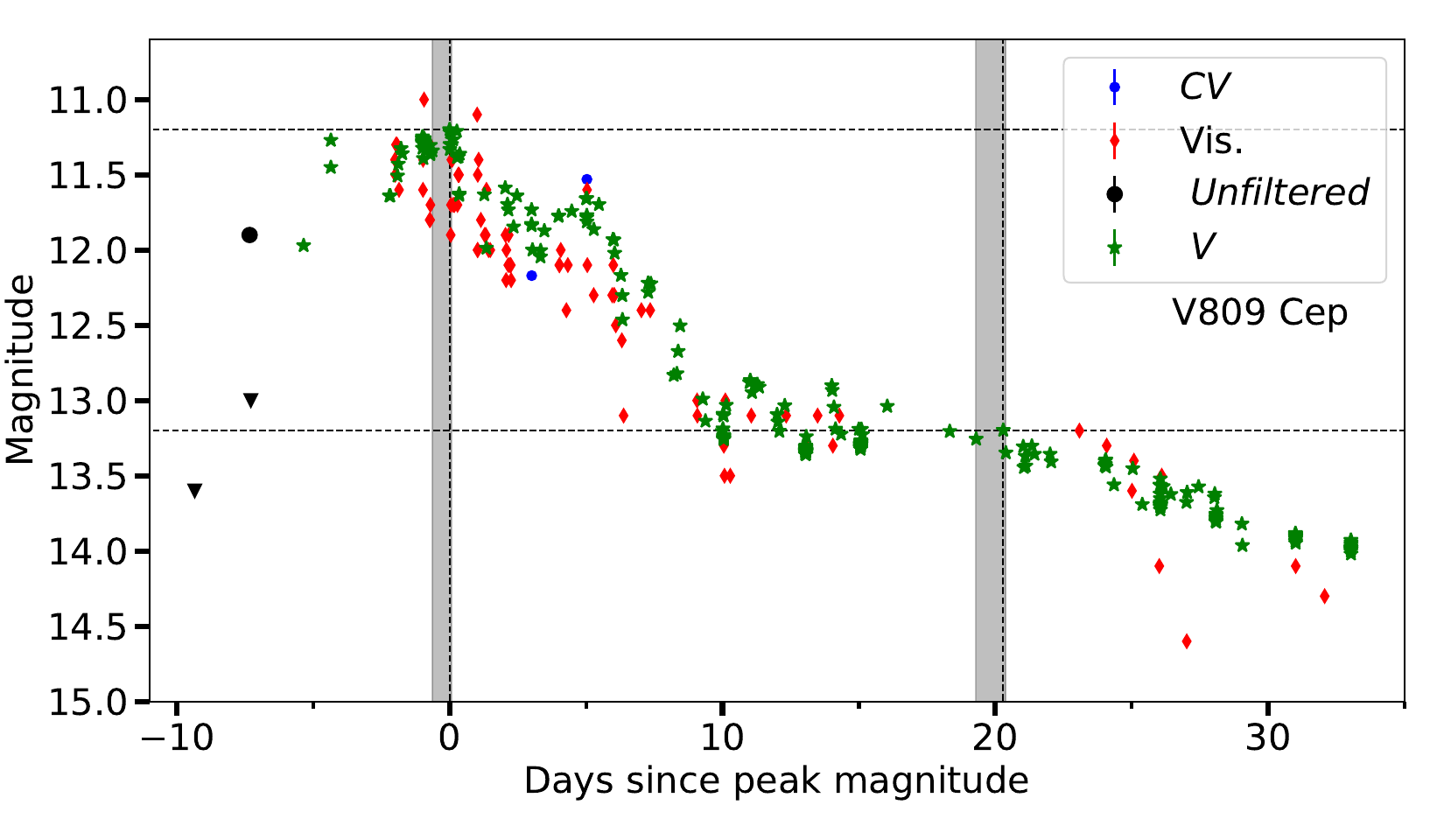}
\caption{The optical light-curve of V809~Cep, plotted against time since peak magnitude ($t_{max}=$ 2013 Feb 04.7). Horizontal and vertical dashed lines and the shaded regions have the same meaning as in Figure~\ref{Fig:V1674_Her_LC}. Pre-discovery measurements of the nova flux well constrain the time of optical peak to around 2013 Feb 04.7 \citep{2013CBET.3397....1J}.}
\label{Fig:V809_Cep_LC}
\end{center}
\end{figure*}


\begin{figure*}
\begin{center}
  \includegraphics[width=0.9\textwidth]{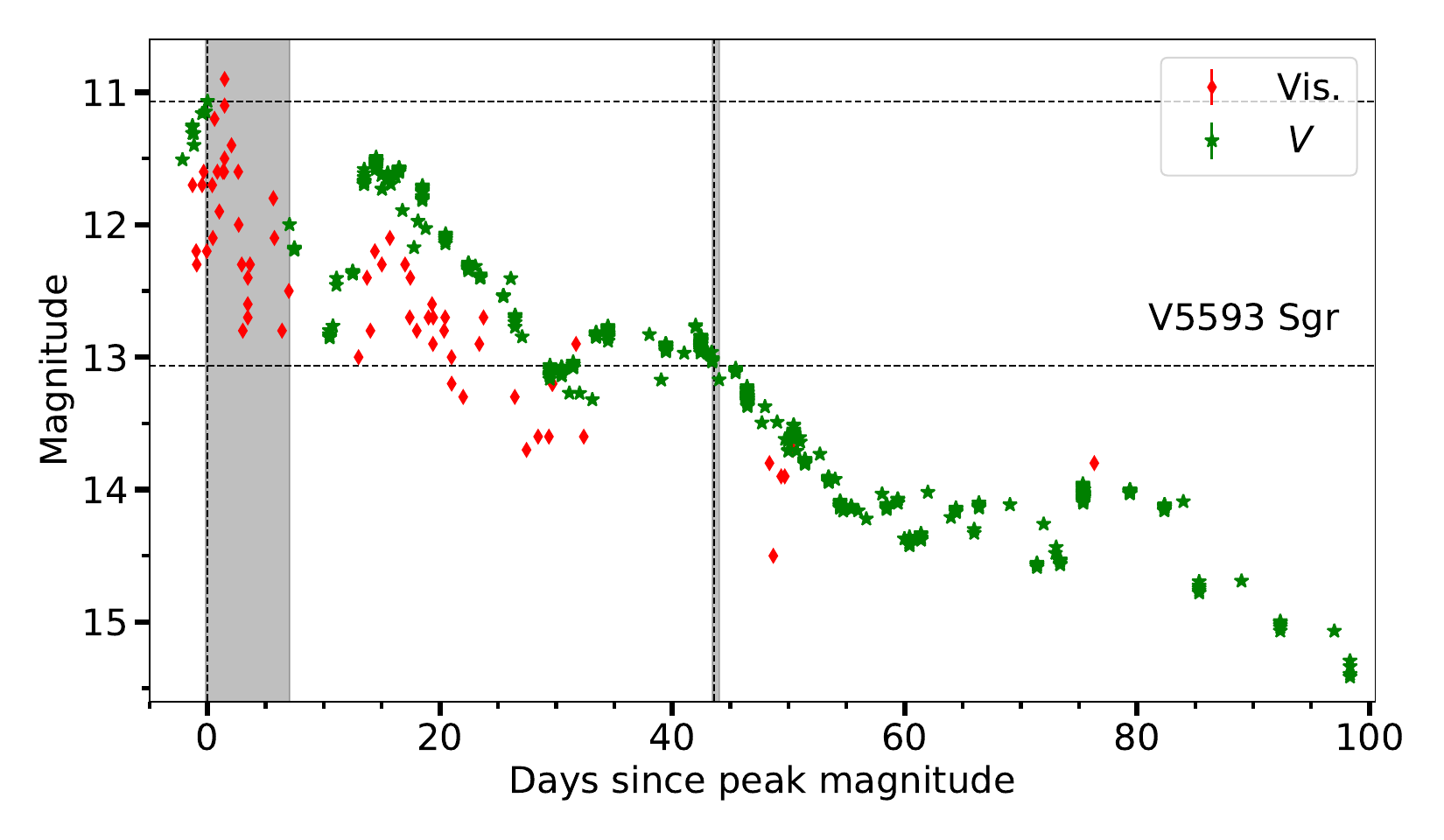}
\caption{The optical light-curve of V5593~Sgr, plotted against time since peak magnitude ($t_{max}=$ 2012 Jul 22.4). Horizontal and vertical dashed lines and the shaded regions have the same meaning as in Figure~\ref{Fig:V1674_Her_LC}. Optical spectroscopy confirms that the time of light curve maximum is around 2012 Jul 22.4.}
\label{Fig:v5593_Sgr_LC}
\end{center}
\end{figure*}


\begin{figure*}
\begin{center}
  \includegraphics[width=0.9\textwidth]{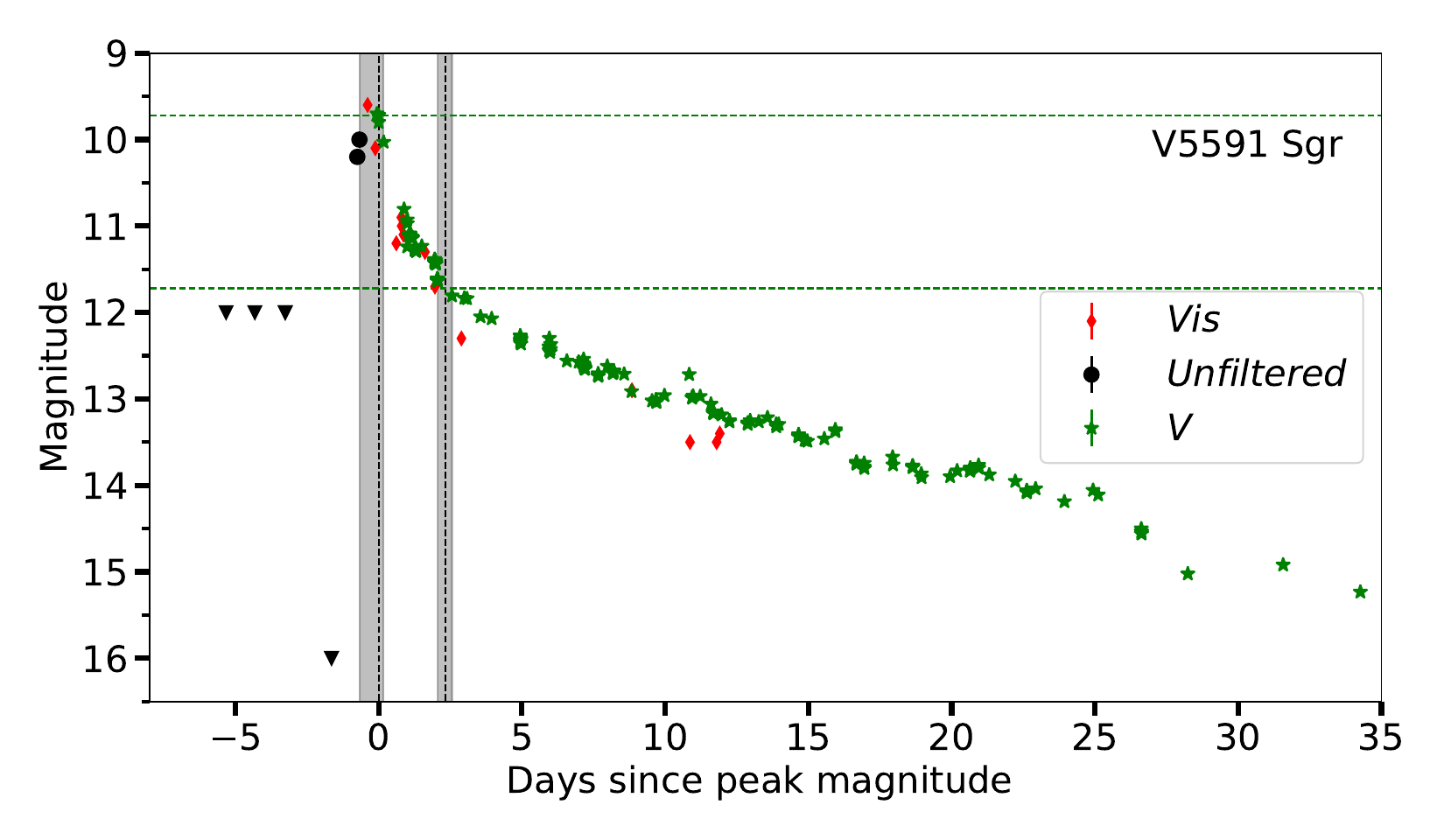}
\caption{The optical light-curve of V5591~Sgr, plotted against time since peak magnitude ($t_{max}=$ 2012 Jun 27.3). Horizontal and vertical dashed lines and the shaded regions have the same meaning as in Figure~\ref{Fig:V1674_Her_LC}. Here, the two horizontal dashed lines are green to indicate that the optical peak is an upper limit, meaning the peak may not have been observed. A pre-eruption limit from 2012 June 25.7 constrains that light curve maximum occurred around 2012 Jun 27.3 \citep{2012CBET.3156....2S}.}
\label{Fig:V5591_Sgr_LC}
\end{center}
\end{figure*}


\begin{figure*}
\begin{center}
  \includegraphics[width=0.9\textwidth]{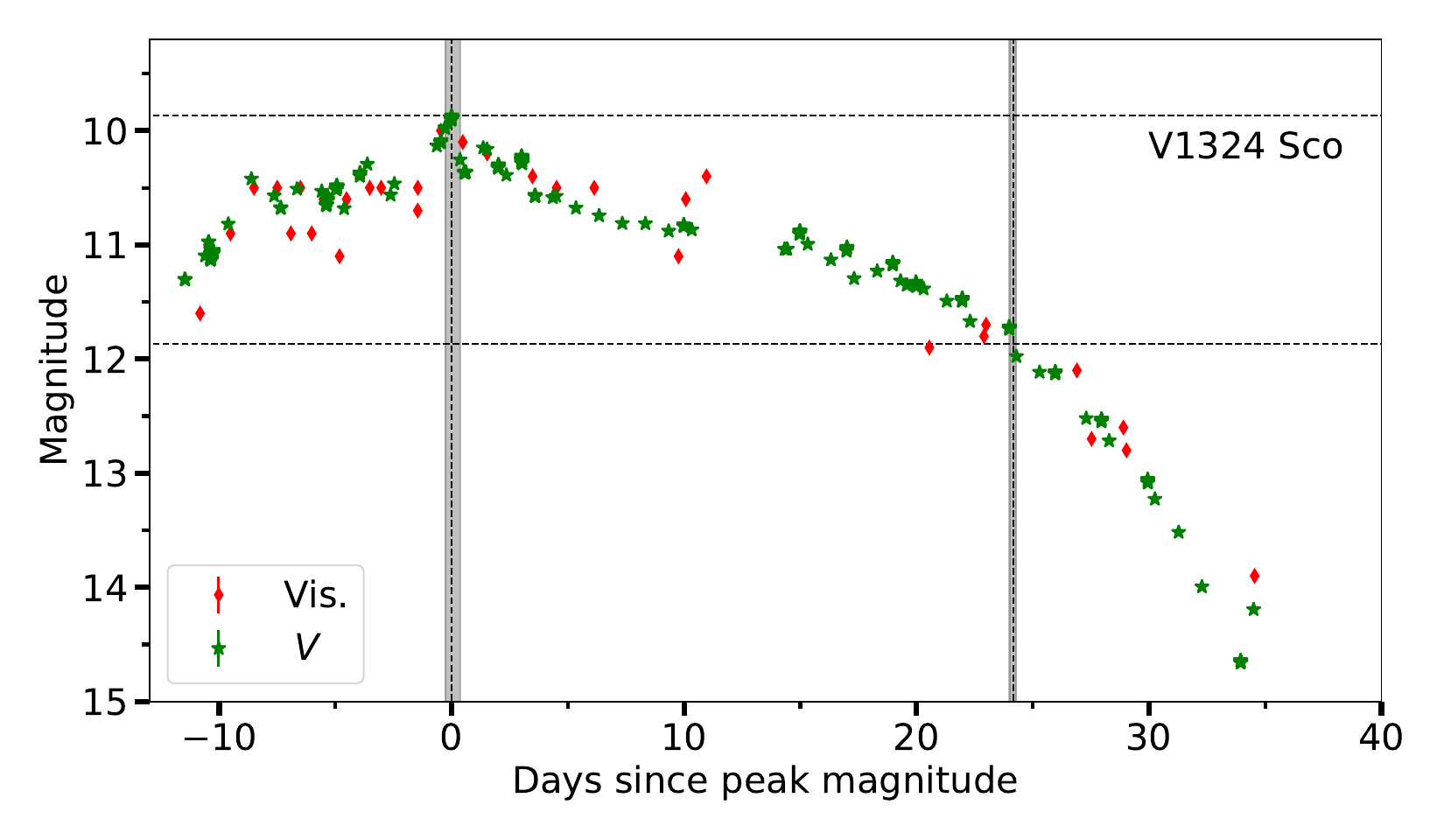}
\caption{The optical light-curve of  the $\gamma$-ray detected nova V1324~Sco, plotted against time since peak magnitude ($t_{max}=$ 2012 Jun 20.0). Horizontal and vertical dashed lines and the shaded regions have the same meaning as in Figure~\ref{Fig:V1674_Her_LC}. The rise to light curve maximum is captured by AAVSO observers.}
\label{Fig:V1324_Sco_LC}
\end{center}
\end{figure*}


\begin{figure*}
\begin{center}
  \includegraphics[width=0.9\textwidth]{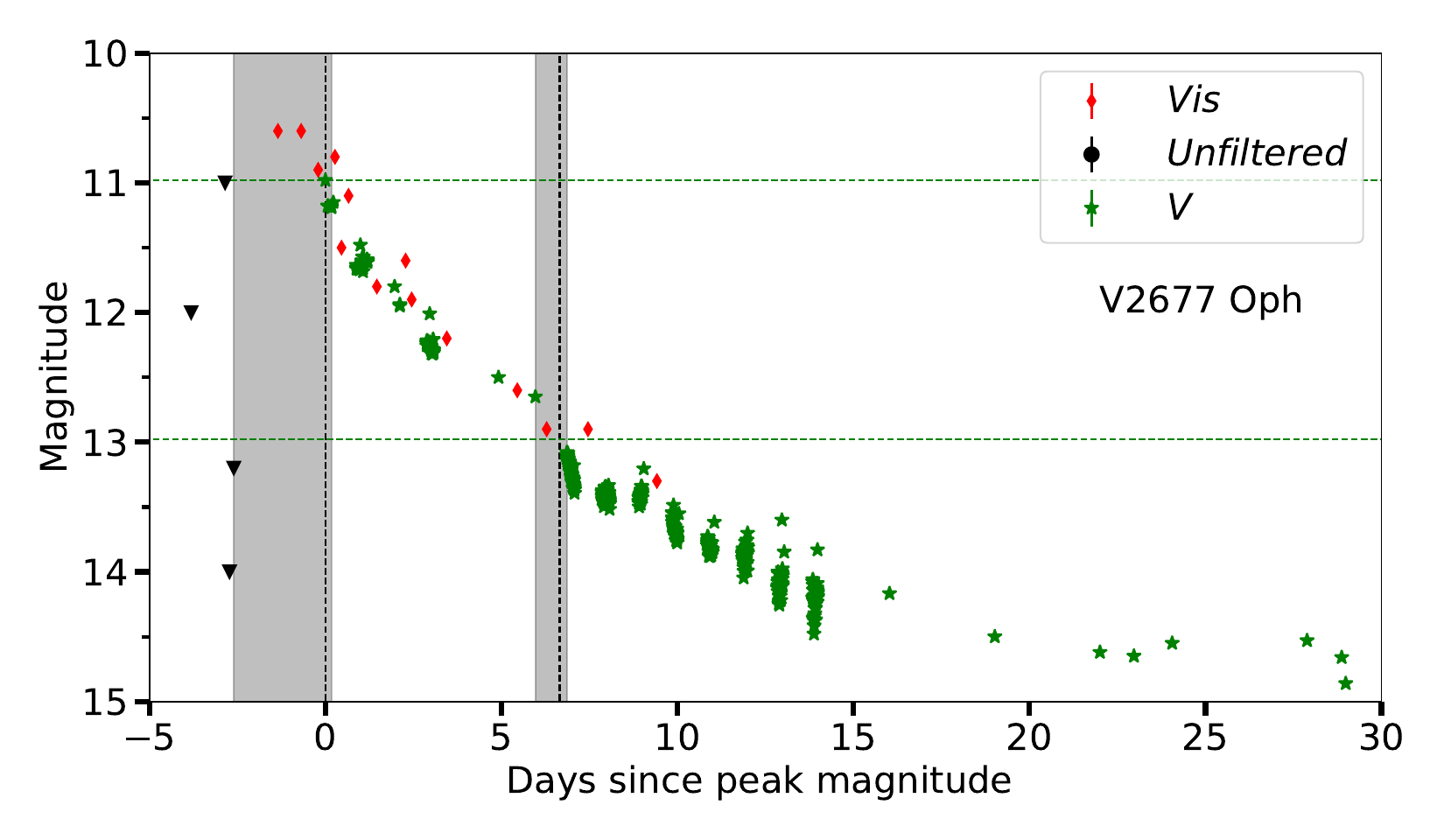}
\caption{The optical light-curve of V2677~Oph, plotted against time since peak magnitude ($t_{max}=$ 2012 May 21.3). Horizontal and vertical dashed lines and the shaded regions have the same meaning as in Figure~\ref{Fig:V1674_Her_LC}. Here, the two horizontal dashed lines are green to indicate that the optical peak is an upper limit, meaning the peak may not have been observed. A pre-eruption limit from 2012 May 18.7 constrains that light curve maximum occurred around 2012 May 21.3 \citep{2012CBET.3124....1W}.}
\label{Fig:V2677_Oph_LC}
\end{center}
\end{figure*}


\begin{figure*}
\begin{center}
  \includegraphics[width=0.9\textwidth]{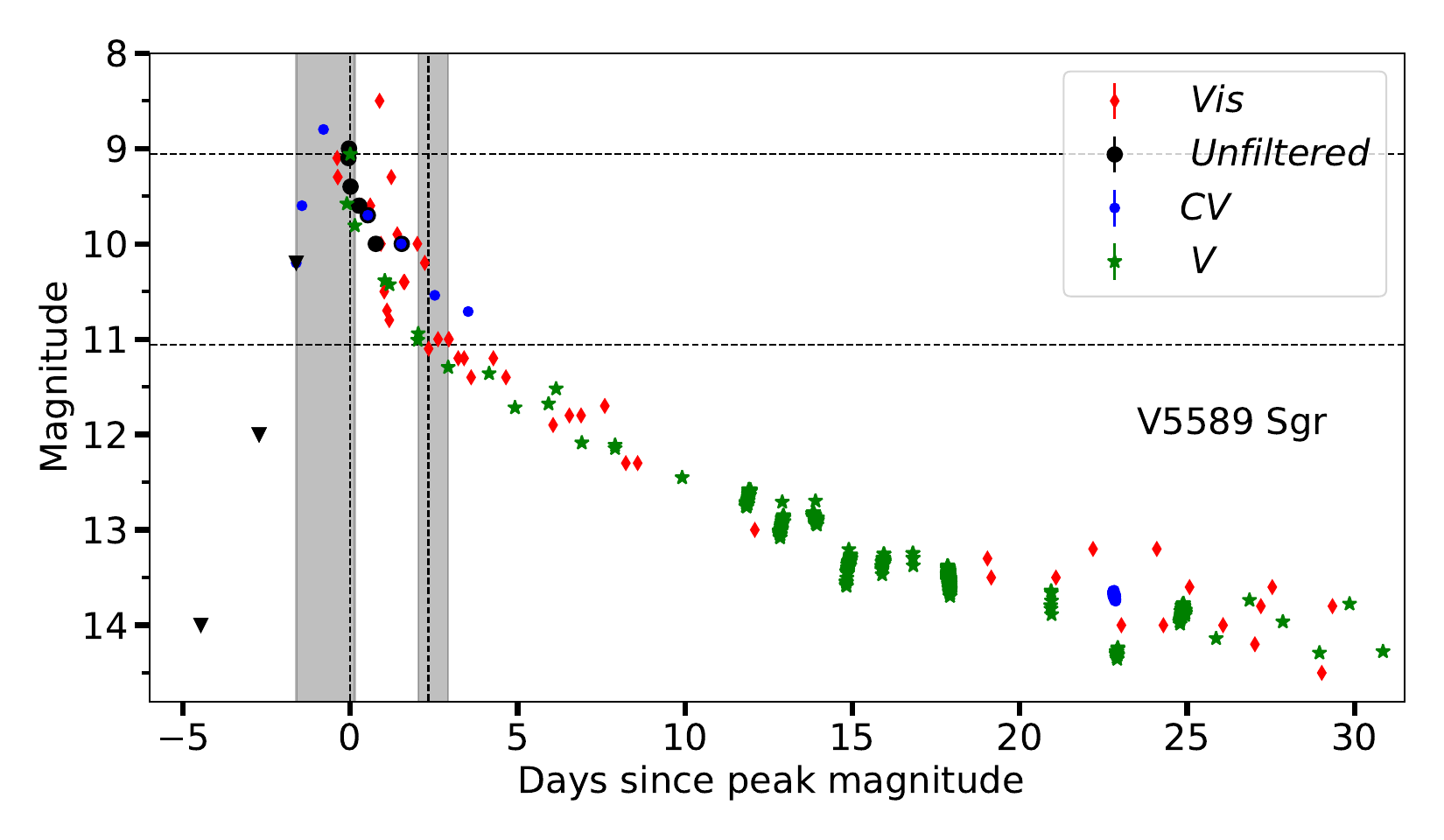}
\caption{The optical light-curve of V5589~Sgr, plotted against time since peak magnitude ($t_{max}=$ 2012 Apr 22.5). Horizontal and vertical dashed lines and the shaded regions have the same meaning as in Figure~\ref{Fig:V1674_Her_LC}. The rise to peak is constrained by pre-eruption limits from \protect\cite{2012CBET.3089....2B}. Light curve peak was captured in the $CV$-band by AAVSO observers.}
\label{Fig:V5589_Sgr_LC}
\end{center}
\end{figure*}


\begin{figure*}
\begin{center}
  \includegraphics[width=0.9\textwidth]{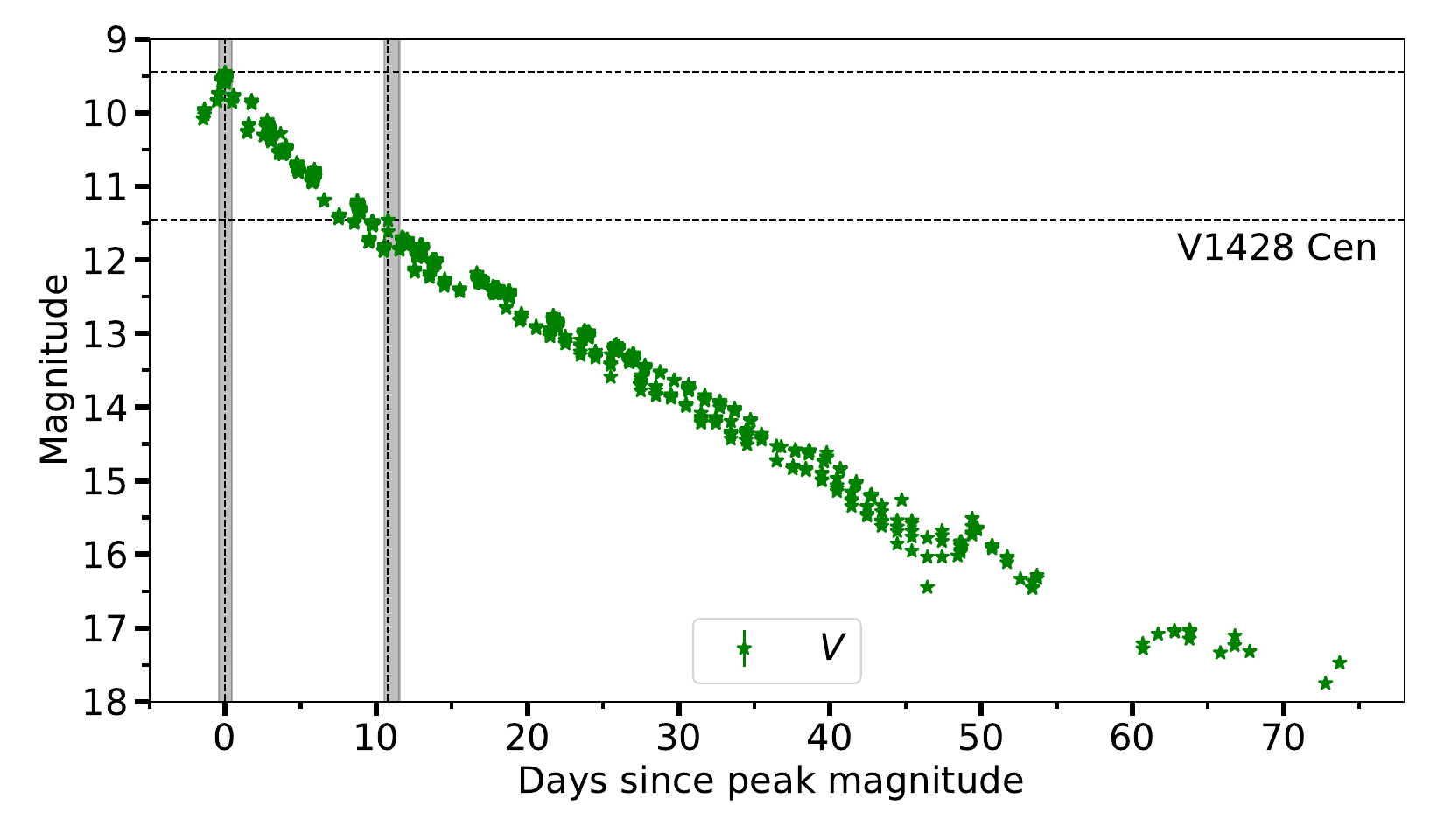}
\caption{The optical light-curve of V1428~Cen, plotted against time since peak magnitude ($t_{max}=$ 2012 Apr 07.8). Horizontal and vertical dashed lines and the shaded regions have the same meaning as in Figure~\ref{Fig:V1674_Her_LC}. Light curve peak was captured in the $V$-band by AAVSO observers.}
\label{Fig:V1428_Cen_LC}
\end{center}
\end{figure*}


\begin{figure*}
\begin{center}
  \includegraphics[width=0.9\textwidth]{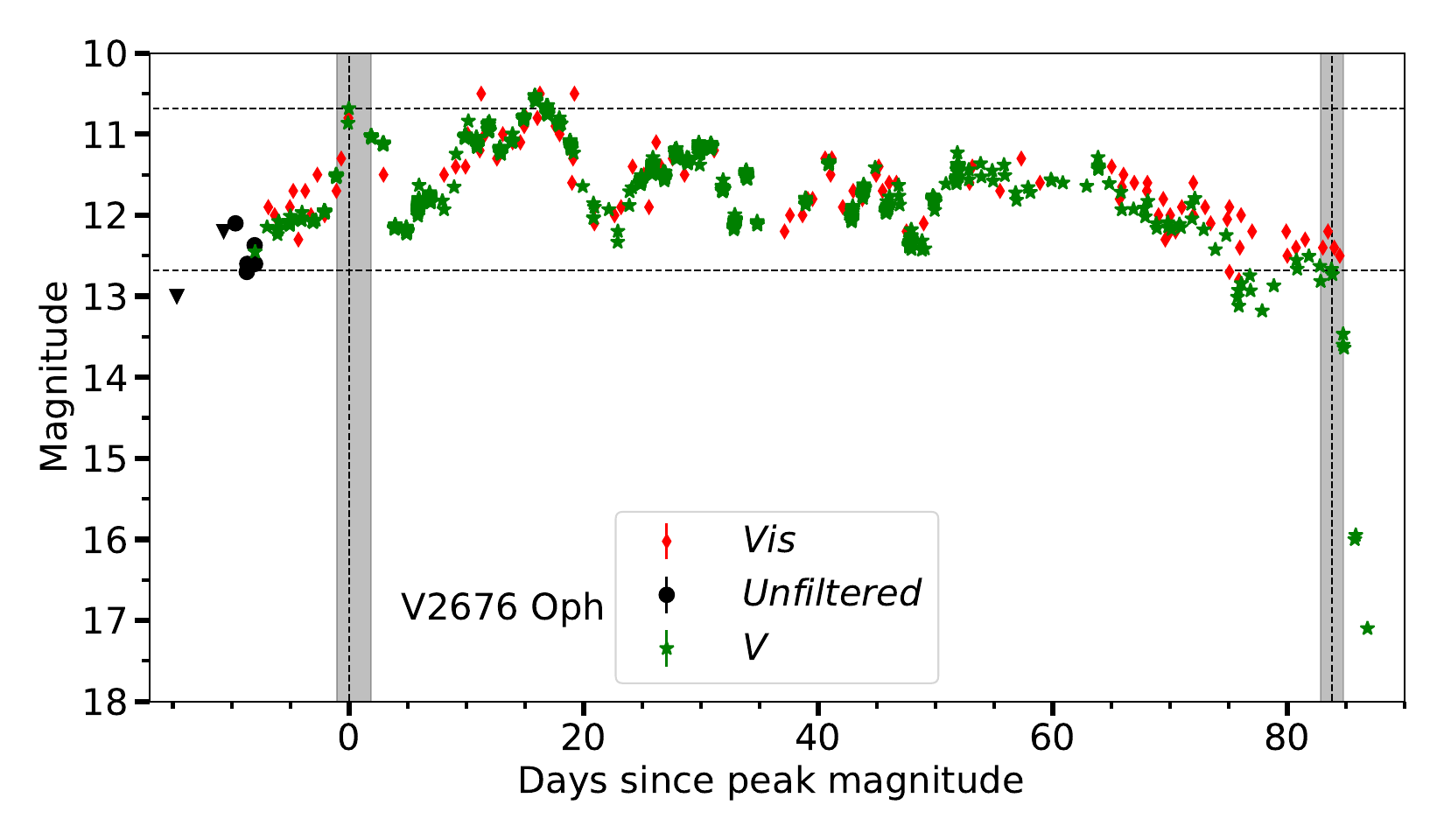}
\caption{The optical light-curve of V2676~Oph, plotted against time since peak magnitude ($t_{max}=$ 2012 Apr 04.5). Horizontal and vertical dashed lines and the shaded regions have the same meaning as in Figure~\ref{Fig:V1674_Her_LC}. Pre-eruption upper limits from \protect\cite{2012CBET.3072....1N} reveal that the AAVSO V-band light curve captured the peak well. The late-time decline of the light curve was observed by \protect\cite{Walter_etal_2012}.}
\label{Fig:V2676_Oph_LC}
\end{center}
\end{figure*}


\begin{figure*}
\begin{center}
  \includegraphics[width=0.9\textwidth]{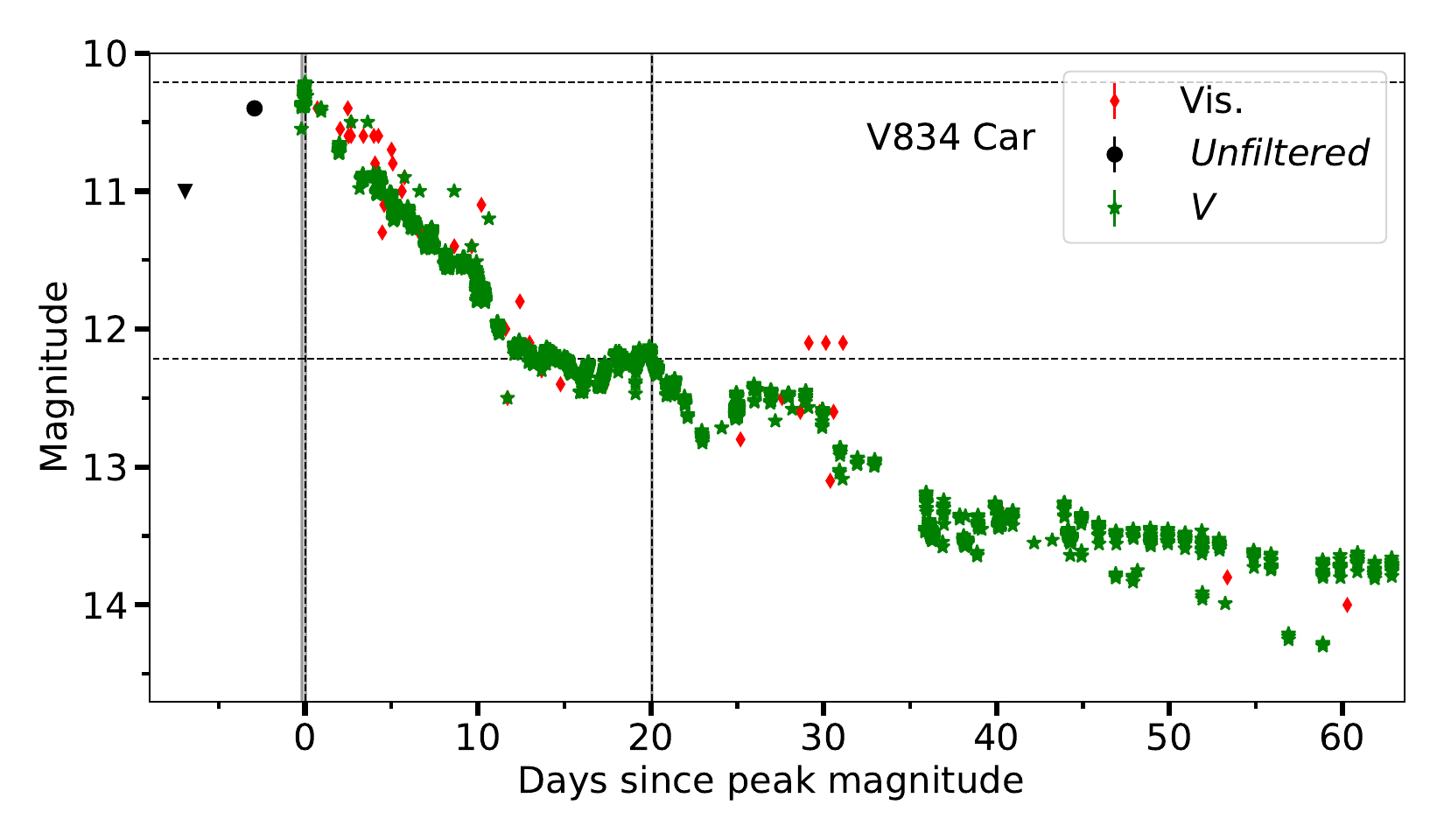}
\caption{The optical light-curve of V834~Car, plotted against time since peak magnitude ($t_{max}=$ 2012 Mar 01.4). Horizontal and vertical dashed lines and the shaded regions have the same meaning as in Figure~\ref{Fig:V1674_Her_LC}. A non-detection on 2012 Feb 23.4 constrains the time of optical maximum \citep{2012CBET.3040....1H}.}
\label{Fig:V834_Car_LC}
\end{center}
\end{figure*}


\begin{figure*}
\begin{center}
  \includegraphics[width=0.9\textwidth]{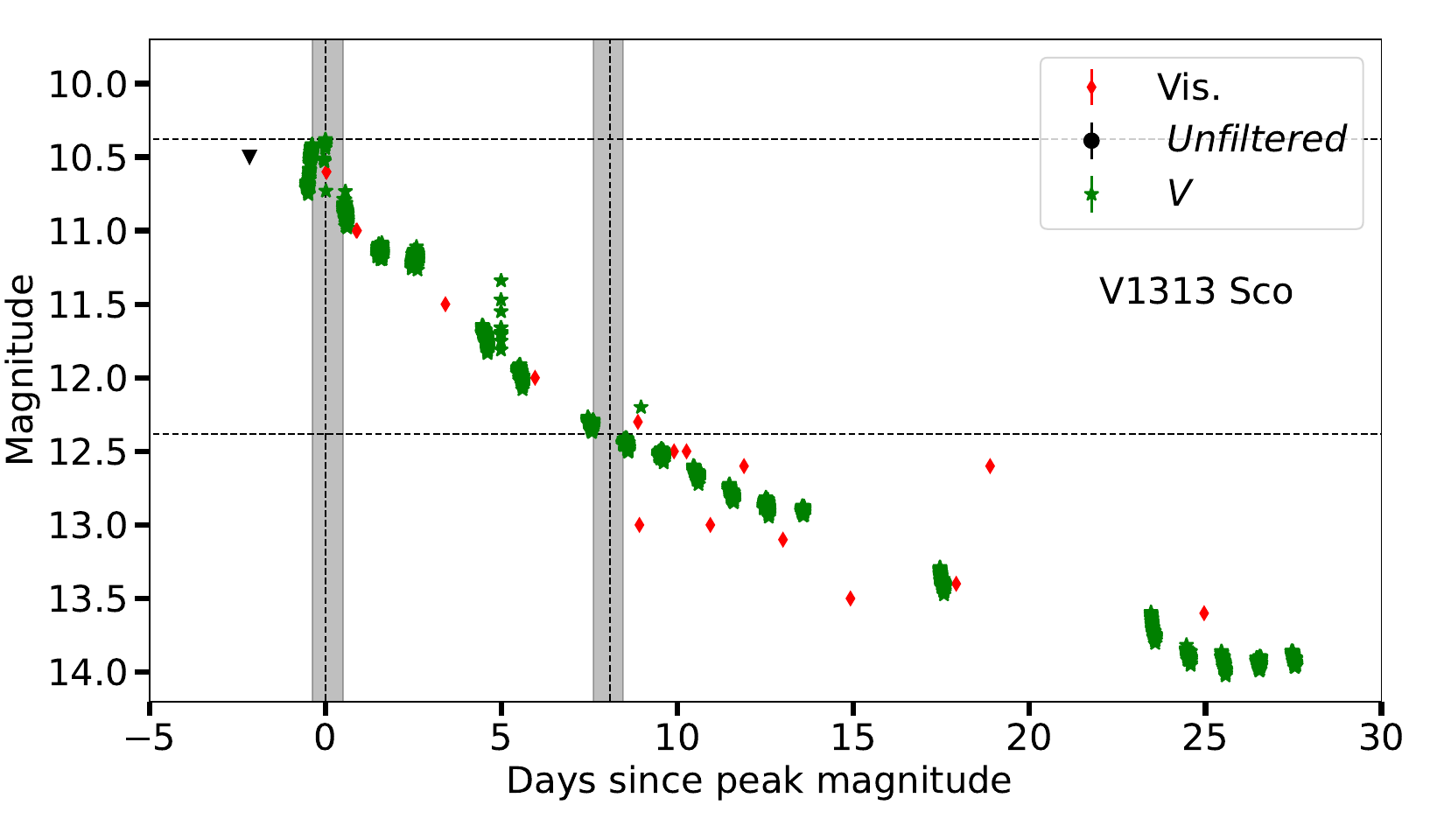}
\caption{The optical light-curve of V1313~Sco, plotted against time since peak magnitude ($t_{max}=$ 2011 Sep 07.5). Horizontal and vertical dashed lines and the shaded regions have the same meaning as in Figure~\ref{Fig:V1674_Her_LC}. A non-detection from 2011 Sep 5.4 constrains the rise to optical maximum \citep{2011CBET.2813....1Y}.}
\label{Fig:V1313_Sco_LC}
\end{center}
\end{figure*}


\begin{figure*}
\begin{center}
  \includegraphics[width=0.9\textwidth]{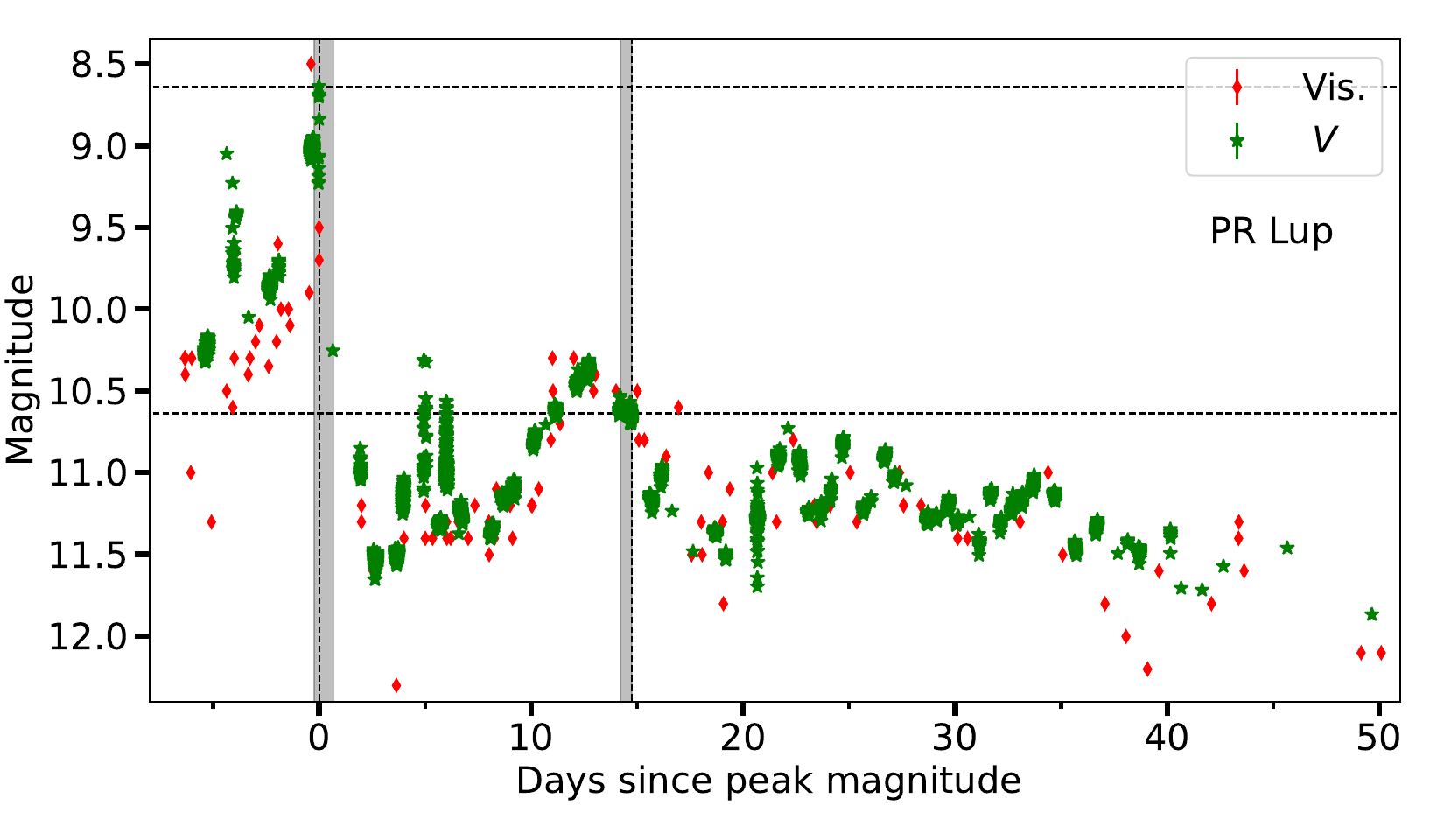}
\caption{The optical light-curve of PR~Lup, plotted against time since peak magnitude ($t_{max}=$ 2011 Aug 14.4). Horizontal and vertical dashed lines and the shaded regions have the same meaning as in Figure~\ref{Fig:V1674_Her_LC}. The rise to light curve peak was captured well by AAVSO observers and \protect\cite{2011CBET.2796....2L}.}
\label{Fig:PR_Lup_LC}
\end{center}
\end{figure*}


\begin{figure*}
\begin{center}
  \includegraphics[width=0.9\textwidth]{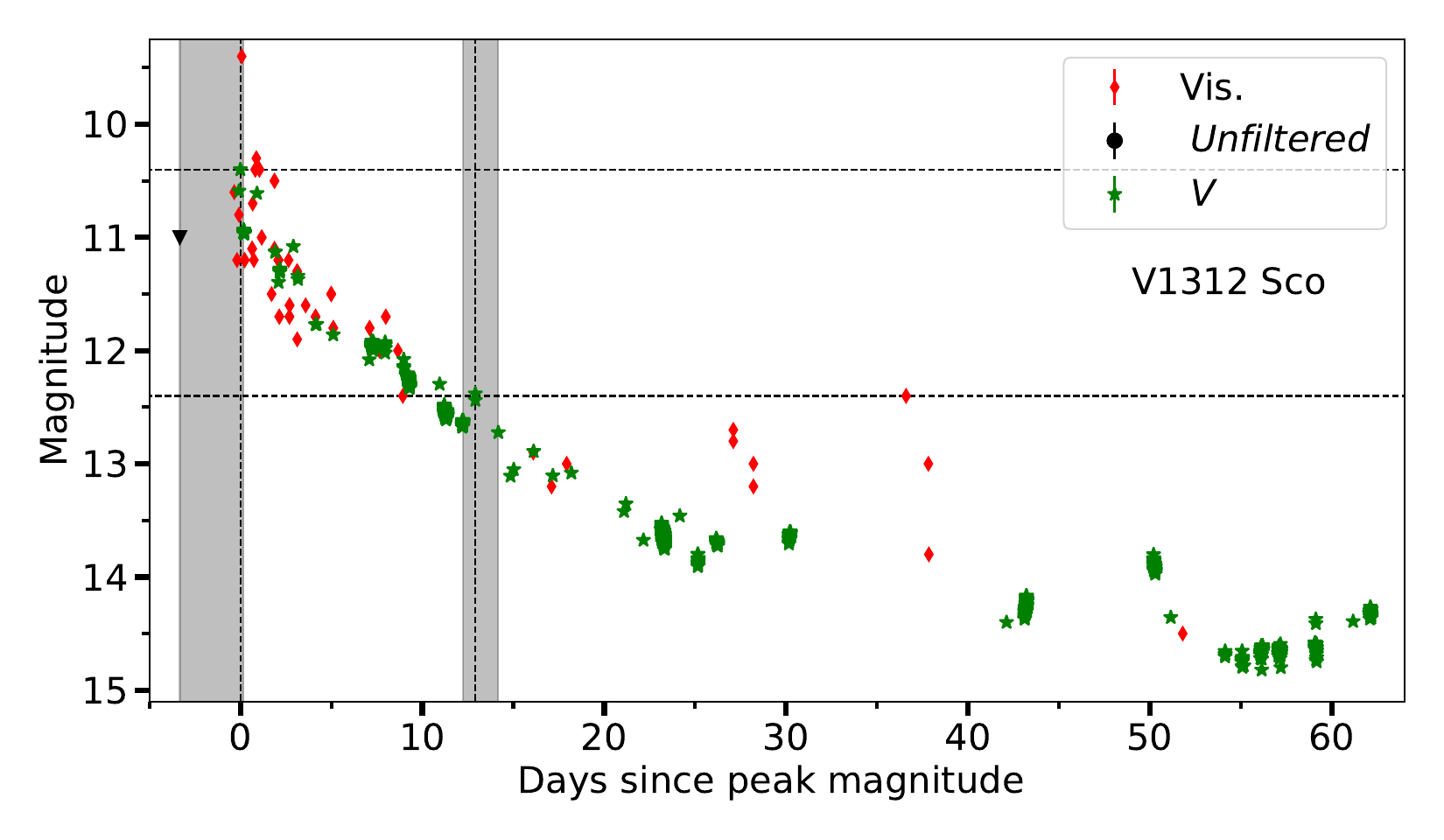}
\caption{The optical light-curve of V1312~Sco, plotted against time since peak magnitude ($t_{max}=$ 2011 Jun 02.3). Horizontal and vertical dashed lines and the shaded regions have the same meaning as in Figure~\ref{Fig:V1674_Her_LC}. A non-detection from 2011 May 30 constrains the time of optical maximum \citep{2011CBET.2735....1G}.}
\label{Fig:V1312_Sco_LC}
\end{center}
\end{figure*}


\begin{figure*}
\begin{center}
  \includegraphics[width=0.9\textwidth]{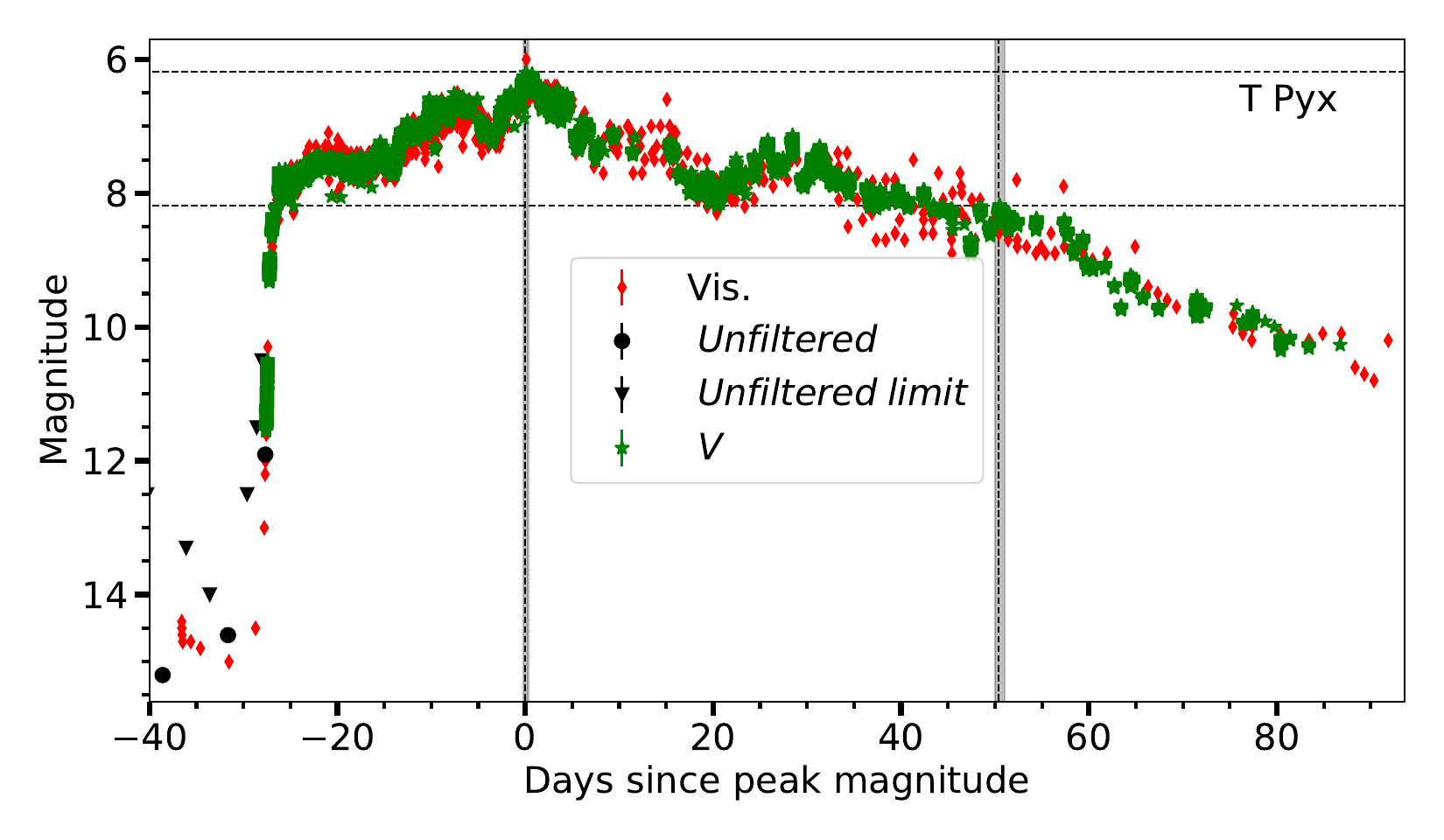}
  \caption{The optical light-curve of T~Pyx, plotted against time since peak magnitude ($t_{max}=$ 2011 May 12.0). Horizontal and vertical dashed lines and the shaded regions have the same meaning as in Figure~\ref{Fig:V1674_Her_LC}. The start of the 2011 eruption was caught promptly, and the rise to light curve maximum is well-sampled by AAVSO observers. Pre-eruptions limits from \protect\cite{2011CBET.2700....3N}, \protect\cite{2011CBET.2700....1W}, and \protect\cite{2011CBET.2700....2C} also help constrain the rise to peak.}
\label{Fig:T_Pyx_LC}
\end{center}
\end{figure*}


\begin{figure*}
\begin{center}
  \includegraphics[width=0.9\textwidth]{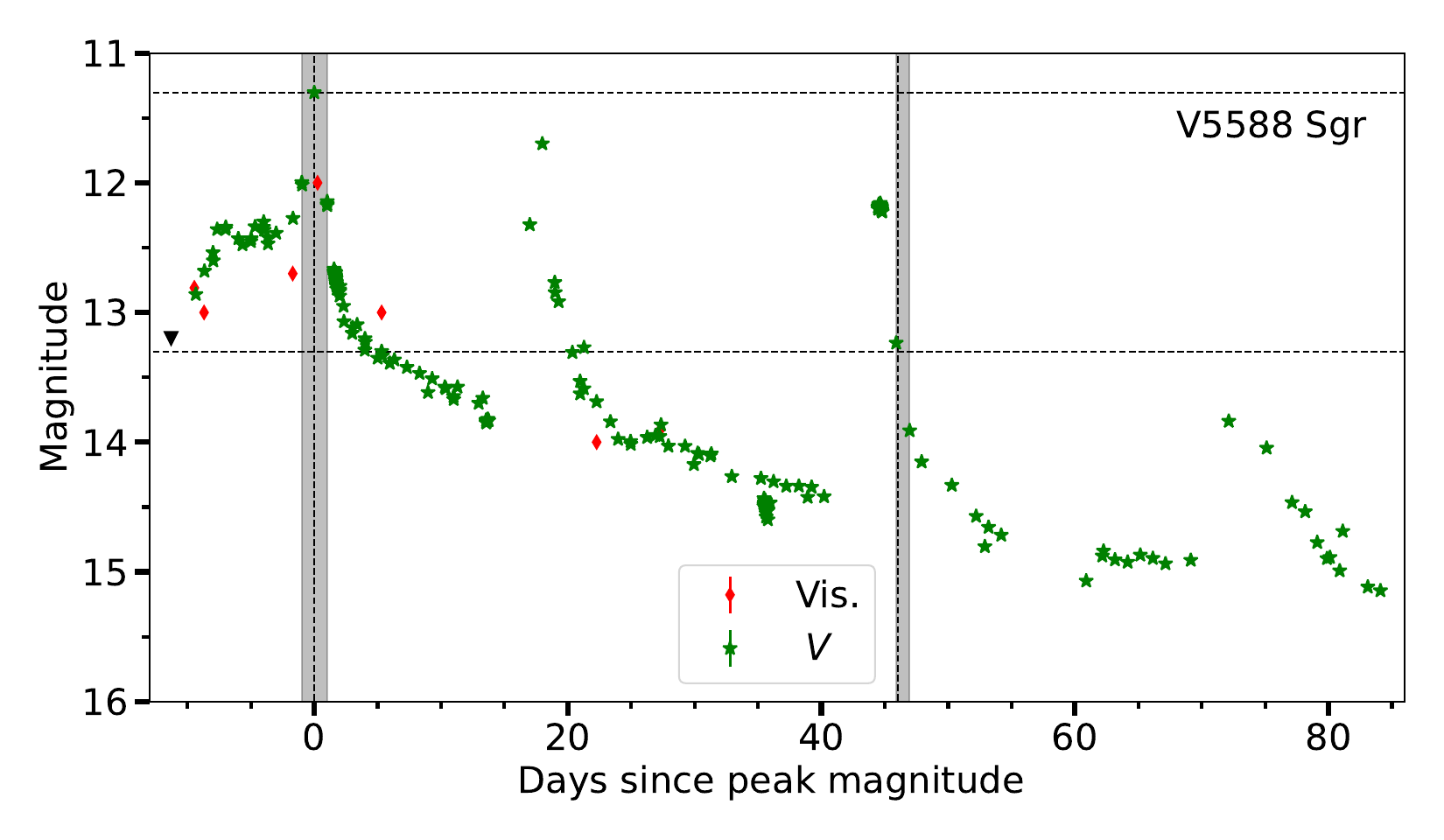}
\caption{The optical light-curve of V5588~Sgr, plotted against time since peak magnitude ($t_{max}=$ 2011 Apr 07.1). Horizontal and vertical dashed lines and the shaded regions have the same meaning as in Figure~\ref{Fig:V1674_Her_LC}. A non-detection on 2011 May 26.8 constrains the rise of the light curve well \citep{2011CBET.2679....1A}.}
\label{Fig:V5588_Sgr_LC}
\end{center}
\end{figure*}


\begin{figure*}
\begin{center}
  \includegraphics[width=0.9\textwidth]{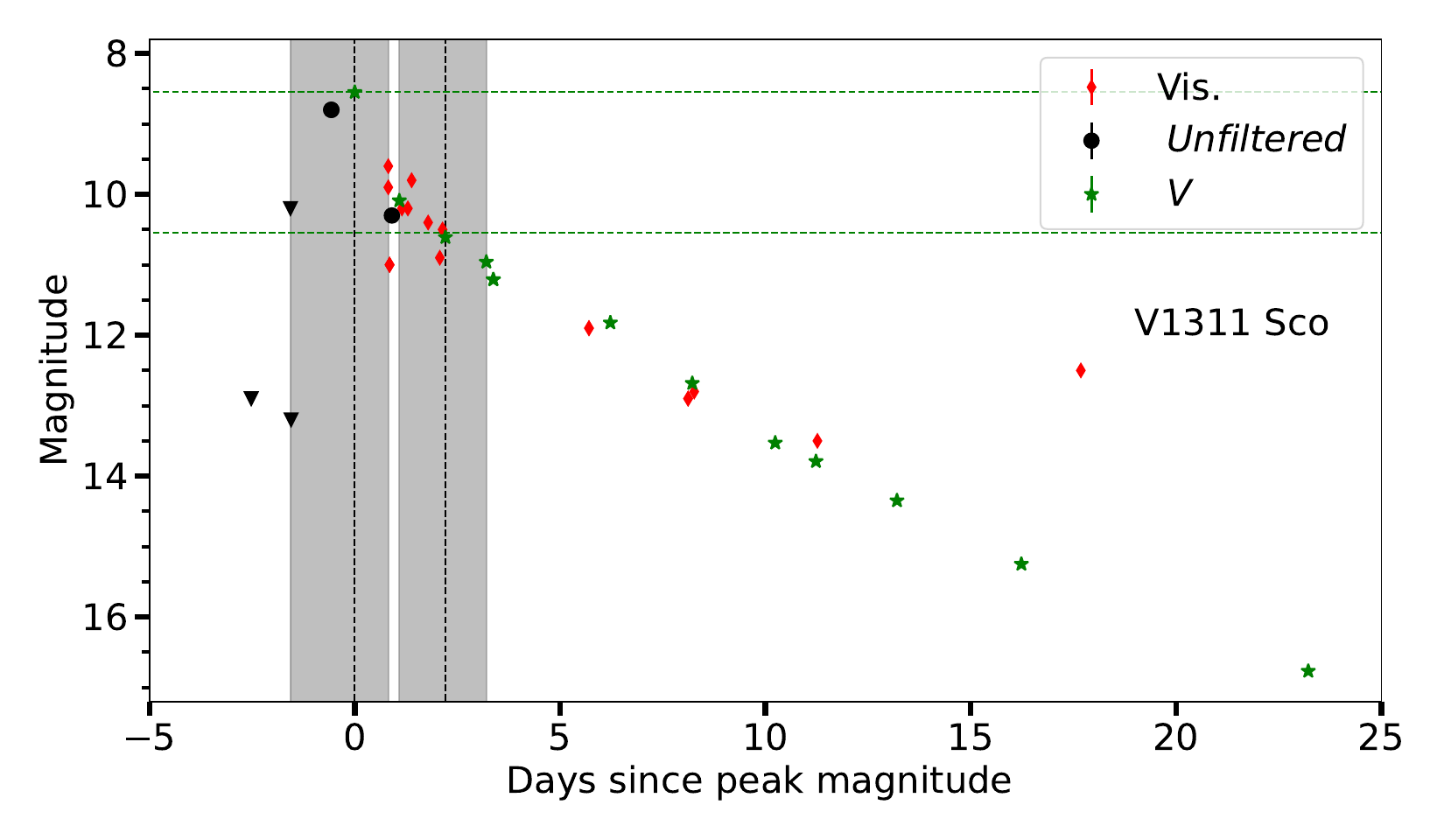}
  \caption{ The optical light-curve of V1311~Sco, plotted against time since peak magnitude ($t_{max}=$ 2010 Apr 26.3). Horizontal and vertical dashed lines and the shaded regions have the same meaning as in Figure~\ref{Fig:V1674_Her_LC}. Here, the two horizontal dashed lines are green to indicate that the optical peak is an upper limit, meaning the peak may not have been observed. Pre-eruption upper limits from \protect\cite{2010IAUC.9142....1N}, \protect\cite{2010CBET.2265....2K}, and \protect\cite{2010CBET.2262....1G} help to constrain the rise to light curve maximum.}
\label{Fig:V1311_Sco_LC}
\end{center}
\end{figure*}

\begin{figure*}
\begin{center}
  \includegraphics[width=0.9\textwidth]{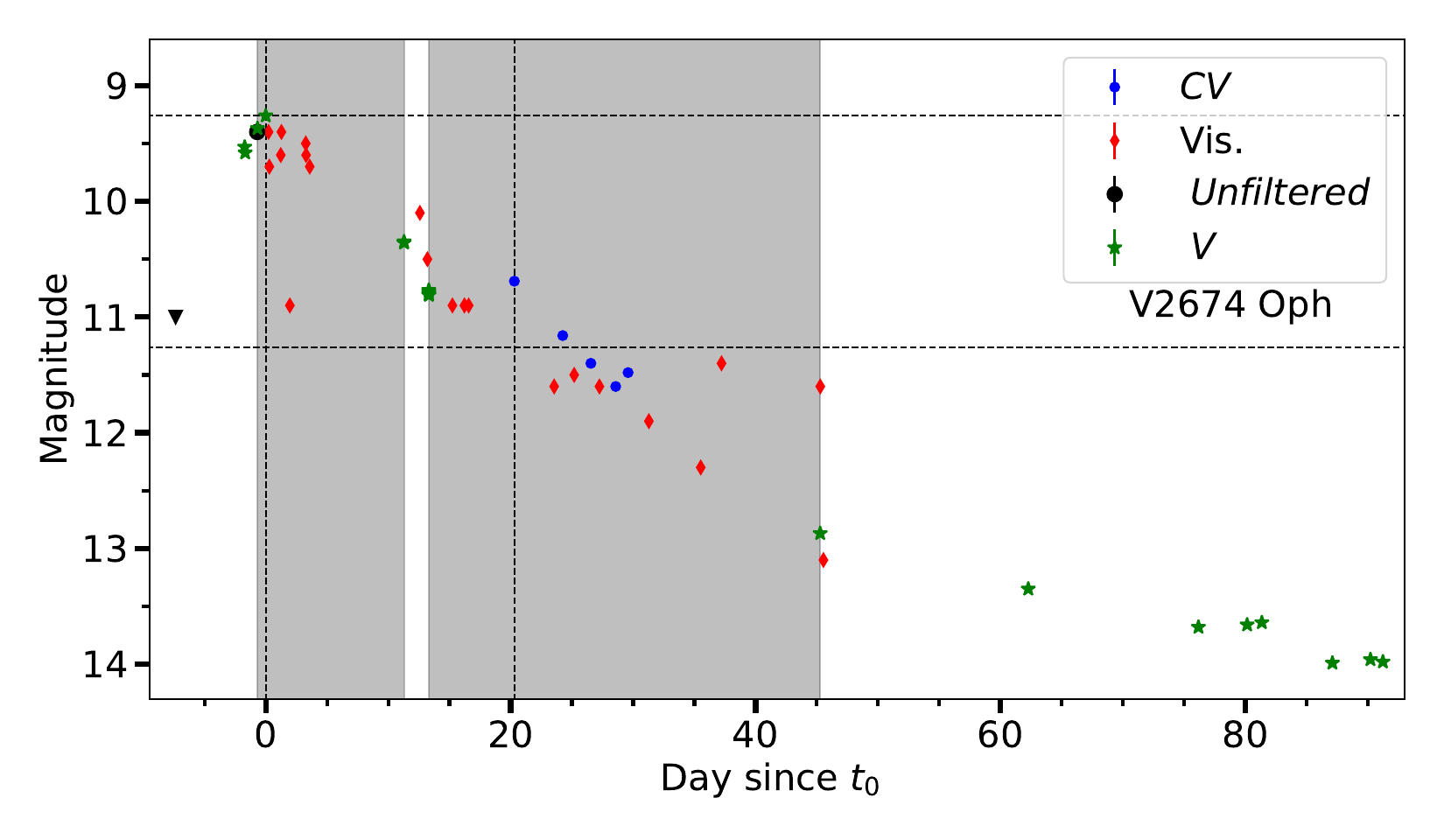}
  \caption{The optical light-curve of V2674~Oph, plotted against time since peak magnitude ($t_{max}=$ 2010 Feb 21.2). Horizontal and vertical dashed lines and the shaded regions have the same meaning as in Figure~\ref{Fig:V1674_Her_LC}. A non-detection from 2010 Feb 13.8 helps to constrain the peak magnitude \citep{2010IAUC.9119....1N}.}
\label{Fig:V2674_Oph_LC}
\end{center}
\end{figure*}

\begin{figure*}
\begin{center}
  \includegraphics[width=0.9\textwidth]{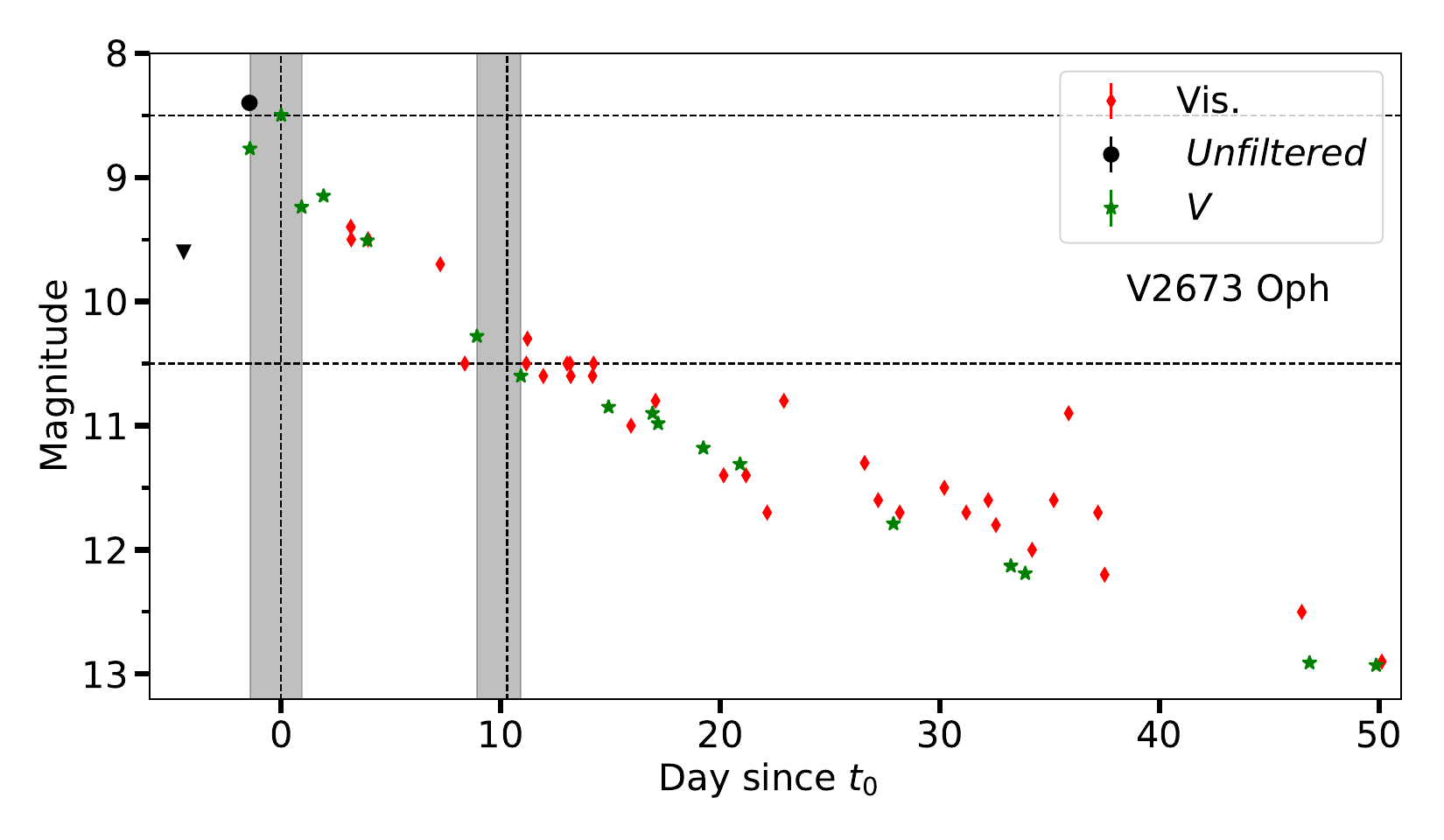}
  \caption{The optical light-curve of V2673~Oph, plotted against time since peak magnitude ($t_{max}=$ 2010 Jan 18.3). Horizontal and vertical dashed lines and the shaded regions have the same meaning as in Figure~\ref{Fig:V1674_Her_LC}. A non-detection from 2010 Jan 13.9 helps to constrain the peak magnitude \citep{2010CBET.2128....1N}.}
\label{Fig:V2673_Oph_LC}
\end{center}
\end{figure*}

\begin{figure*}
\begin{center}
  \includegraphics[width=0.9\textwidth]{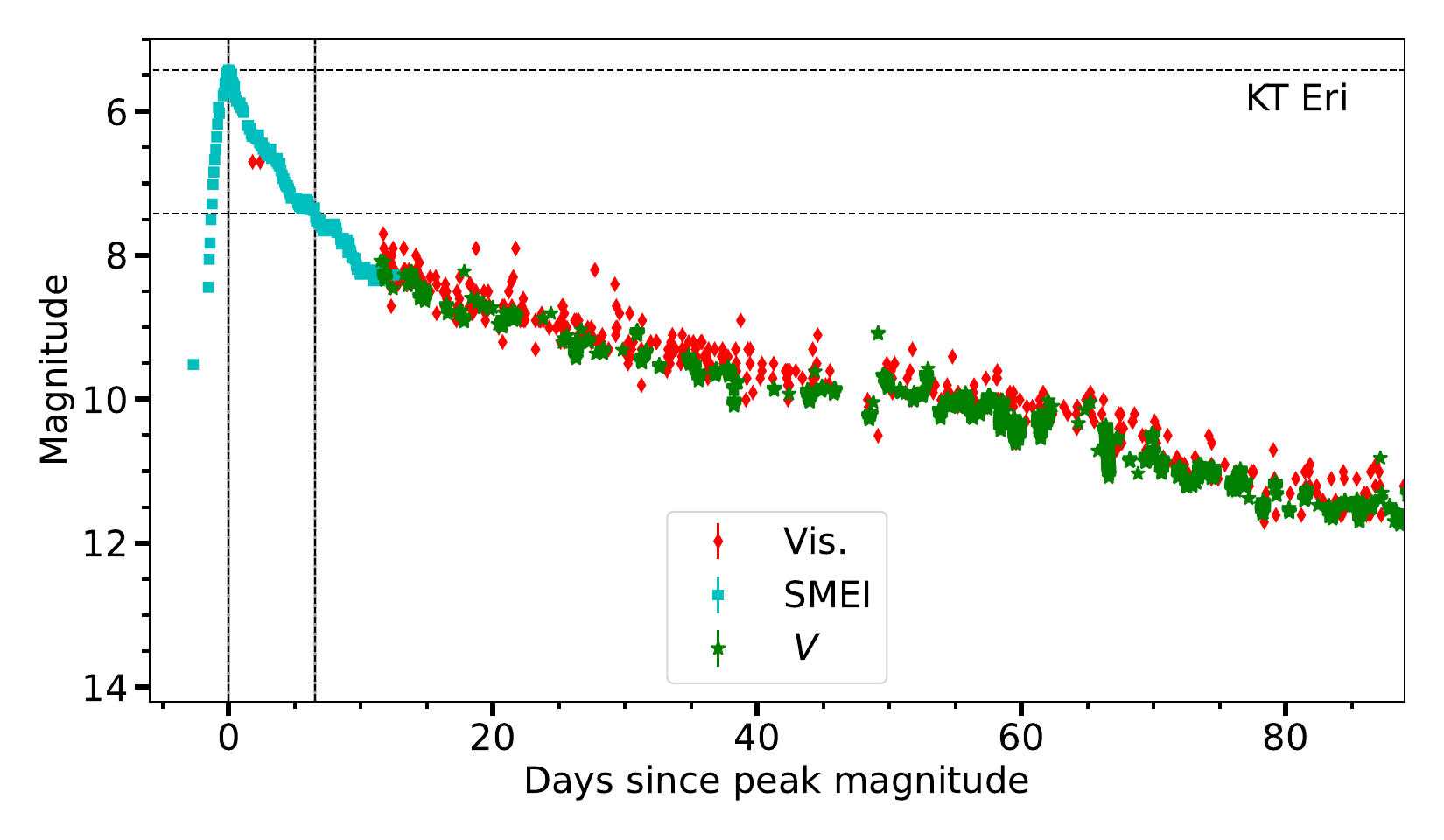}
  \caption{The optical light-curve of KT~Eri, plotted against time since peak magnitude ($t_{max}=$ 2009 Nov 14.6). Horizontal and vertical dashed lines and the shaded regions have the same meaning as in Figure~\ref{Fig:V1674_Her_LC} The light curve maximum is well-resolved by SMEI \citep{2016ApJ...820..104H}.}
\label{Fig:KT_Eri_LC}
\end{center}
\end{figure*}


\begin{figure*}
\begin{center}
  \includegraphics[width=0.9\textwidth]{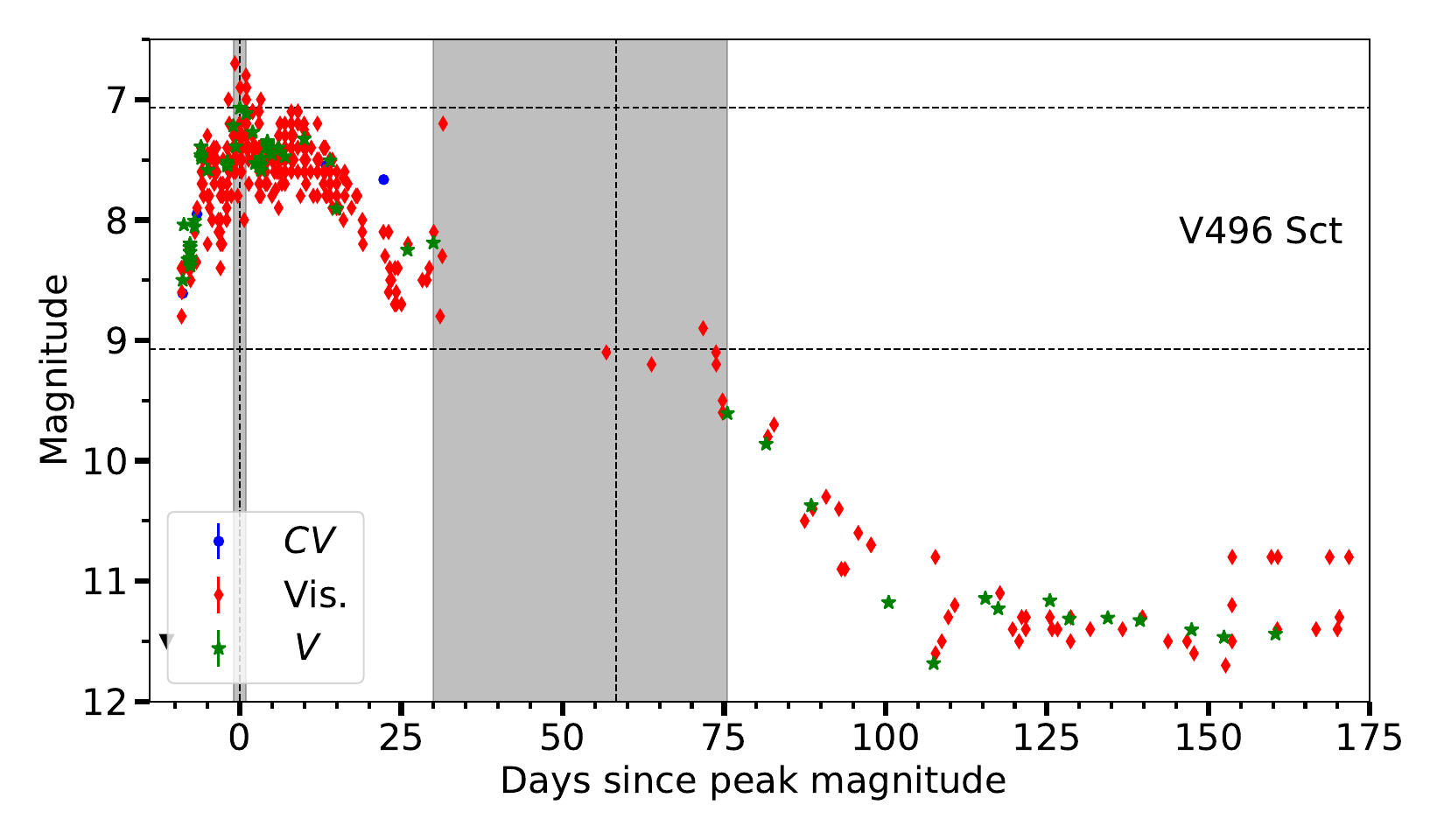}
  \caption{The optical light-curve of V496~Sct, plotted against time since peak magnitude ($t_{max}=$ 2009 Nov 18.5). Horizontal and vertical dashed lines and the shaded regions have the same meaning as in Figure~\ref{Fig:V1674_Her_LC}. A non-detection on 2009 Nov. 7.4 constrains the rise of the light curve well \citep{2009CBET.2008....1N}. Light curve maximum is well-measured by the ANS Collaboration \citep{2012MNRAS.425.2576R}.}
\label{Fig:V496_Sct_LC}
\end{center}
\end{figure*}


\begin{figure*}
\begin{center}
  \includegraphics[width=0.9\textwidth]{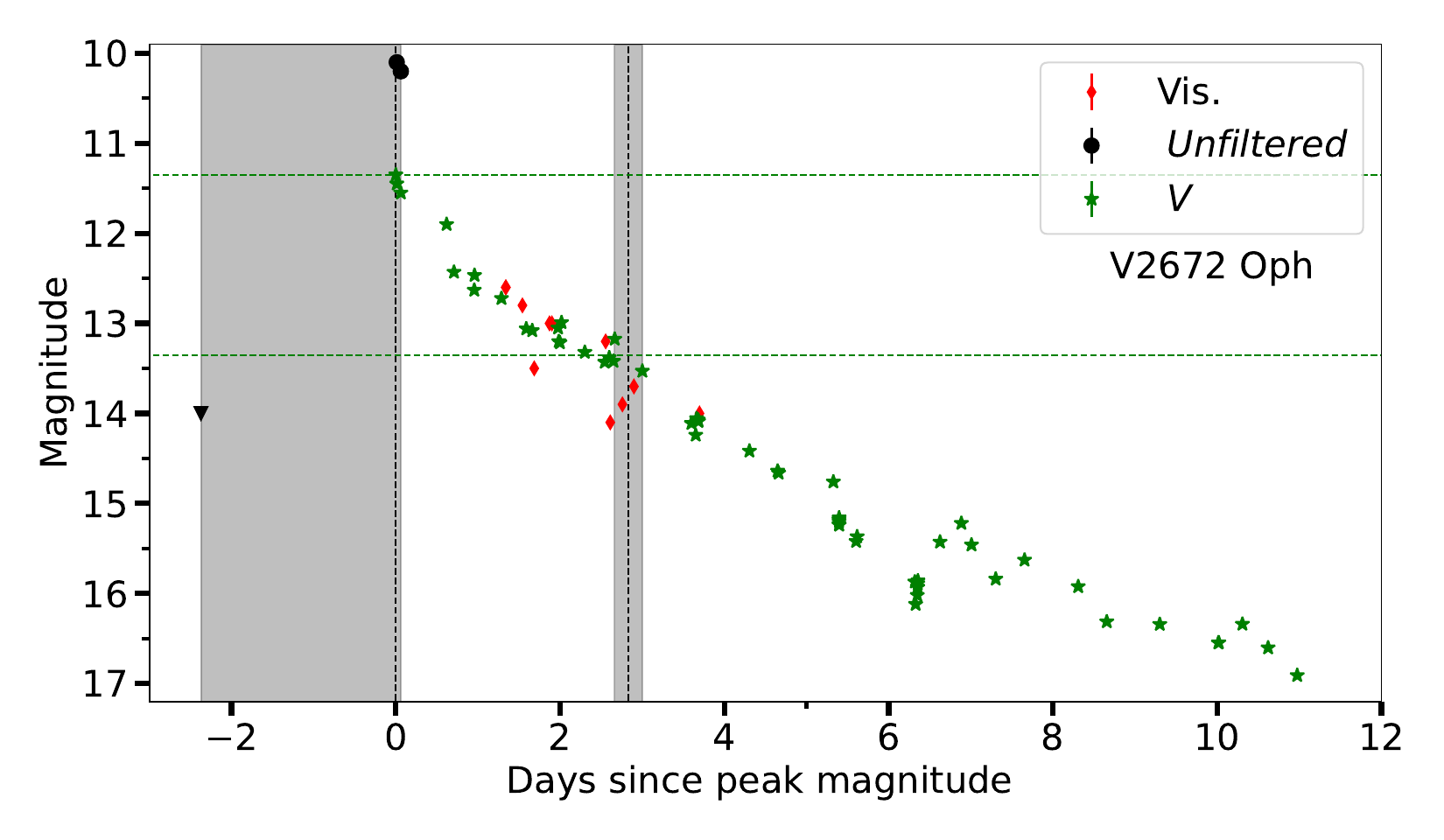}
\caption{The optical light-curve of V2672~Oph, plotted against time since peak magnitude ($t_{max}=$ 2009 Aug 16.5). Horizontal and vertical dashed lines and the shaded regions have the same meaning as in Figure~\ref{Fig:V1674_Her_LC}. Here, the two horizontal dashed lines are green to indicate that the optical peak is an upper limit, meaning the peak may not have been observed. An ASAS-SN non-detection on 2009 Aug 14.1 constrains the rise of the light curve well \citep{2009IAUC.9064....1N}.} \label{Fig:V2672_Oph_LC}
\end{center}
\end{figure*}


\begin{figure*}
\begin{center}
  \includegraphics[width=0.9\textwidth]{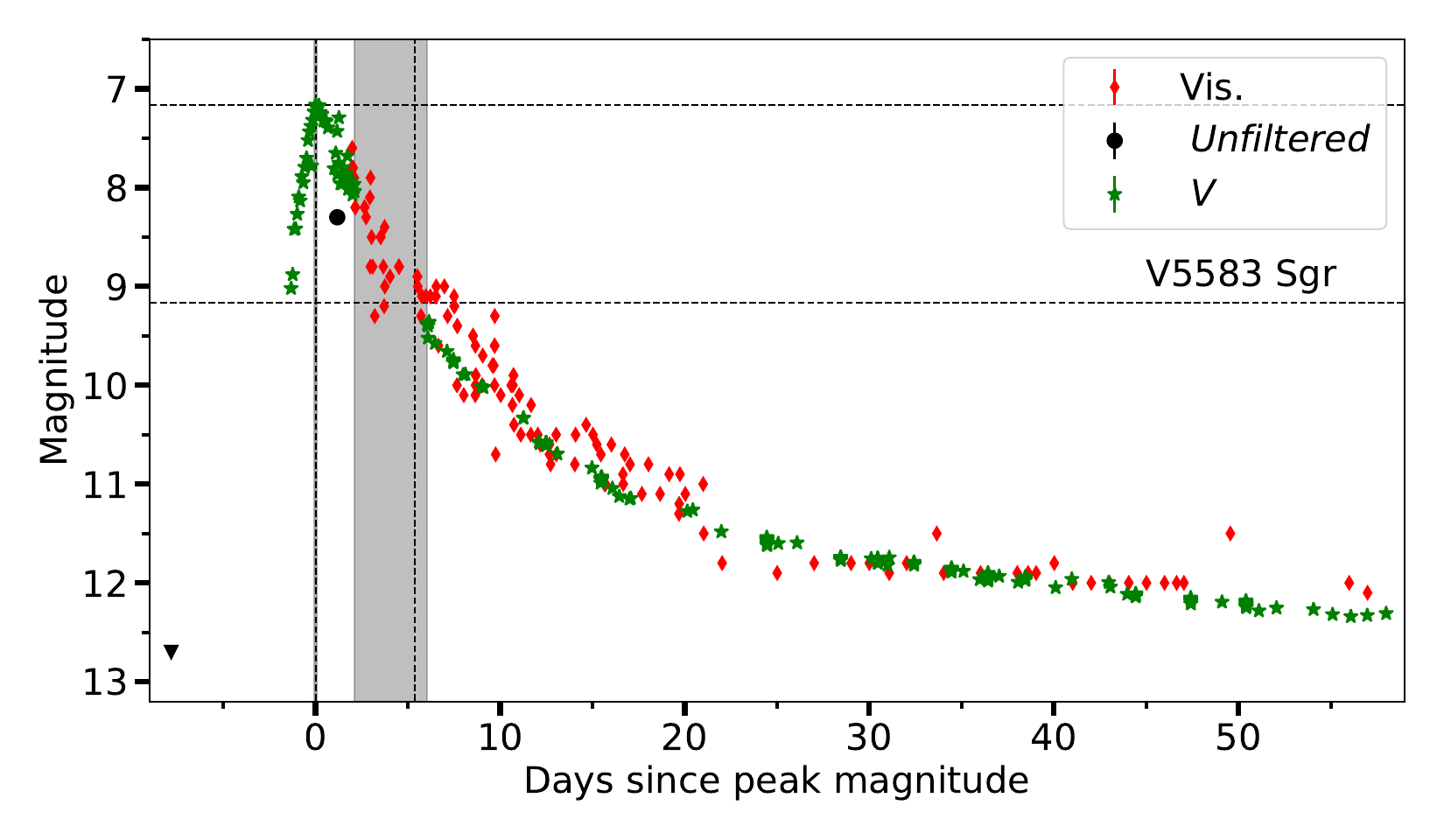}

\caption{The optical light-curve of V5583~Sgr, plotted against time since peak magnitude ($t_{max}=$ 2009 Aug 06.4). Horizontal and vertical dashed lines and the shaded regions have the same meaning as in Figure~\ref{Fig:V1674_Her_LC}. The light curve rise and maximum were well observed by the \emph{STEREO}/HI satellite \citep{2014MNRAS.438.3483H}.}\label{Fig:V5583_Sgr_LC}
\end{center}
\end{figure*}


\begin{figure*}
\begin{center}
  \includegraphics[width=0.9\textwidth]{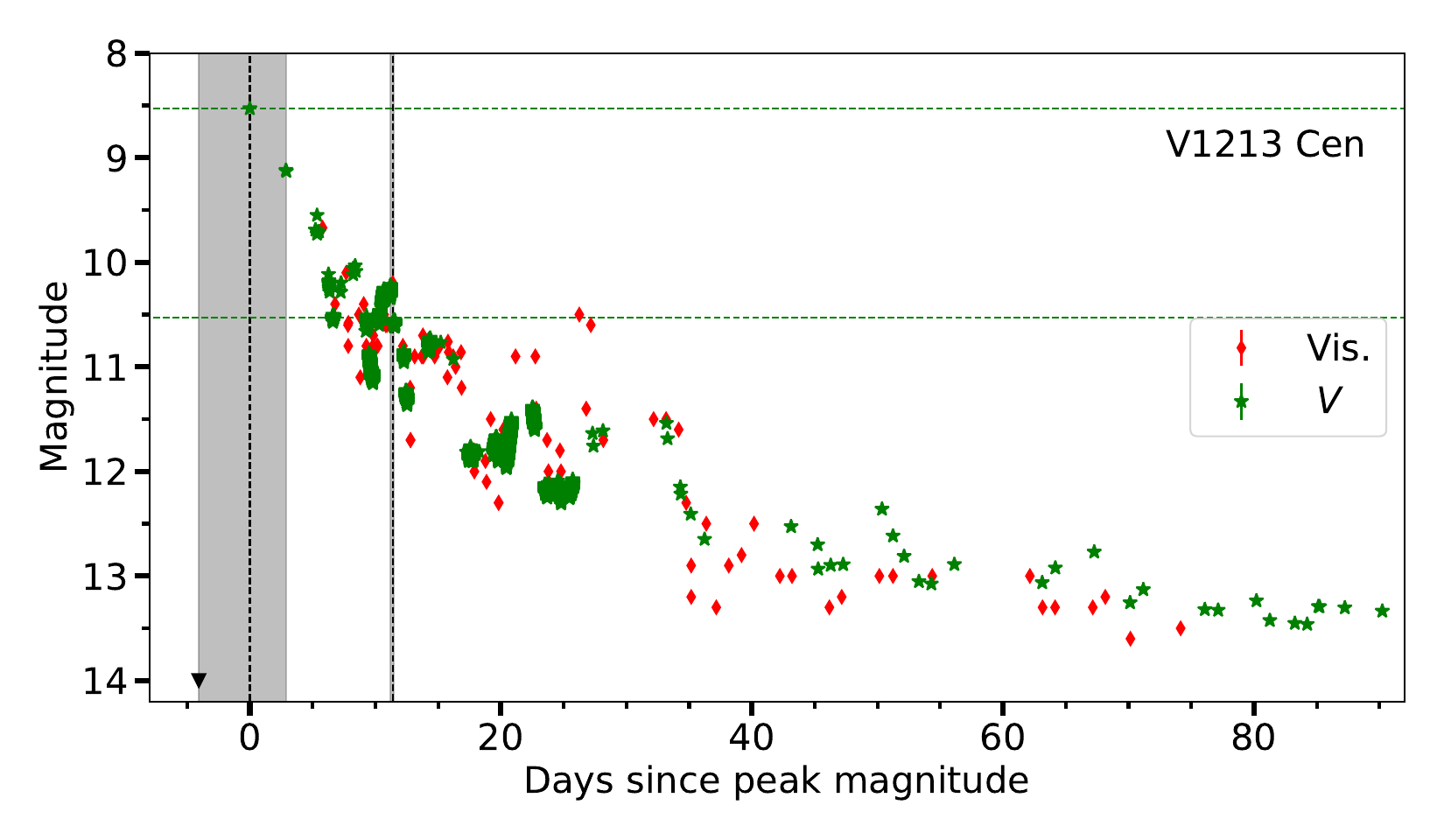}

\caption{The optical light-curve of V1213~Cen, plotted against time since peak magnitude ($t_{max}=$ 2009 May 08.2). Horizontal and vertical dashed lines and the shaded regions have the same meaning as in Figure~\ref{Fig:V1674_Her_LC}. Here, the two horizontal dashed lines are green to indicate that the optical peak is an upper limit, meaning the peak may not have been observed. A non-detection from ASAS-3 on 2009 May 4.1 constrains the time of optical maximum \citep{2009CBET.1800....1T}.} \label{Fig:V1213_Cen_LC}
\end{center}
\end{figure*}


\begin{figure*}
\begin{center}
  \includegraphics[width=0.9\textwidth]{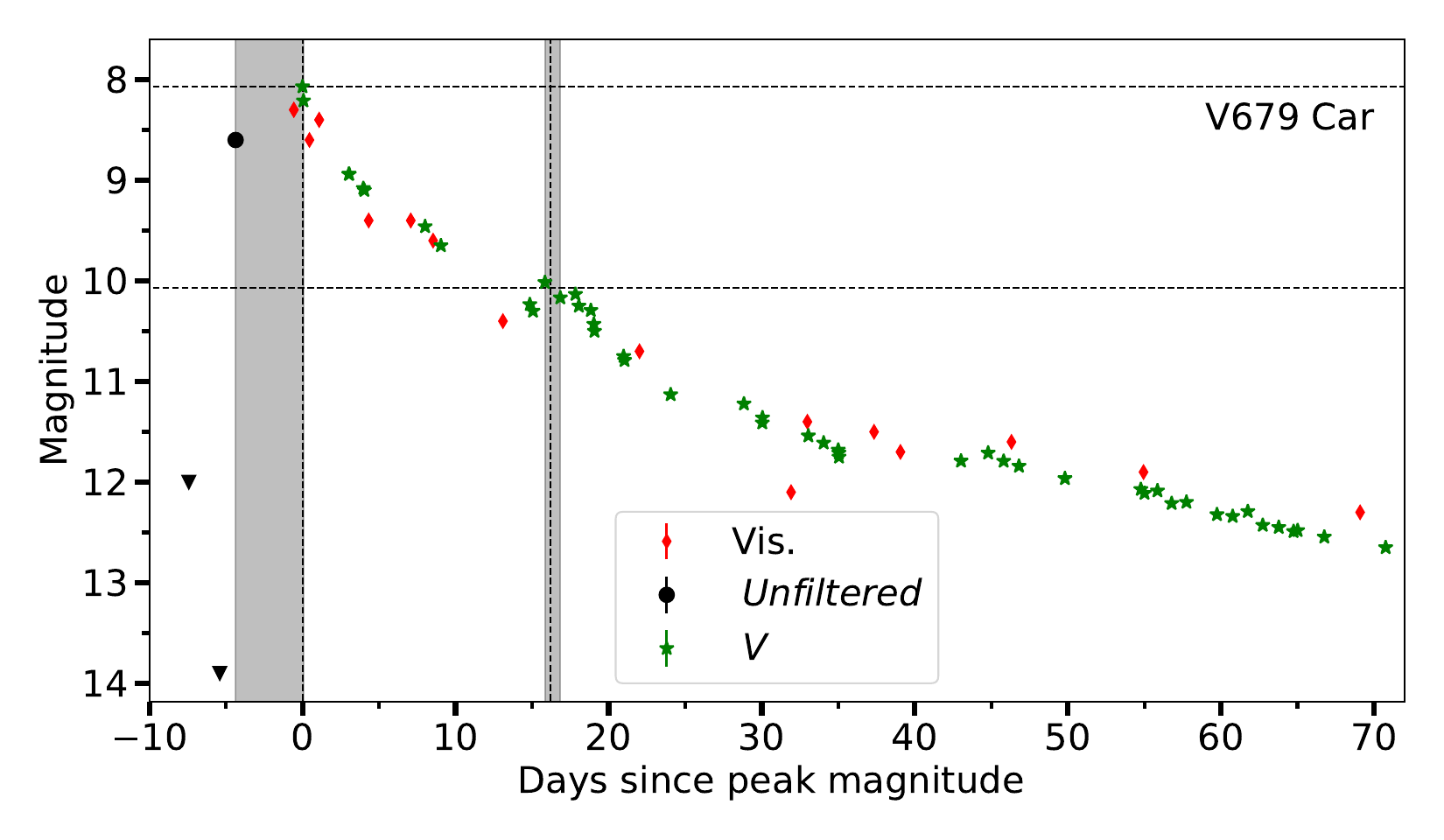}

\caption{The optical light-curve of V679~Car, plotted against time since peak magnitude ($t_{max}=$ 2008 Nov 29.7). Horizontal and vertical dashed lines and the shaded regions have the same meaning as in Figure~\ref{Fig:V1674_Her_LC}. An upper limit of $>$13.9 mag on 2008 Nov 24.3 constrains the time of optical maximum \citep{2008IAUC.8999....3A}.} \label{Fig:V679_Car_LC}
\end{center}
\end{figure*}


\begin{figure*}
\begin{center}
  \includegraphics[width=0.9\textwidth]{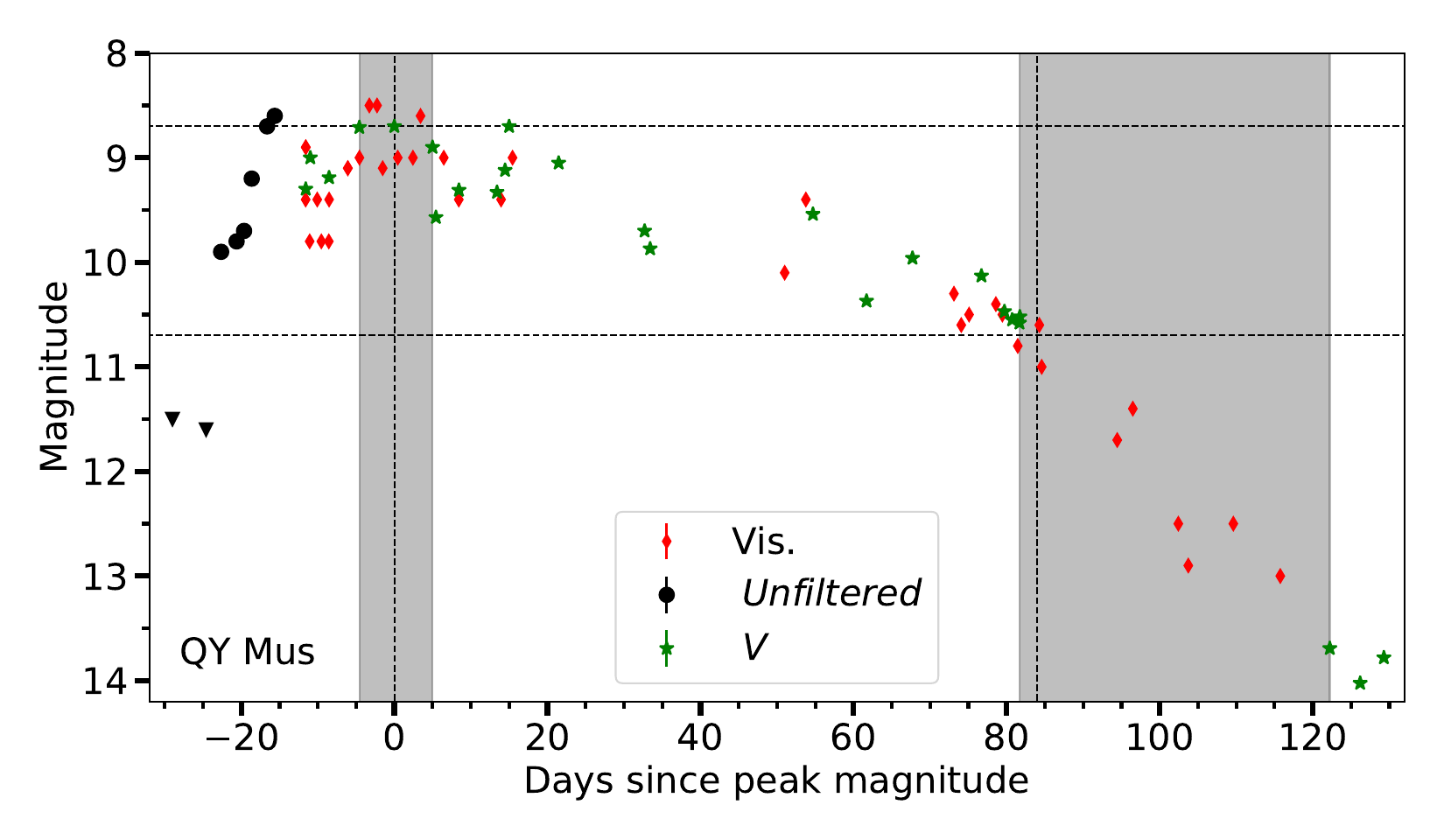}

\caption{The optical light-curve of QY~Mus, plotted against time since peak magnitude ($t_{max}=$ 2008 Oct 14.0). Horizontal and vertical dashed lines and the shaded regions have the same meaning as in Figure~\ref{Fig:V1674_Her_LC}. Photometry from V.\ Tabur presented in \citet{2008IAUC.8990....2L} constrains the rise to optical maximum.}\label{Fig:QY_Mus_LC}
\end{center}
\end{figure*}

\clearpage




\section{Spectral logs and plots}

\begin{table*}[t]
\setlength{\tabcolsep}{3pt}
\captionsetup{justification=centering, singlelinecheck=false} 
\caption{Spectroscopic log of novae detected in $\gamma$-rays.}\label{table:spec_log_1}
\centering
\def\arraystretch{1.0}
\begin{tabular}{lcccccccc}

\hline
\hline

\rule{0pt}{2ex} Name & $t_{\mathrm{max}}$ & $t_1 - t_{\text{peak}}$ & Source & $R$ 
& $t_2 - t_{\text{peak}}$ & Source & $R$& Ref.\tablenotemark{a}\\
& (UT date) & (days) & & & (days) & & &\\
\hline
V1674~Her & 2021 Jun 13.0$^{+0.1}_{-0.002}$ & 0 & ARAS & 11,000 & 2 & ARAS & 11,000 & 1\\
V1405~Cas & 2021 May 10.3$^{+0.2}_{-0.2}$ & $-16$ & ARAS & 10,000 & 6 & ARAS & 10,0000 & 1\\ 
YZ~Ret & 2020 Jul 17.1$^{+4.2}_{-8.9}$ &  8 & SALT-HRS & 14,000 & 24 & SALT-HRS & 14,000 &  2\\
V1707~Sco & 2019 Sep 15.6$^{+0.7}_{-0.3}$ & 0 & SOAR-Goodman & 5000 & 1.8 & SALT-HRS  & 14,000 & 3\\
V392~Per & 2018 Apr 29.8$^{+0.1}_{-8.3}$ & 0 & ARAS & 1000 & 1 & ARAS & 1000 & 1\\
V906~Car & 2018 Mar 28.5$^{+1.0}_{-0.9}$ & $-4$ & VLT-UVES & 59,000 & 2 & VLT-UVES & 59,000 & 3\\
V357~Mus & 2018 Jan 16.4$^{+1.1}_{-11.1}$ &  14 & SALT-HRS & 67,000 & 29 & SALT-HRS & 67,000 & 3\\
V549~Vel & 2017 Oct 17.7$^{+4.0}_{-2.3}$ &  $-13$ & SOAR-Goodman & 1000 & 41 & SOAR-Goodman & 5000 & 3\\
V5856~Sgr & 2016 Nov 08.0$^{+0.4}_{-0.2}$ &  $-8$ & Astrosurf & 1000 & 8 & Astourf & 1000 & 1\\
V5855~Sgr & 2016 Oct 24.4$^{+1.0}_{-1.0}$ &  $-3$ & Astrosurf & 1000 & 7 & Astourf & 1000 & 1\\
V5668~Sgr & 2015 Mar 21.4$^{+0.3}_{-0.3}$ & $-2$ & VLT-Pucheros & 20,000 & 6 & VLT-Pucheros & 20,000 & 3\\
V1369~Cen & 2013 Dec 14.7$^{+2.0}_{-1.9}$ & $-1$ & VLT-Feros & 48,000 & 3 & VLT-Feros & 48,000 & 3\\
V339~Del & 2013 Aug 16.7$^{+0.1}_{-0.5}$ & 0 & ARAS & 10,000 & 4 & ARAS & 10,000 & 1\\
V1324~Sco & 2012 Jun 20.0$^{+0.4}_{-0.3}$ &  $-2$ & SOAR-Goodman & 1000 & 13 & SOAR-Goodman & 1000 & 1\\
V679~Car & 2008 Nov 29.7$^{+0.1}_{-4.4}$ & -- & -- & -- & -- &  -- & -- & -- \\
\hline
\tablenotetext{a}{\justifying References for spectral data: 1= this work; 2 = \citet{Aydi_etal_2024} 3 = \citet{Aydi_etal_2020b}}
\end{tabular}
\end{table*}

\begin{table*}[t]
\setlength{\tabcolsep}{3pt}
\captionsetup{justification=centering, singlelinecheck=false} 
\caption{Spectroscopic log of novae not detected in $\gamma$-rays.}\label{table:spec_log_2}
\centering
\def\arraystretch{1.0}
\begin{tabular}{lcccccccc}

\hline
\hline

\rule{0pt}{2ex} Name & $t_{\mathrm{max}}$ & $t_1 - t_{\text{peak}}$ & Source & $R$ 
& $t_2 - t_{\text{peak}}$ & Source & $R$& Ref.\tablenotemark{a}\\
& (UT date) & (days) & & & (days) & & &\\
\hline
V1710~Sco & 2021 Apr 13.6$^{+0.5}_{-0.2}$ & -- & -- & -- & 10 & ARAS & 1000 & 1 \\
V6595~Sgr & 2021 Apr 06.2$^{+0.4}_{-0.5}$ & $-2$ & SALT-HRS & 67,000 & 1 & SOAR-Goodman & 4000 & 1 \\
V1112~Per & 2020 Nov 28.2$^{+0.1}_{-0.01}$ & $-1$ & ARAS & 10,000 & 2 & ARAS & 10,000 & 1 \\
V659~Sct & 2019 Oct 31.1$^{+0.7}_{-0.3}$ & $-2$ & ARAS & 1000 & 3 & SOAR-Goodman & 5000 & 1 \\
V3666~Oph &  2018 Aug 11.9$^{+0.9}_{-0.1}$ & $-5$ & ARAS & 1000 & 13 & ARAS & 10,000 & 1 \\
V408~Lup & 2018 Jun 03.6$^{+0.9}_{-12.3}$ & 0 & SALT-HRS & 67,000 & 24 & ARAS & 1000 & 1 \\
FM~Cir & 2018 Jan 28.3$^{+1.0}_{-0.4}$ & $-0.5$ & SALT-HRS & 67,000 & 8 & SALT-HRS & 67,000 & 3 \\
V612~Sct & 2017 Jul 30.0$^{+0.9}_{-1.0}$ & $-1$ & ARAS &10,000 & 12 & ARAS & 10,000 & 1 \\
V407~Lup & 2016 Sep 25.1$^{+0.9}_{-1.1}$ & $-1$ & VLT-Pucheros & 20,000 & 5 & VLT-Pucheros & 20,000 & 2 \\
V5669~Sgr & 2015 Sep 28.5$^{+0.9}_{-0.4}$ & $-1$ & ARAS & 11,000 & 6 & ARAS & 1400 & 1 \\
V2944~Oph & 2015 Apr 13.6$^{+1.0}_{-1.0}$ & $-5$ & ARAS & 1000 & 6 & ARAS & 1000 & 1 \\
V5667~Sgr & 2015 Feb 15.7$^{+2.0}_{-1.3}$ & -- & -- & -- & 3 & ARAS & 1000 & 1 \\
V2659~Cyg & 2014 Apr 10.5$^{+0.5}_{-0.7}$ & 0 & ARAS & 2000 & 7 & ARAS & 2000 & 1 \\
V962~Cep & 2014 Mar 13.9$^{+0.5}_{-0.8}$ & $-3$ & ARAS & 1000 & 8 & ARAS & 1000 & 1 \\
V1533~Sco & 2013 Jun 04.5$^{+0.7}_{-3.8}$ & -- & -- & -- & 54 & ARAS & 1500 & 1 \\
V809~Cep & 2013 Feb 04.7$^{+0.1}_{-0.6}$ & $-1$ & ARAS & 1000 & 2 & ARAS & 2000 & 1 \\
V5593~Sgr & 2012 Jul 22.4$^{+7.1}_{-0.2}$ $0$ & ARAS & 1000 & 21 & ARAS & 1000 & 1 \\
V5591~Sgr & 2012 Jun 27.3$^{+0.2}_{-0.7}$ & -- & -- & -- & 2 & ARAS & 1000 & 1 \\
V2677~Oph & 2012 May 21.3$^{+0.2}_{-2.6}$ & -- & -- & -- & 3 & ARAS & 1000 & 1 \\
V5589~Sgr & 2012 Apr 22.5$^{+0.1}_{-1.6}$ & -- & -- & -- & 1 & ARAS & 1000 & 1 \\
V1428~Cen & 2012 Apr 07.8$^{+0.5}_{-0.4}$ & -- & -- & -- & 9 & SMARTS & 17,000 & 1 \\
V2676~Oph & 2012 Apr 04.5$^{+1.9}_{-1.0}$ & -- & -- & -- & -- & -- & --\\
V834~Car & 2012 Mar 01.4$^{+0.1}_{-0.2}$ & -- & -- & -- & 6 & SMARTS & 17,000 & 1 \\
V1313~Sco & 2011 Sep 07.5$^{+0.5}_{-0.4}$ & -- & -- & -- & 1 & SMARTS & 1000 & 1 \\
PR~Lup & 2011 Aug 14.4$^{+0.6}_{-0.2}$ & -- & -- & -- & -- & -- & --\\
V1312~Sco & 2011 Jun 02.3$^{+0.2}_{-3.3}$ & -- & -- & -- & -- & -- & --\\
T~Pyx & 2011 May 12.0$^{+0.3}_{-0.2}$ & $-10$ & astrosurf & 1000 & 44 & astrosurf & 1000 & 1 \\
V5588~Sgr & 2011 Apr 07.1$^{+1.0}_{-1.0}$ & -- & -- & -- & 47 & astrosurf & 1000 & 1 \\
V1311~Sco & 2010 Apr 26.3$^{+0.8}_{-1.6}$ & -- & -- & -- & 8 & SMARTS & 1000 & 1 \\
V2674~Oph & 2010 Feb 21.2$^{+11.3}_{-0.7}$ & -- & -- & -- & -- & -- & --\\
V2673~Oph & 2010 Jan 18.3$^{+0.9}_{-1.4}$ & -- & -- & -- & -- & -- & --\\
KT~Eri & 2009 Nov 14.6$^{+0.1}_{-0.1}$ & -- & -- & -- & 16 & SMARTS & 1000 & 1 \\
V496~Sct & 2009 Nov 18.5$^{+1.0}_{-1.0}$ & $-9$ & SMARTS & 17,000 & -- & -- & --\\
V2672~Oph & 2009 Aug 16.5$^{+0.1}_{-2.4}$ & -- & -- & -- & 2 & SMARTS & 17,000 & 1 \\
V5583~Sgr & 2009 Aug 06.4$^{+0.1}_{-0.1}$ & -- & -- & -- & 2 & astourf & 1000 & 1 \\
V1213~Cen & 2009 May 08.2$^{+2.9}_{-4.1}$ & -- & -- & -- & -- & -- & --\\
QY~Mus & 2008 Oct 14.0$^{+5.0}_{-4.6}$ & -- & -- & -- & 73 & SMARTS & 17,000 & 1 \\
\hline
\tablenotetext{a}{\justifying References for spectral data: 1= this work; 2 = \citet{Aydi_etal_2024} 3 = \citet{Aydi_etal_2020b}}
\end{tabular}
\end{table*}

\begin{figure*}
\begin{center}
  \includegraphics[width=\textwidth]{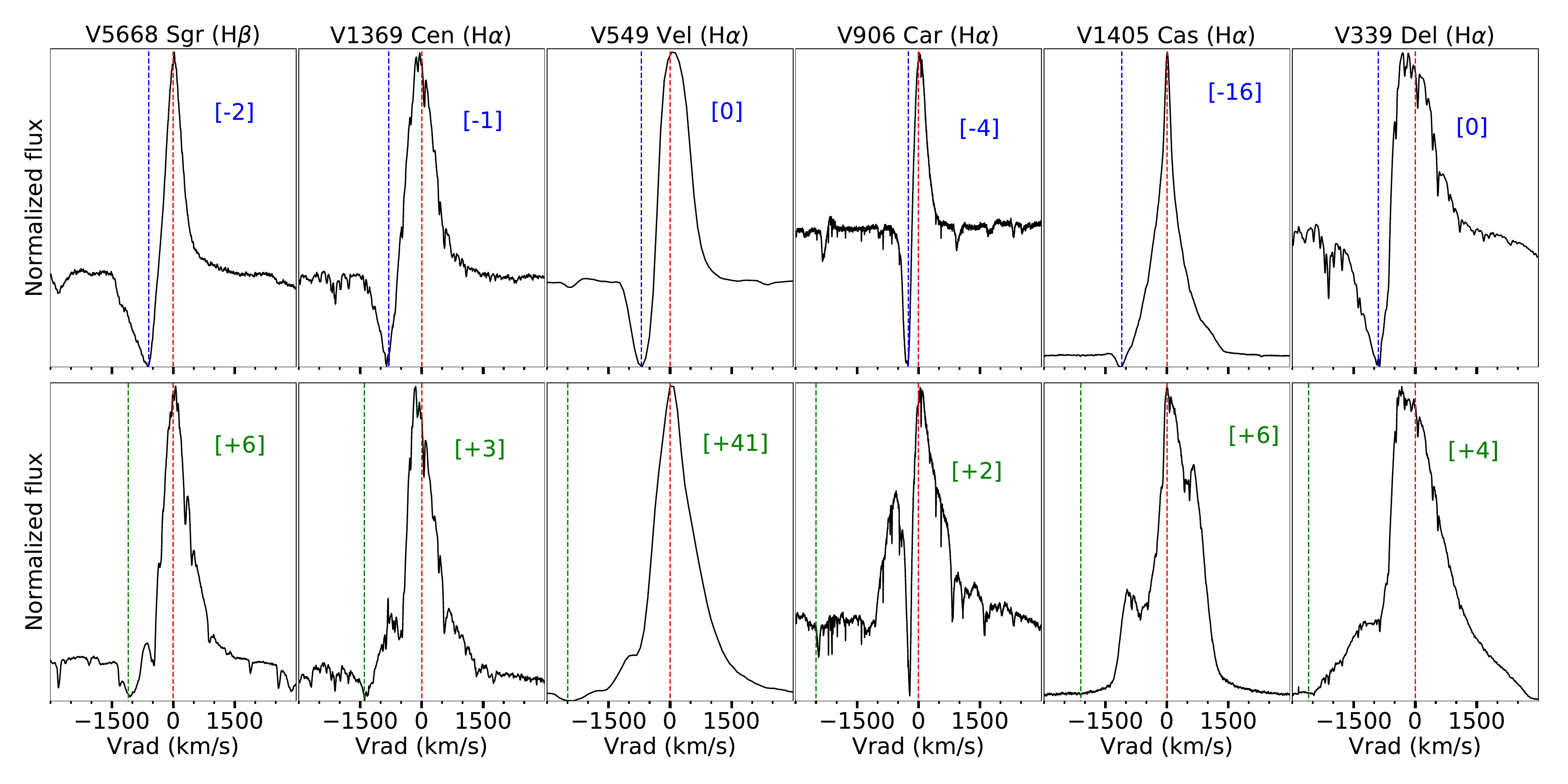}
\caption{The line profiles of H$\alpha$ or H$\beta$ before (\textit{top}) and after (\textit{bottom}) optical peak for the $\gamma$-ray detected novae V5668~Sgr, V1369~Cen, V549~Vel, V906~Car, V1405~Cas, and V339~Del. The red dashed lines represent rest velocity ($v_{\mathrm{rad}}$ = 0\,km\,s$^{-1}$). The blue and green dashed lines represent the velocities of the slow ($v_1$) and fast components ($v_2$), respectively; they are centered at the minima of the absorption features or the edge of the broad emission lines. The numbers in brackets are the day of observation relative to the optical peak ($t_{\mathrm{1}} - t_{\mathrm{max}}$ in the top row and $t_{\mathrm{2}} - t_{\mathrm{max}}$ in the bottom row). Heliocentric corrections are applied to the radial velocities in all the plots throughout the paper.}
\label{Fig:line_profiles_spec_gamma_slow}
\end{center}
\end{figure*}

\begin{figure*}
\begin{center}
  \includegraphics[width=\textwidth]{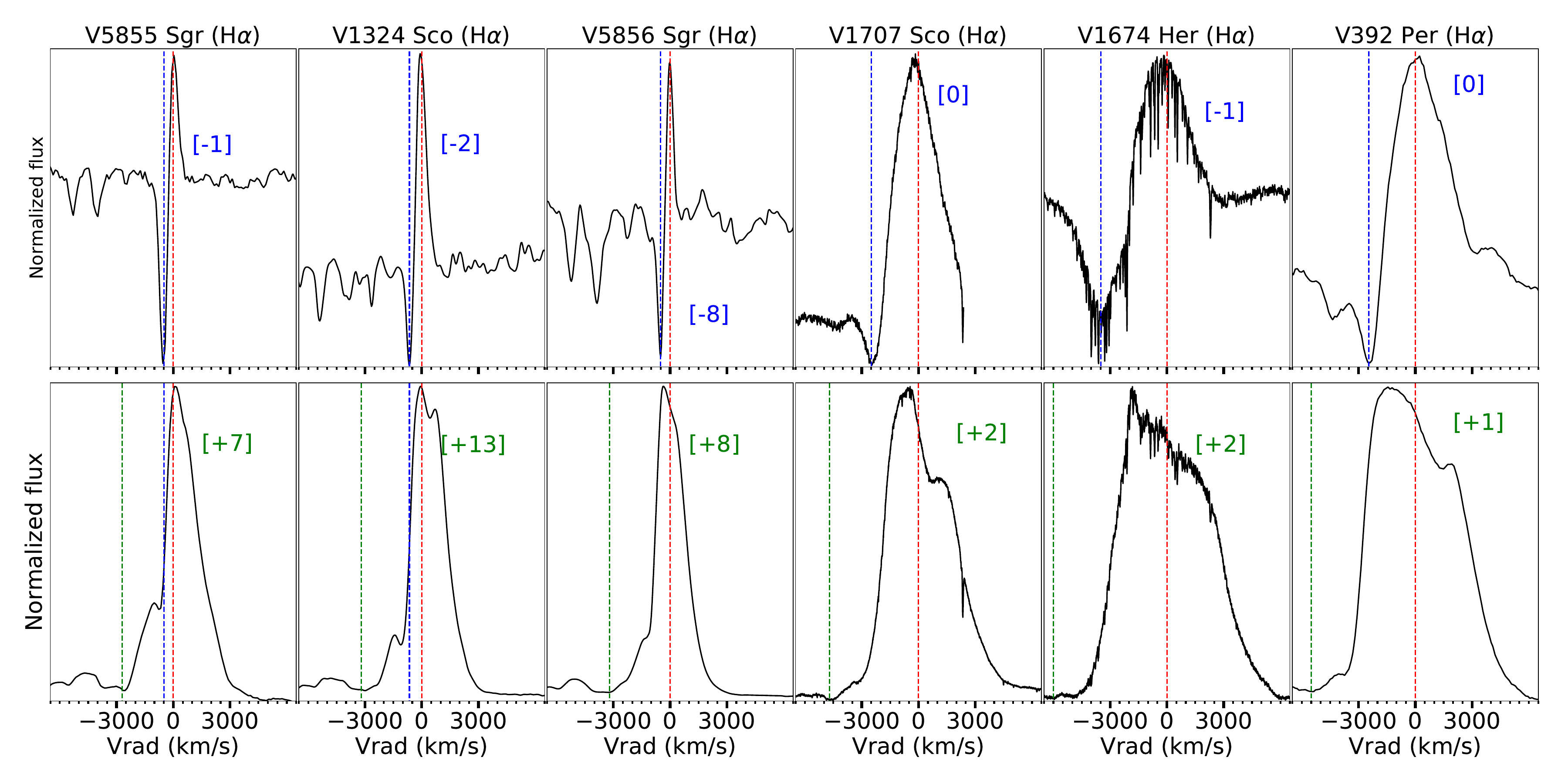}
\caption{same as Figure~\ref{Fig:line_profiles_spec_gamma_slow} but for the $\gamma$-ray detected novae V5855~Sgr, V1324~Sco, V5856~Sgr, V1707~Sco, V1674~Her, V392~Per.}
\label{Fig:line_profiles_spec_gamma_fast}
\end{center}
\end{figure*}

\begin{figure*}
\begin{center}
  \includegraphics[width=\textwidth]{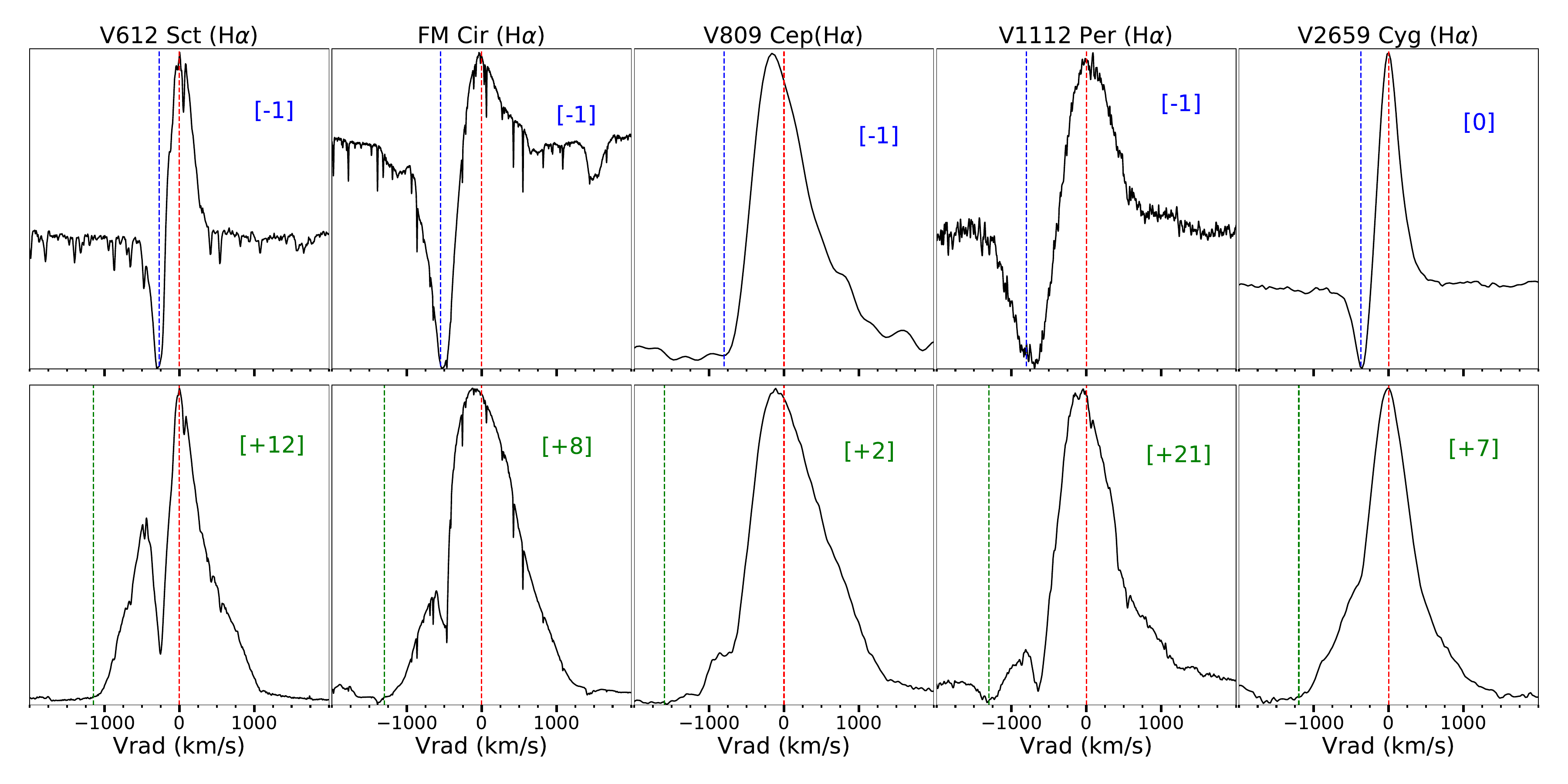}
\caption{same as Figure~\ref{Fig:line_profiles_spec_gamma_slow} but for $\gamma$ ray non-detected novae V612~Scr, FM~Cir, V809~Cep, V1112~Per, and V2659~Cyg.}
\label{Fig:line_profiles_spec_non_gamma_slow}
\end{center}
\end{figure*}

\begin{figure*}
\begin{center}
  \includegraphics[width=\textwidth]{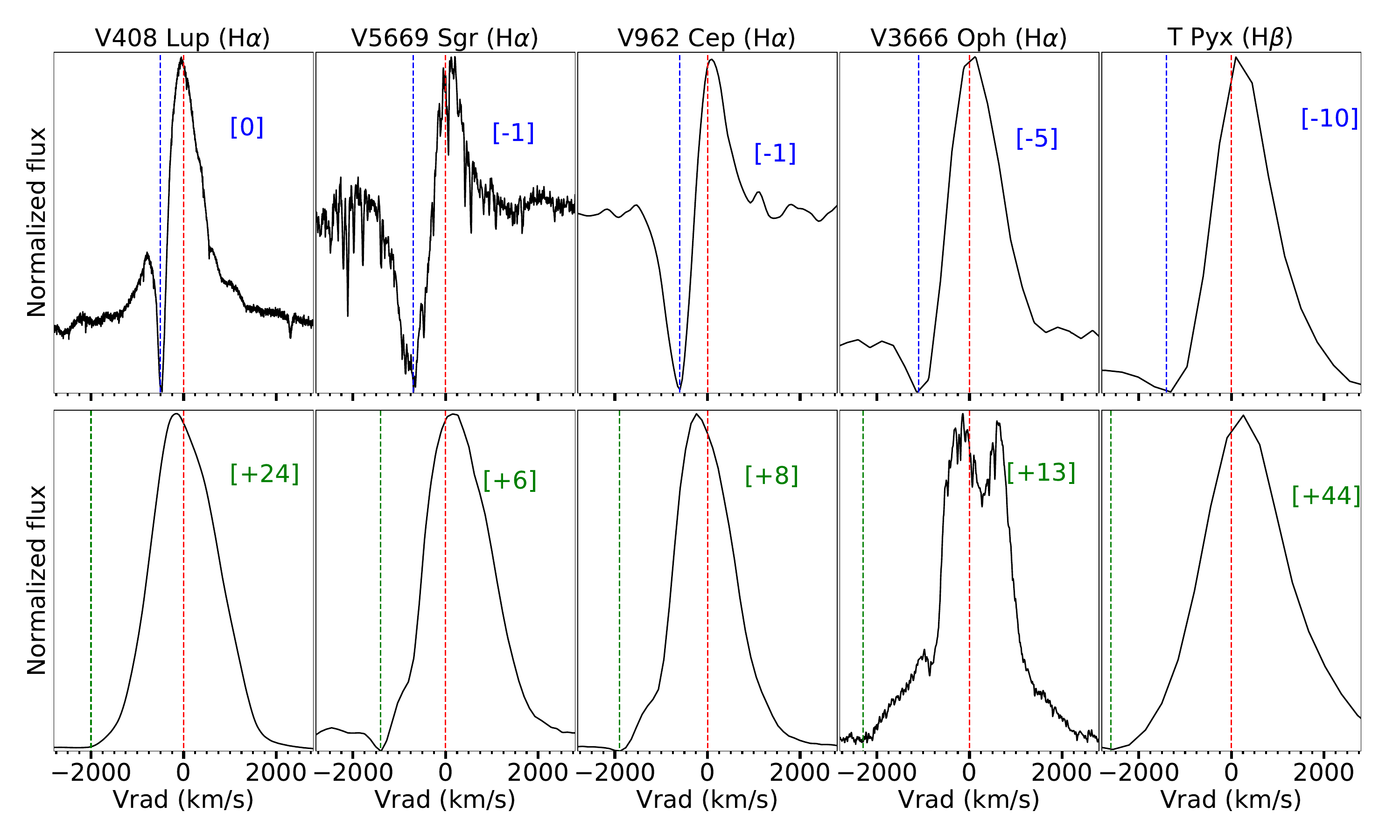}
\caption{same as Figure~\ref{Fig:line_profiles_spec_gamma_slow} but for $\gamma$-ray non-detected novae V408~Lup, V5669~Sgr, V962~Cep, V3666~Oph, and T~Pyx.}
\label{Fig:line_profiles_spec_non_gamma_mod}
\end{center}
\end{figure*}

\begin{figure*}
\begin{center}
  \includegraphics[width=\textwidth]{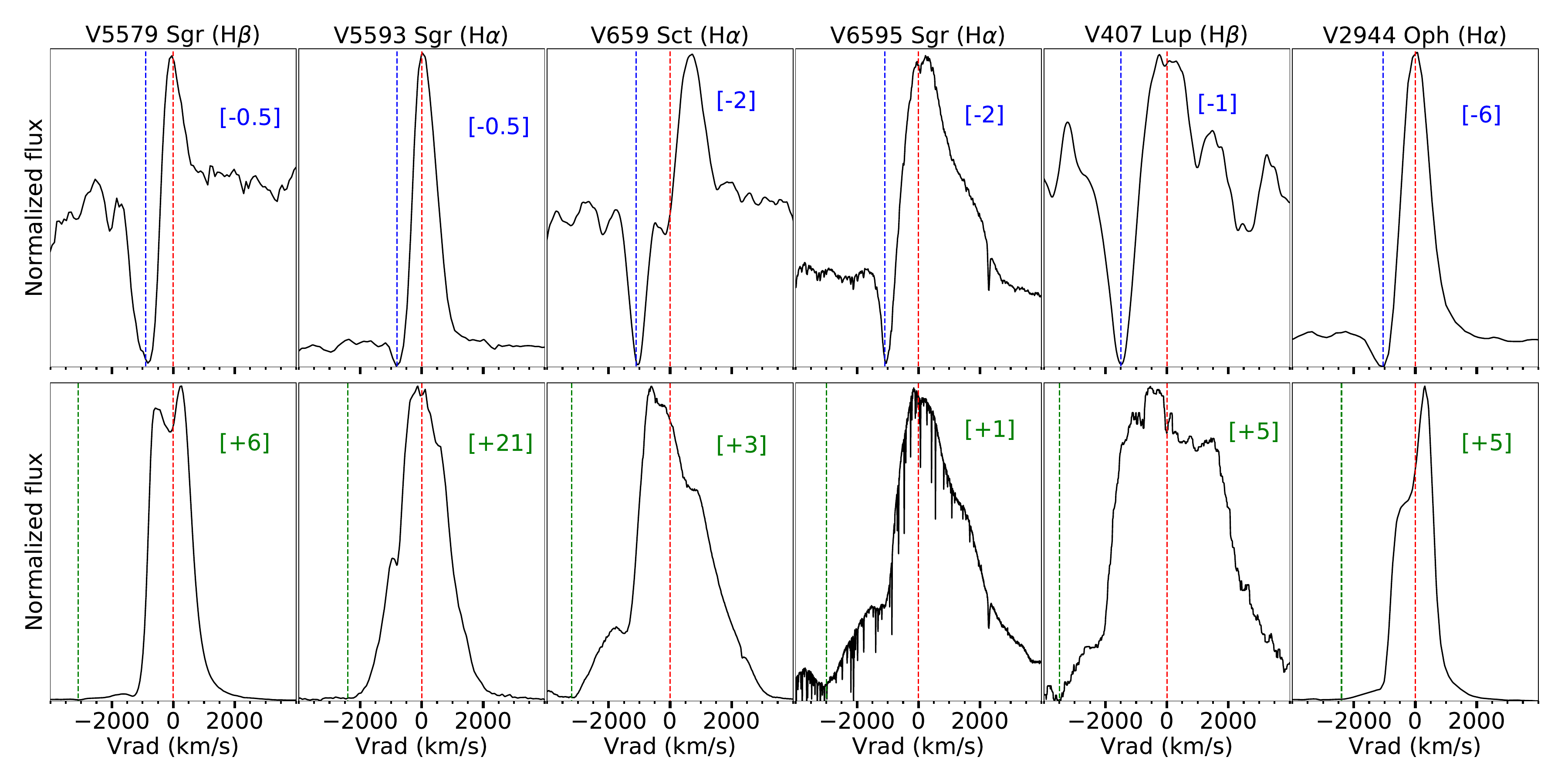}
\caption{same as Figure~\ref{Fig:line_profiles_spec_gamma_slow} but for $\gamma$-ray non-detected novae V5579~Sgr, V5593~Sgr, V659~Sct, V6595~Sgr, V407~Lup, and V2944~Oph.}
\label{Fig:line_profiles_spec_non_gamma_fast}
\end{center}
\end{figure*}

\begin{figure*}
\begin{center}
  \includegraphics[width=\textwidth]{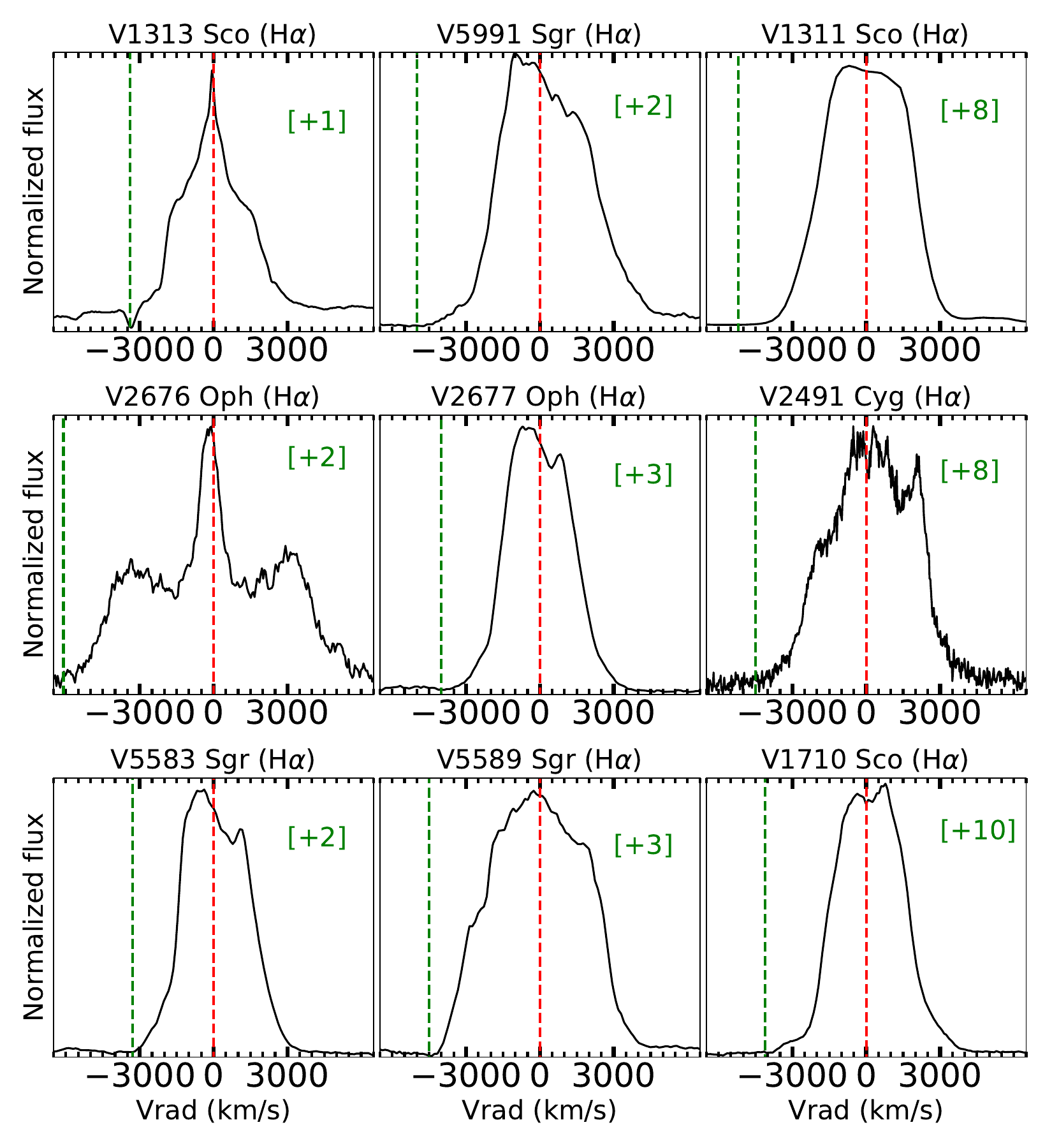}
\caption{The line profiles of H$\alpha$ or H$\beta$ after optical peak for the $\gamma$-ray non-detected novae V1313~Sco, V5591~Sgr, V1311~Sco, V2676~Oph, V2677~Oph, V2491~Cyg, V5583~Sgr, V5589~Sgr, and V1710~Sco. The red dashed lines represent rest velocity ($v_{\mathrm{rad}}$ = 0\,km\,s$^{-1}$). The green dashed lines represent the velocity of the fast component; they are centered at the minima of the absorption features or the edge of the broad emission. The numbers in brackets are the day of observation relative to the optical peak ($t_{\mathrm{2}} - t_{\mathrm{max}})$.}
\label{Fig:line_profiles_non_gamma_cyan_1}
\end{center}
\end{figure*}   

\begin{figure*}
\begin{center}
  \includegraphics[width=\textwidth]{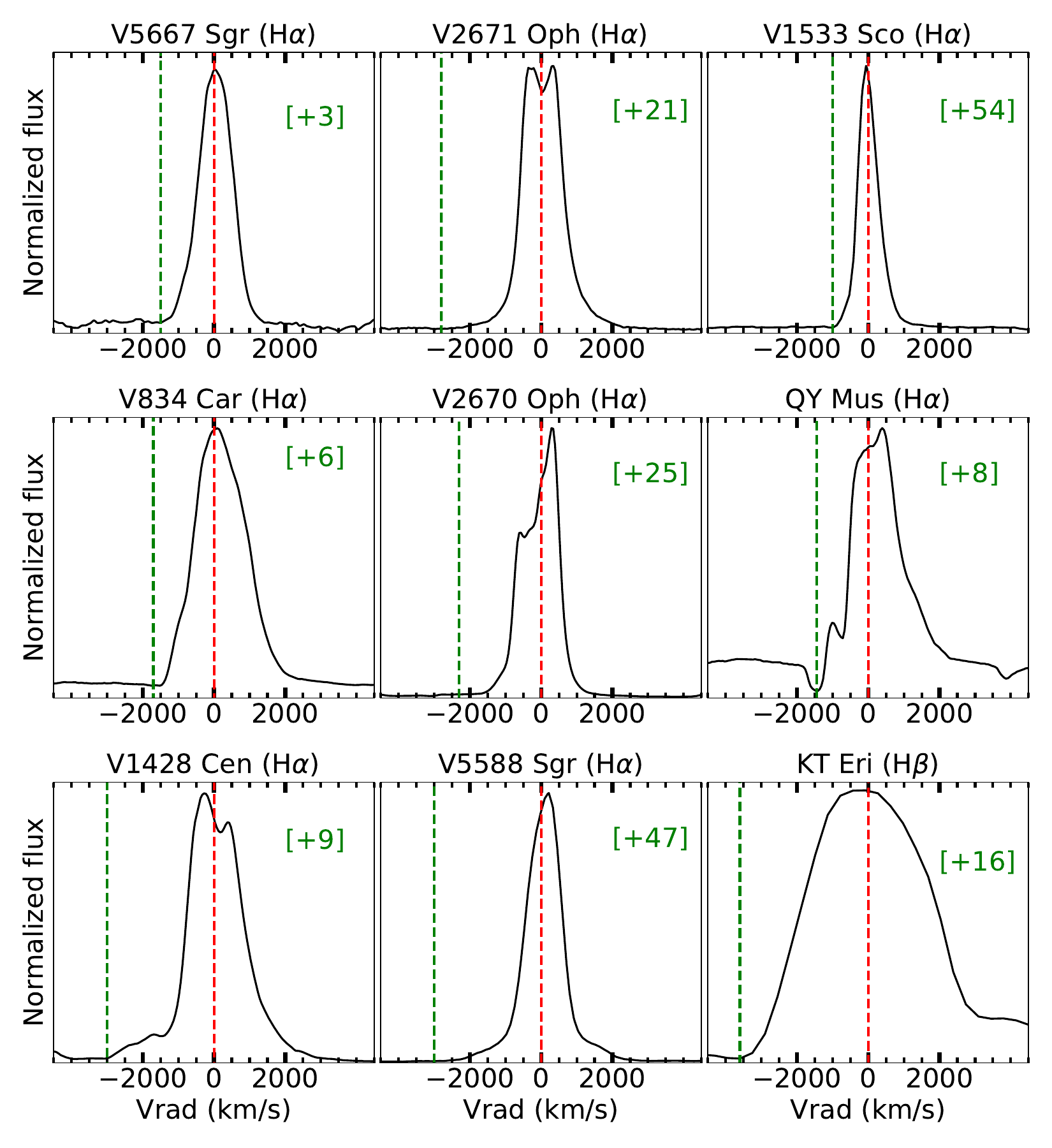}
\caption{Same as Figure~\ref{Fig:line_profiles_non_gamma_cyan_1} but for the $\gamma$-ray non-detected novae V5667~Sgr, V2671~Oph, V1533~Sco, V834~Car, V2670~Oph, QY~Mus, V1428~Cen, V5588~Sgr, and KT~Eri.}
\label{Fig:line_profiles_non_gamma_cyan_2}
\end{center}
\end{figure*}

\begin{figure*}
\begin{center}
  \includegraphics[width=\textwidth]{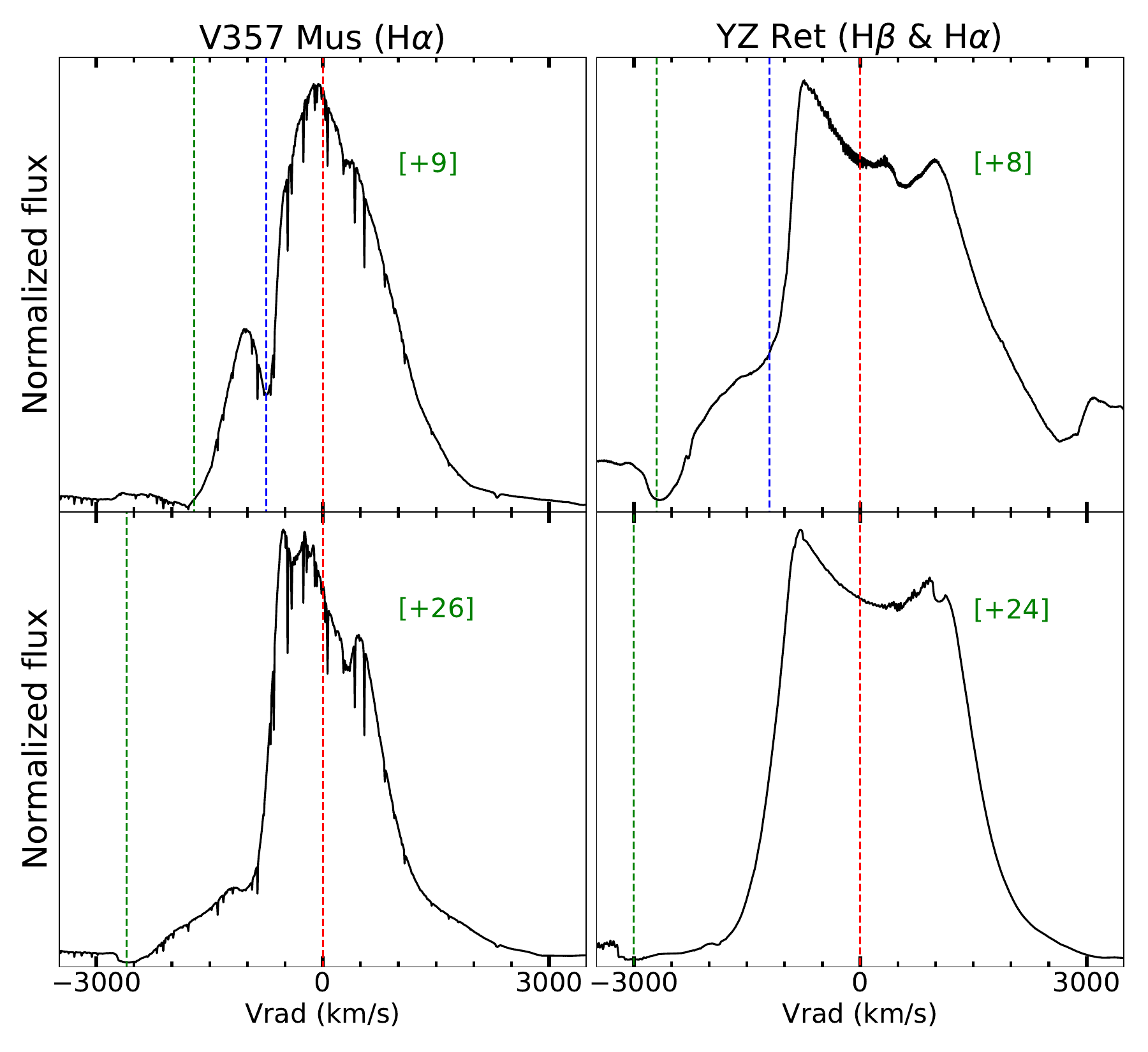}
\caption{same as Figure~\ref{Fig:line_profiles_spec_gamma_slow} but for $\gamma$-ray-detected novae V357~Mus and YZ~Ret, with spectra taken after peak (see text for more details).}
\label{Fig:line_profiles_Mus_Ret}
\end{center}
\end{figure*}

\bsp     
\label{lastpage}

\end{document}